\documentclass[11pt]{article}
\usepackage[a4paper]{geometry}
\usepackage[utf8]{inputenc}
\usepackage{graphicx}
\usepackage{dsfont}
\usepackage{fancyhdr}
\usepackage{booktabs}
\usepackage[colorlinks=true, allcolors=blue,pagebackref=true]{hyperref}
\usepackage{url}
\usepackage{xcolor}
\usepackage{colortbl}
\usepackage{subfig}
\usepackage{mathtools}
\usepackage[T1]{fontenc}
\usepackage{comment}
\usepackage{natbib}
\usepackage{eurosym}
\usepackage{babel}
\usepackage{amsmath}
\allowdisplaybreaks
\usepackage{xcolor}
\usepackage{enumitem}
\usepackage{float}
\usepackage{here}
\usepackage{wrapfig}
\usepackage{lscape}
\usepackage{bbold}
\usepackage{stmaryrd}
\usepackage[export]{adjustbox}
\usepackage{tabularx}
\usepackage{amssymb}
\usepackage{mathtools}
\usepackage{multirow,makecell, framed}
\usepackage[onelanguage, boxruled, vlined]{algorithm2e}
\SetAlFnt{\footnotesize}
\usepackage{amsthm, bm}
\usepackage{lscape}
\usepackage{ulem}
\usepackage{vmargin}
\usepackage{siunitx}

\newcommand{\dint}{\displaystyle\int}
\newcommand{\dsum}{\displaystyle\sum}

\newtheorem{lemma}{Lemma}

\newtheorem{theorem}{Theorem}
\newtheorem{definition}{Definition}

\theoremstyle{remark}

\usepackage{authblk}

\renewcommand*{\backref}[1]{}
\renewcommand*{\backrefalt}[4]{}

\setmarginsrb{2cm}{1cm}{2cm}{1cm}{0.5cm}{0.5cm}{0.5cm}{0.5cm}

\title{A signature-based spatial scan statistic for functional data}
\author[,1]{Camille Frévent\thanks{Corresponding author: \texttt{camille.frevent@univ-lille.fr}}}
\affil[1]{Univ. Lille, CHU Lille, ULR 2694 - METRICS: Évaluation des technologies de santé et des pratiques médicales, F-59000 Lille, France}

\date{}

\begin{document}

\maketitle

\hrule
\section*{Abstract}
We have developed a new signature-based spatial scan statistic for functional data (SigFSS). This scan statistic can be applied to both univariate and multivariate functional data. In a simulation study, SigFSS almost always performed better than the literature approaches and yielded more precise clusters in geographic terms. Lastly, we used SigFSS to search for spatial clusters of abnormally high or abnormally low mortality rates in mainland France. \\[0.2cm]
\textbf{Keywords:} Cluster detection, Functional data, Signature, Spatial scan statistics \vspace{0.5cm}
\hrule

\section{Introduction}

In many applications, researchers seek to determine the presence or absence of aggregations of spatial units (namely spatial clusters) that behave unusually. 
Various categories of spatial cluster detection tests have been developed. For example, non-focused cluster detection tests can usefully detect statistically significant spatial clusters without the \textit{a priori} need for information about the latter's location. Within this category, spatial scan statistics use a scanning window to detect statistically significant spatial clusters of various sizes and shapes without pre-selection bias. \\
Accordingly, spatial scan statistics have been applied to many fields of applied research, such as epidemiology \citep{costa2024space, jiao2024spatial, wang2025spatiotemporal}, environmental sciences \citep{social2, social1}, and criminology \citep{saravag2024application, de2024application}. \\
 
These methods were initially introduced by \cite{spatialdisease} and \cite{spatialscanstat} for Bernoulli and Poisson data distributions but have since been extended to many other spatial data distributions, such as Gaussian \citep{normalkulldorff}, ordinal \citep{jung2007spatial} and zero-inflated \citep{Canado2011AZP, de2015spatial, canccado2017bayesian} models. In the context of multivariate data, \cite{kulldorffmulti} developed a parametric approach based on the combination of independent univariate scan statistics. This method does not, however, take account of correlations between variables. This obstacle was circumvented by \cite{a_multivariate_gaussian} and \cite{cucala2019multivariate}, who developed spatial scan statistics designed specifically for multivariate data by taking account of correlations between variables. \\

Thanks to advances in sensing technology and data storage capacities, data are increasingly being measured continuously over time. This prompted the introduction of functional data analysis by \cite{ramsaylivre} and the adaptation of classical statistical methods (such as clustering \citep{jacques2014functional, zhang2023review}, principal component analysis (PCA) \citep{fpc_boente, berrendero2011principal} and regression \citep{reg_cuevas,  reg_ferraty, reg_chiou}) to a functional framework. \\
In domains where data naturally involve a spatial component (e.g., demographics,
environmental science, and agricultural science \citep{Hung_16}), the emergence of functional data has led to the introduction of spatial functional data and the development of new methods for clustering \citep{giraldo2012hierarchical, vandewalle2022clustering}, kriging \citep{giraldo2011ordinary, ignaccolo2014kriging}, PCA \citep{liu2017functional, si2025principal} or regression \citep{attouch2011robust, bernardi2017penalized}.  \\

In the field of spatial scan statistics, the detection of spatial clusters characterized by abnormal observed values of a stochastic process is an active research topic.
\cite{frevent2021detecting} and \cite{smida2022wilcoxon} first developed parametric and nonparametric approaches for the detection of spatial clusters on univariate functional data; the methods were based on an analysis of variance, Student's t-test, and Wilcoxon-Mann-Whitney test statistics, respectively. Recently, \cite{smida2025hotelling} developed a new approach based on a Hotelling test statistic.
Furthermore, \cite{frevent2023investigating} developed three new approaches suited to use with a multivariate, functional data framework. \\

In recent years, signatures - initially defined by \cite{chen1957integration, chen1977iterated} for smooth paths and rediscovered in the context of rough path theory \citep{lyons1998differential, friz2010multidimensional} - have been widely used in many domains, such as character recognition \citep{graham2013sparse,liu2017ps,7995142}, medicine \citep{perez2018signature, morrill2020utilization}, and finance \citep{gyurko2013extracting, arribas2018derivatives, perez2020signatures}. \cite{fermanian2022functional} recently investigated their use in the context of a linear regression model with functional covariates. \cite{frevent2025signature} and \cite{frevent2024multivariate} extended this approach to a spatial autoregressive model with functional covariates, in the context of an univariate outcome and a multivariate outcome, respectively. \\
For the analysis of functional data, signatures offer several advantages over classical approaches: they are (i) applicable to a wide range of processes that are not necessarily square-integrable, (ii) naturally suited to use with multivariate functions, and (iii) suitable for encoding a process's geometric properties \citep{fermanian2022functional}. \\

Here, we describe a new, signature-based spatial scan statistic for use with both univariate and multivariate functional data. \\
Section \ref{sec:sig} serves as an introduction to the notion of signatures. Section \ref{sec:method} describes our new approach to the detection of spatial clusters in functional data. Section \ref{sec:simu} presents both the design and the results of simulation studies in which our new approach was compared with existing methods in univariate and multivariate functional contexts. The results of application of our method to a real dataset are presented in Section \ref{sec:appli}. Lastly, our results are discussed in Section \ref{sec:discussion}.

\section{The concept of signatures} \label{sec:sig}

This section provides a brief introduction to the concept of signatures. For a more comprehensive description, we refer the reader to \cite{lyons2007differential,friz2010multidimensional}. It should be borne in mind that although signatures were initially formulated for smooth paths, we use them in the context of functional data and thus, for ease of understanding, employ the associated notations.

Let $X: \mathcal{T} \rightarrow \mathbb{R}^p$ be a $p$-dimensional continuous function of bounded variation, that is
$$
\left\| X\right\|_{TV} = \sup_{(t_0,...,t_k)\in \mathcal{I}} \sum_{i=1}^k \left\| X(t_i) - X(t_{i-1}) \right\| < +\infty,
$$
where $\left\| .\right\|_{TV}$ denotes the total variation distance, $\left\| .\right\|$ denotes the Euclidean norm on $\mathbb{R}^p$, and $\mathcal{I}$ denotes the set of all finite partitions of $\mathcal{T}$. 

\begin{definition}
The signature of $X$ is defined by
$$ Sig(X) = (1, \bm{X}^1, \dots, \bm{X}^d, \dots) $$
where $$ \bm{X}^d = \underset{\substack{t_1 < \dots < t_d \\ t_1, \dots, t_d \in \mathcal{T}}}{\dint \dots \dint} \ \text{d}X(t_1) \otimes \dots \otimes \text{d}X(t_d) \in \left(\mathbb{R}^p \right)^{\otimes d}. $$

\end{definition}

\begin{definition}
The truncated signature of $X$ at order $D \ge 1$ is
$$ Sig^D(X) = (1, \bm{X}^1, \dots, \bm{X}^D). $$
\end{definition}

\begin{definition}
The truncated signature coefficient vector of $X$ at order $D$ is defined as 
$$
S^D(X)=\left(1,S_{(1)}(X), \dots, S_{(p)}(X), S_{(1,1)}(X), S_{(1,2)}(X), \dots, S_{\underbrace{ \scriptstyle(p, \ldots, p)}_{\text{length } D}}(X) \right),
$$
and the truncated shifted-signature coefficient vector of $X$ at order $D$ is defined as follows
$$
\widetilde{S}^D(X)=\left(S_{(1)}(X), \dots, S_{(p)}(X), S_{(1,1)}(X), S_{(1,2)}(X), \dots, S_{\underbrace{ \scriptstyle(p, \ldots, p)}_{\text{length } D}}(X)\right),
$$
where for all $d \ge 1$ and for all multi-index , $(i_1, \dots, i_d) \subset\{1, \dots, p\}^d$, 
$$
S_{(i_1, \dots, i_d)}(X) = \underset{\substack{t_1 < \dots < t_d \\ t_1, \dots, t_d \in \mathcal{T}}}{\dint \dots \dint} \ \text{d}X^{(i_1)}(t_1) \dots \text{d}X^{(i_d)}(t_d) \in \mathbb{R}.
$$
Next, the truncated shifted-signature coefficient vector of $X$ at order $D$ is a vector of length  
$$
\widetilde{s}_p(D) = \sum_{d = 1}^D p^d = \frac{p^{D+1}-p}{p-1} \text{ if } p \ge 2 \text{ and } \widetilde{s}_p(D) = D \text{ if } p = 1.
$$
\end{definition}

\begin{lemma}
Let $\Psi: \mathcal{T} \rightarrow \mathcal{T}$ be a non-decreasing surjection and 
\begin{align*}
Y: \mathcal{T} & \longrightarrow \mathbb{R}^p \\
t & \longmapsto X({\Psi(t)}).
\end{align*}
Next, $Sig(X) = Sig(Y)$: the signature of $X$ is invariant by time reparameterization.
\end{lemma}

\begin{lemma}
Let 
\begin{align*}
Y: \mathcal{T} & \longrightarrow \mathbb{R}^p \\
t & \longmapsto X(t) + a, a \neq 0.
\end{align*}
Next, $Sig(X) = Sig(Y)$: the signature of $X$ is invariant by translation.
\end{lemma}

\begin{lemma} \label{lemma:avoidinvariance}
It is possible to (i) circumvent the invariance by translation by adding an observation point taking the value 0 at the beginning of $X$, and (ii) avoid the invariance by time reparameterization by considering the time-augmented function 
\begin{align*}
\widetilde{X}: \mathcal{T} & \longrightarrow \mathbb{R}^{p+1} \\
t & \longmapsto \left(X(t)^\top, t\right)^\top
\end{align*}
instead of $X$.
\end{lemma}

\section{Methodology} \label{sec:method}

\subsection{General principle}

Let $X: \mathcal{T} \rightarrow \mathbb{R}^p$ be a $p$-dimensional continuous stochastic process of bounded variation and $s_1, \dots, s_n$ be $n$ non-overlapping spatial locations of an observation domain $\mathcal{S} \subset \mathbb{R}^2$.
Let $X_1, \dots, X_n$ be the observations of $X$ in $s_1, \dots, s_n$. Hereafter, all observations are considered to be independent; this is a classical assumption in scan statistics. 

Spatial scan statistics are designed to detect spatial clusters and then test the latter's statistical significance. 
Hence, one tests a null hypothesis $\mathcal{H}_0$ (the absence of a
cluster) against a composite alternative hypothesis $\mathcal{H}_1$ (the presence of at least one spatial cluster $w \subset \mathcal{S}$ presenting abnormal values of $X$, compared with the rest of the domain).

According to \cite{cressie}, a scan statistic is the maximum of a concentration index over a set of potential clusters $\mathcal{W}$. 
In the following and without loss of generality, we focus on the variable-size circular clusters introduced by \cite{spatialscanstat}. The set of potential circular clusters $\mathcal{W}$ is defined by the following set of discs $w_{i,j}$ centered on $s_i$ and passing through $s_j$:
$$ \mathcal{W} = \{ w_{i,j} / 1 \le |w_{i,j}| \le n, 1 \le i,j \le n \}, $$ where $|w_{i,j}|$ is the number of sites in $w_{i,j}$.
Thus, as recommended by \cite{spatialdisease}, a cluster cannot cover more than 50\% of the studied region.

\subsection{The signature-based spatial scan statistic for functional data}

\subsubsection{Computation of the signatures}

In the signature-based spatial scan statistic for functional data (SigFSS), we first calculated the truncated signature of $X$ in each spatial location. The objective is to detect the spatial clusters in which the distribution of $X$ is abnormal; temporal concordance is important but the height of the curves (relative to the others) is also important. Thus, invariance by translation and by time reparameterization are not desirable, and we used the strategy described in Lemma \ref{lemma:avoidinvariance} to avoid them.

Furthermore, in order to encode the maximum amount of available information on $X_1, \dots, X_n$, the truncation order is important. Here and in the following, we shall set a truncation order $D$ such that $\widetilde{s}_{p+1}(D)$ is at most $10^4$, similar to \cite{frevent2025signature}.

At the end of this step, the dataset consisted of $\widetilde{S}^D(\widetilde{X}_1), \dots, \widetilde{S}^D(\widetilde{X}_n)$ where 
$\widetilde{X}_i$ is defined as
\begin{align*}
\widetilde{X}_i: \mathcal{T} & \longrightarrow \mathbb{R}^{p+1} \\
t & \longmapsto \left(X_i(t)^\top, t\right)^\top,
\end{align*}
after adding the value 0 at the beginning of $X_i$.

\subsubsection{Reduction of dimension}

Encoding the functions $X_1, \dots, X_n$ with truncated signatures transformed them into vectors of size $\widetilde{s}_{p+1}(D) \le 10^4$. This value is generally very large and greater than $n$. Hence, it is therefore mathematically impossible to apply spatial scan statistics to the vectors without reducing their dimension.

Thus, the second step in the SigFSS approach was to reduce the dimension of $\widetilde{S}^D(\widetilde{X}_1), \dots, \widetilde{S}^D(\widetilde{X}_n)$ using a PCA. \\

The data $\widetilde{S}^D(\widetilde{X}_1), \dots, \widetilde{S}^D(\widetilde{X}_n)$ is thus transformed to $\zeta_1^{(D,K)}, \dots, \zeta_n^{(D,K)}$ where 
$$ \zeta_i^{(D,K)} = \left(\zeta_{i,1}^{(D)}, \dots, \zeta_{i,K}^{(D)}\right)^\top$$ is the vector of the $K$ first scores computed for $\widetilde{S}^D(\widetilde{X}_i)$ on the $K$ first axes. \\

The number $K$ of principal components was chosen as a function of the cumulative inertia criterion. 
Two approaches can be considered: $K$ is determined such that the cumulative inertia (i) reaches a fixed threshold (95\%, for example, as considered in \cite{ahmed2020functional}), or (ii) shows a marginal increase towards 1 \citep{smida2025hotelling}.

\subsubsection{Computing the spatial scan statistic}

Once the size of the signatures has been reduced through the use of principal scores, we consider the following test hypotheses:
\begin{center}
$\mathcal{H}_0$: $\zeta_1^{(D,K)}, \dots, \zeta_n^{(D,K)}$ are identically distributed,
whatever their associated locations $s_1, \dots, s_n$ \\
vs. \\
$\mathcal{H}_1$: There exists $w \in \mathcal{W}$ such that the distribution of $\zeta^{(D,K)}$ in $w$ differs from that of $\zeta^{(D,K)}$ outside $w$. \\
\end{center}

In the context of multivariate data, \cite{cucala2019multivariate} developed a nonparametric spatial scan statistic that relies on the extension of the Wilcoxon-Mann-Whitney test statistic for multivariate data, as defined by \cite{oja2004multivariate}.

Let
$ \text{sgn}_K(\cdot) $ be the multivariate sign function defined as
$$ \begin{array}{rccl}
\text{sgn}_K: & \mathbb{R}^K & \rightarrow & \mathbb{R}^K \\
& z & \mapsto & ||z||_2^{-1} z \ \mathds{1}_{z \neq \bm{0}_K}
\end{array} $$
with $\bm{0}_K$ the $K$-dimensional vector composed only of 0, and let 
$$ R_i = \dfrac{1}{n} \dsum_{j=1}^n \text{sgn}_K\left[A_{\zeta^{(D,K)}} \left( \zeta_i^{(D,K)} - \zeta_j^{(D,K)} \right) \right] $$ be the multivariate ranks, where $A_{\zeta^{(D,K)}}$ is a data-based transformation matrix that makes the multivariate ranks behave as though they were spherically distributed in the unit $K$-sphere:
$$ \dfrac{K}{n} \dsum_{i=1}^n R_i R_i^\top = \dfrac{1}{n} \dsum_{i=1}^n R_i^\top R_i I_K. $$

To test
\begin{center}
$\mathcal{H}_0: $ The distribution of $\zeta^{(D,K)}$ is the same inside and outside $w$ \\
vs. \\
$\mathcal{H}_1: $ The distribution of $\zeta^{(D,K)}$ in $w$ differs from that of $\bm{Z}$ outside $w$,
\end{center}
we can consider the following test statistic developed by \cite{oja2004multivariate}: 
$$ W^{(w)} = \dfrac{Kn}{\dsum_{i=1}^n R_i^\top} \left( |w| || \bar{R}_w ||_2^2 + |w^\mathsf{c}| || \bar{R}_{w^\mathsf{c}} ||_2^2 \right) $$

where $\bar{R}_g = \dfrac{1}{|g|} \dsum_{\substack{i=1 \\ s_i \in g}}^n R_i, g \in \{w, w^\mathsf{c}\} $.

Next, in the same way as \cite{cucala2019multivariate}, we considered $W^{(w)}$ to be a concentration index for cluster detection and we defined the SigFSS statistic as
$$ \lambda_\text{SigFSS} = \underset{w \in \mathcal{W}}{\max} \ W^{(w)}. $$

The most likely cluster (MLC) was the potential cluster for which this maximum is obtained:
$$ \text{MLC}_\text{SigFSS} = \underset{w \in \mathcal{W}}{\arg \max} \ W^{(w)}. $$

\subsubsection{Statistical significance}

Once the MLC has been identified, its statistical significance must be evaluated. The distribution of the scan statistic is intractable under $\mathcal{H}_0$, due to the overlap between the two concentration indices $W^{(w)}$ and $W^{(w')}$ when $w \cap w' \neq \emptyset$. \\

Thus, we employed an approach based on random permutations (commonly referred to as ``random labelling'') to estimate the p-value associated with $\lambda_\text{SigFSS}$.

Let $P$ denote the number of random permutations performed on the original dataset, and $\lambda_\text{SigFSS}^{(1)}, \dots, \lambda_\text{SigFSS}^{(P)}$ denote the spatial scan statistics obtained from these permuted datasets. According to \cite{dwass}, the p-value for the scan statistic observed in the original data can be estimated as follows:
$$ \widehat{p} = \dfrac{1 + \dsum_{p'=1}^P \mathds{1}_{\lambda_\text{SigFSS}^{(p')} \ge \lambda_\text{SigFSS}}}{1+P}. $$ 
 
The MLC was considered to be statistically significant when its associated estimated p-value was smaller than the type I error.

\subsubsection{Secondary clusters}

In practice, it is sometimes necessary to detect secondary clusters. Here, we considered the approach of \cite{spatialscanstat}, which defines secondary clusters as geographical areas corresponding to a second maximum of the concentration index, then a third maximum, and so on. The p-value of the secondary clusters is calculated by comparing the values of their associated concentration index with the value of $\lambda_\text{SigFSS}^{(p')}$. It should be noted that this approach is rather conservative because for a secondary cluster to be declared statistically significant, it must be able to reject $\mathcal{H}_0$ \textit{per se}, as if it were the MLC itself. Lastly, only statistically significant secondary clusters that did not overlap with a more likely cluster were considered. 

\section{The simulation study} \label{sec:simu}

In a simulation study, we compared the new SigFSS with the DFFSS, the PFSS \citep{frevent2021detecting}, the URBFSS \citep{frevent2022r}, the NPFSS \citep{smida2022wilcoxon} and the HFSS \citep{smida2025hotelling} in the univariate case and with the MDFFSS, the MPFSS, the MRBFSS \citep{frevent2023investigating} and the NPFSS \citep{smida2022wilcoxon} in the multivariate case.

\subsection{Design of the simulation study}

Artificial datasets were generated for the geographical locations of the 94 \textit{départements} (administrative counties) in mainland France (see Figure \ref{fig:simulatedcluster} in the Supplementary Materials). The location of each \textit{département} was defined by its centroid. For each artificial dataset, a spatial cluster $w$ composed of eight \textit{départements} in the Paris region was simulated (the area highlighted in red in Figure \ref{fig:simulatedcluster} in Supplementary Materials). \\

We considered the cases $p = 1$ and $p = 2$.
The $X_i$ were simulated according to the following model considered in \cite{frevent2023investigating}, and were measured at 101 equally spaced times on $[0,1]$:
$$
X_i(t) = \mu + \Delta(t) \mathds{1}_{s_i \in w} + \varepsilon_{i}(t), \ t \in [0,1], 
$$
where 
$\mu = \sin{(2\pi t^2)}^5$ if $p=1$ and $\mu = (\sin{(2\pi t^2)}^5 , 1 + 2.3 t + 3.4 t^2 + 1.5 t^3)^\top$ if $p=2$,
$\varepsilon_{i}(t) = \dsum_{k=1}^{100} Z_{i,k} \sqrt{1.5 \times 0.2^k} \theta_k(t)$, with $\theta_k(t) = \left\{ \begin{array}{ll}
1 & \text{ if } k = 1 \\
\sqrt{2} \sin{[k\pi t]} & \text{ if } k \text{ even} \\
\sqrt{2} \cos{[(k-1)\pi t]} & \text{ if } k \text{ odd and } k > 1
\end{array} \right.$. \\

By denoting $\Sigma = \begin{pmatrix} 1 & \rho \\ \rho & 1 \end{pmatrix}$ the covariance matrix of the $Z_{i,k}$ when $p=2$, and considering $\Sigma = 1$ when $p=1$, three distributions of the $Z_{i,k}$ were used: (i) a normal distribution: $Z_{i,k} \sim \mathcal{N}(0,\Sigma)$, (ii) a standardized Student distribution: $Z_{i,k} = U_{i,k} \left( \dfrac{V_{i,4}}{4} \right)^{-0.5}$ where the  $U_{i,k}$ are independent $\mathcal{N}(0, \Sigma/2)$ variables and the $V_{i,4}$ are independent $\chi_2(4)$ variables, and (iii) a standardized exponential distribution: $Z_{i,k} = \left[ U_{i,k} - \begin{pmatrix} 2 \\ 2 \end{pmatrix} \right] / 2$ where the $U_{i,k}$ are independent and $U_{i,k} \sim \mathcal{E}(\frac{1}{2},\Sigma)$. \\

Three values of $\rho$ ($\rho = 0.2,0.5,0.8$) were tested for $p=2$, and four types of clusters (whose intensity was controlled by a parameter $\alpha > 0$) were studied: $\Delta_1(t) = \alpha (t , t)^\top$, $\Delta_2(t) = \alpha ( t (1-t) , t (1-t) )^\top$, $\Delta_3(t) = \alpha (\exp{( - 100 (t-0.5)^2 )}/3 , \exp{( - 100 (t-0.5)^2 )}/3)^\top$, and $\Delta_4(t) = \alpha (\cos{(4 \pi (t-0.5))} , \cos{(4 \pi (t-0.5))})^\top$, together with the respective univariate versions when $p=1$.  \\

Different values of the parameter $\alpha$ were considered for each $\Delta$: $\alpha \in \{ 0,0.25,0.5,0.75,1 \}$ for $\Delta_1$, $\alpha \in \{ 0,1,2,3,4 \}$ for $\Delta_2$, $\alpha \in \{ 0,0.5,1,1.5,2 \}$ for $\Delta_3$, $\alpha \in \{ 0,0.125,0.25,0.375,0.5 \}$ for $\Delta_4$. 

Note that $\alpha = 0$ was also tested, in order to evaluate the conservation of the nominal type I error. An example of the data for $p=1$ and for $p=2$ with $\rho = 0.2$ (when considering a Gaussian distribution for the $Z_{i,k}$) is given in Supplementary Materials (Figures \ref{fig:examplesimu_uni} and \ref{fig:examplesimu_multi}).

\subsection{Comparison of the methods}

\noindent In both the univariate and multivariate cases, and for each $Z_{i,k}$ distribution, we simulated each $\Delta$, each $\rho$ (in the multivariate case), and each value of $\alpha$ in 200 artificial datasets. A total of 999 samples were generated by random permutations of the data, and the type I error was set to 0.05. 
The methods' respective performance levels were compared with regard to four criteria: the power, the true positive rate, the false positive rate, and the positive predictive value. 

The power was estimated as the proportion of simulations leading to the rejection of $\mathcal{H}_0$, depending on the type I error. Among the simulated datasets leading to the rejection of $\mathcal{H}_0$, the true positive rate is the mean proportion of sites correctly detected among the sites in $w$, the false positive rate is the mean proportion of sites in $w^\mathsf{c}$ that were included in the detected cluster, and the positive predictive value corresponds to the mean proportion of sites in $w$ within the detected cluster. \\

For the SigFSS and the HFSS, we also compared the performance levels by choosing a value of $K$ such that the cumulative inertia showed a marginal increase towards 1 (graphic approach) or reached a fixed threshold of 95\% or 99\%.

\subsection{Results of the simulation study}

\begin{figure}[h!]
\begin{minipage}{0.49\linewidth}
\centering $\Delta_1$
\includegraphics[width=\linewidth]{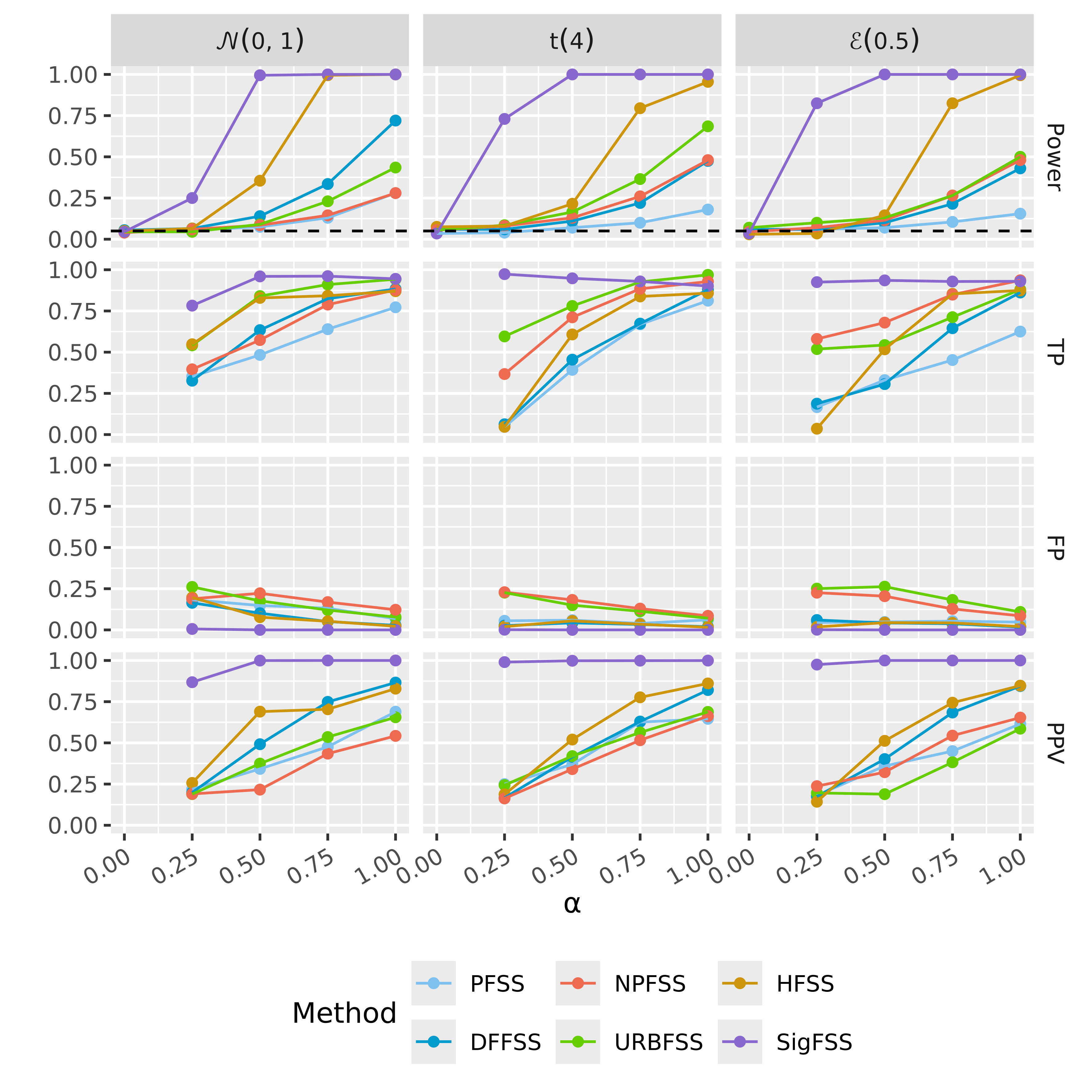}
\end{minipage}
\begin{minipage}{0.49\linewidth}
\centering $\Delta_2$
\includegraphics[width=\linewidth]{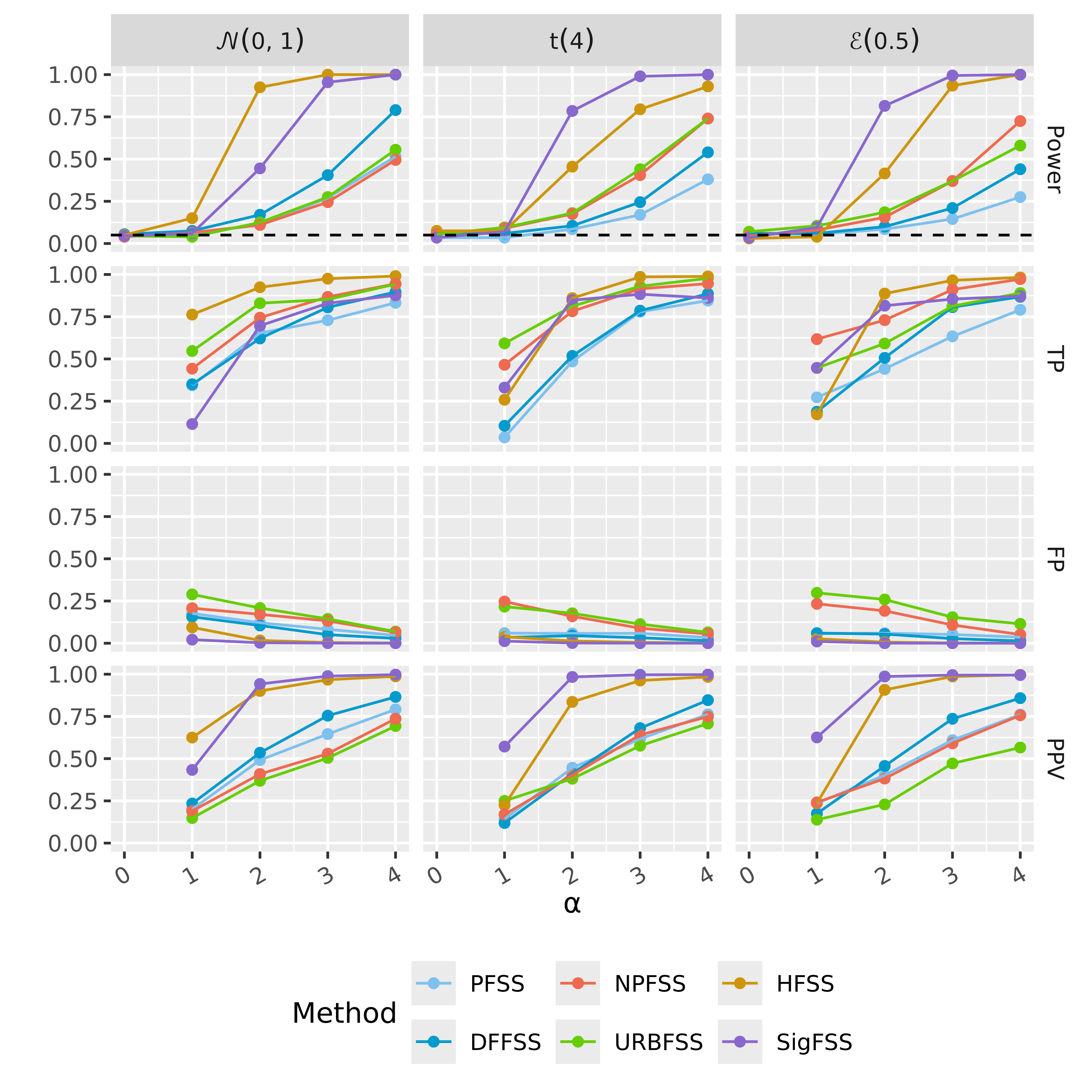}
\end{minipage}
\begin{minipage}{0.49\linewidth}
\centering $\Delta_3$
\includegraphics[width=\linewidth]{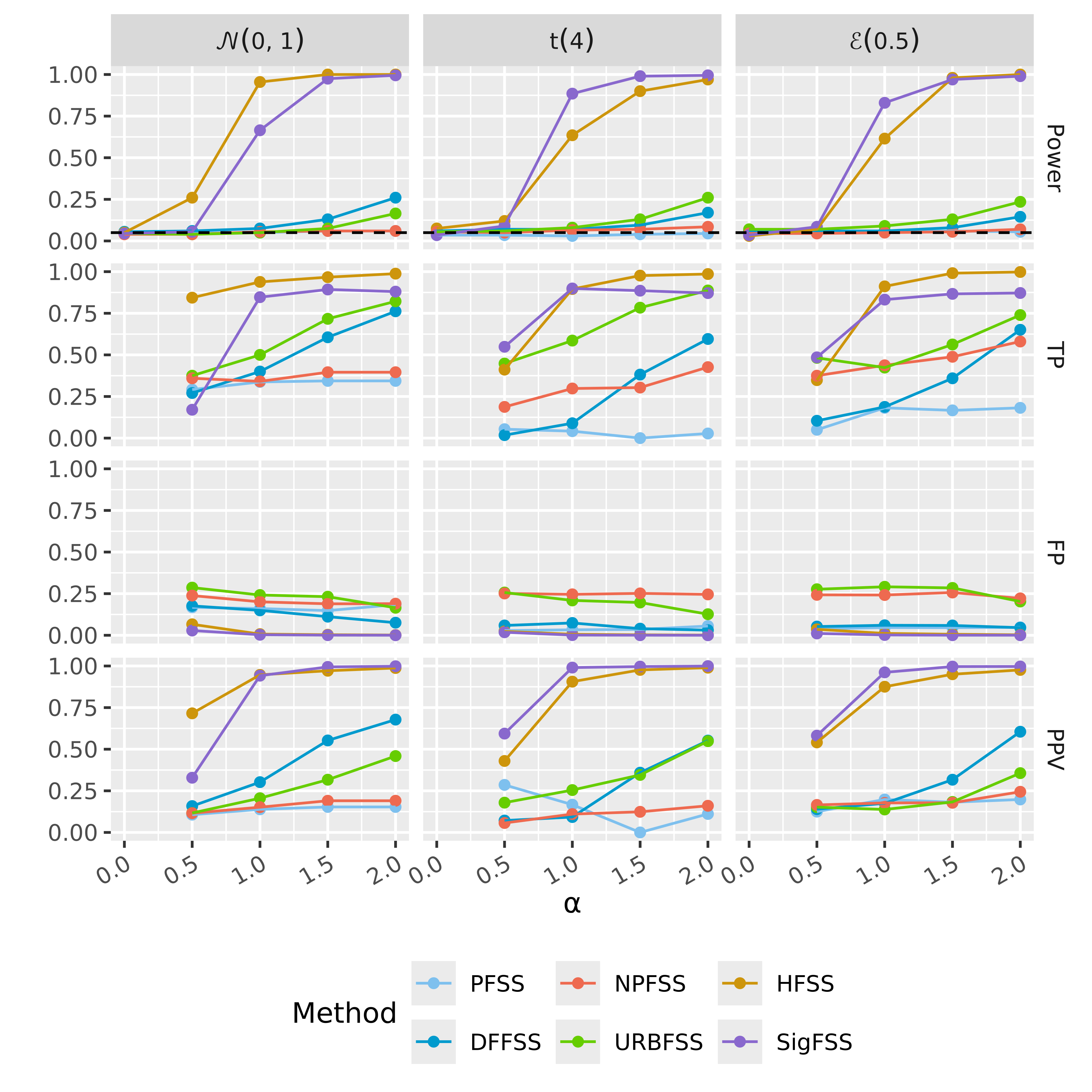}
\end{minipage}
\begin{minipage}{0.49\linewidth}
\centering $\Delta_4$
\includegraphics[width=\linewidth]{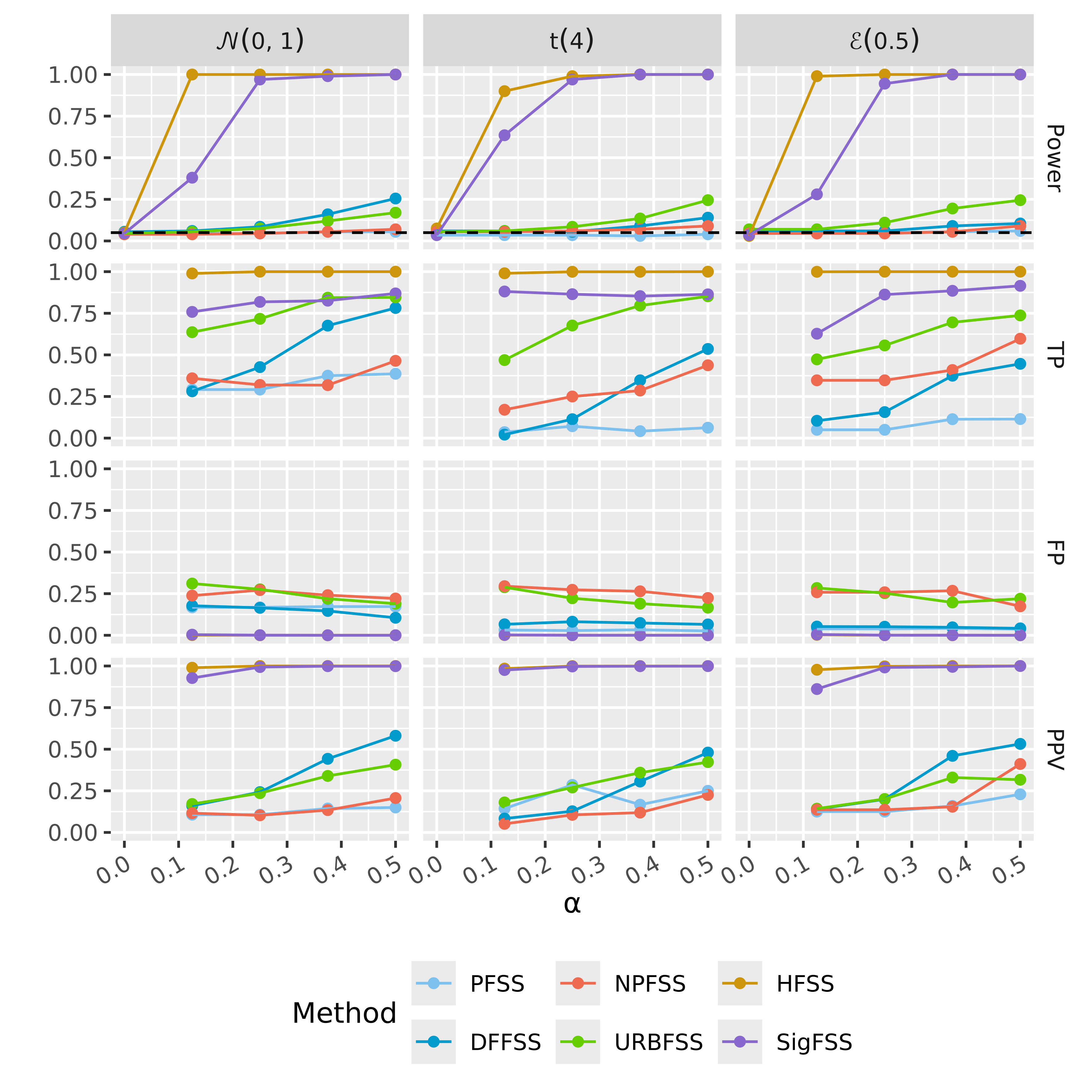}
\end{minipage}
\caption{The simulation study (univariate case): a comparison of the SigFSS, HFSS, NPFSS, DFFSS, URBFSS, and PFSS methods for detection of the spatial cluster as the MLC, when considering a graphical approach for choosing $K$ in the SigFSS and the HFSS. $\alpha$ is the parameter that controls the cluster intensity.}
\label{fig:uninothre}
\end{figure}

\begin{figure}[h!]
\begin{minipage}{0.49\linewidth}
\centering $\Delta_1$
\includegraphics[width=\linewidth]{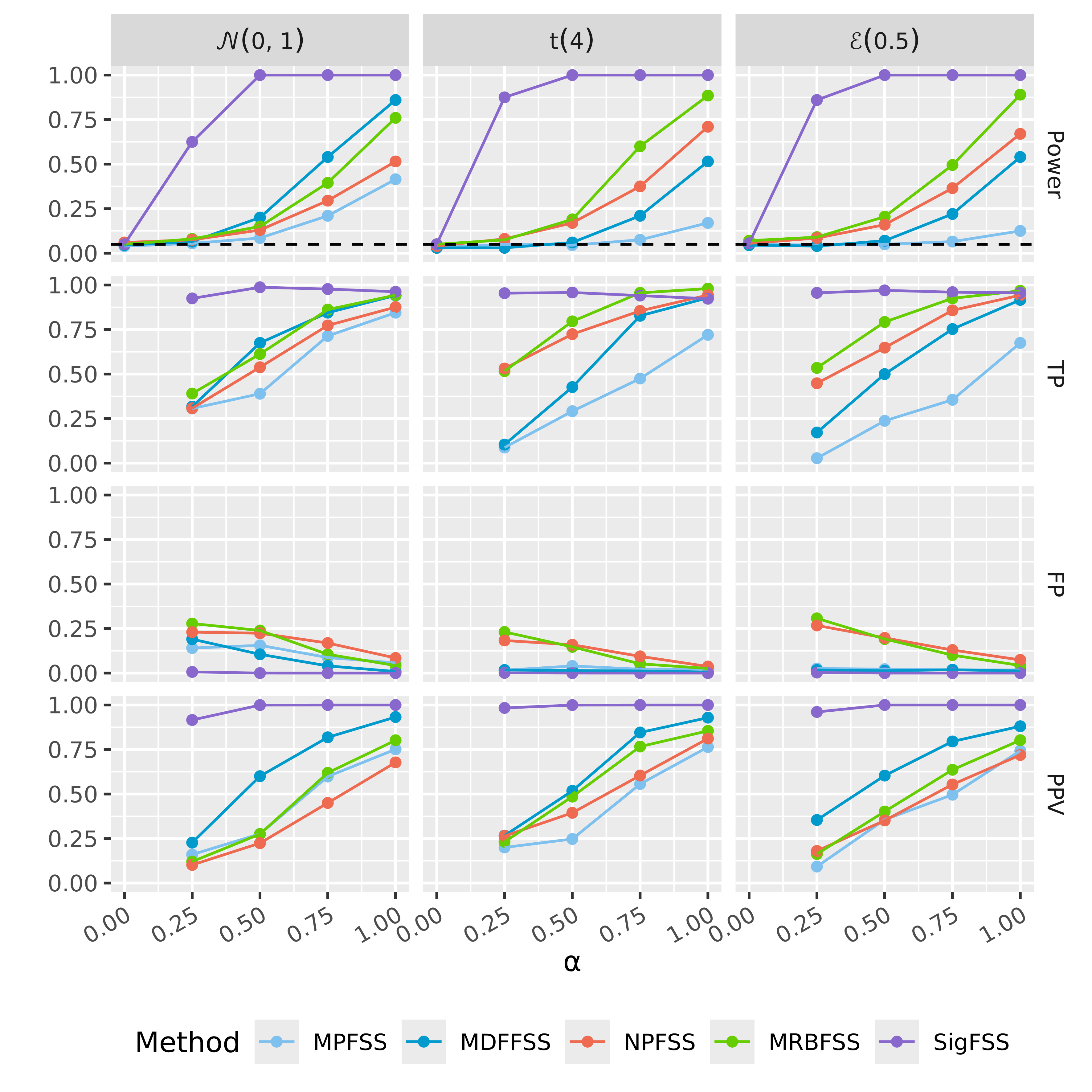}
\end{minipage}
\begin{minipage}{0.49\linewidth}
\centering $\Delta_2$
\includegraphics[width=\linewidth]{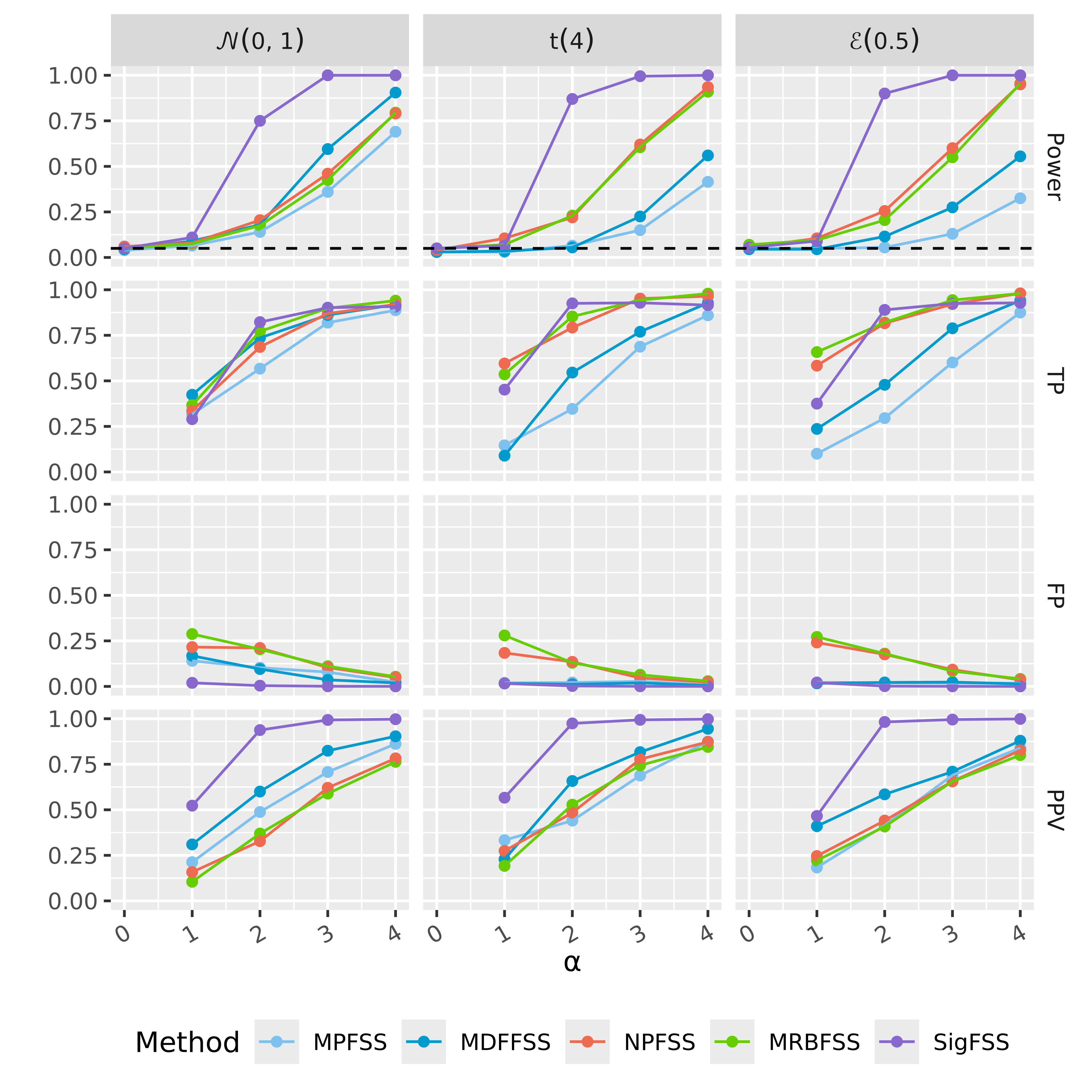}
\end{minipage}
\begin{minipage}{0.49\linewidth}
\centering $\Delta_3$
\includegraphics[width=\linewidth]{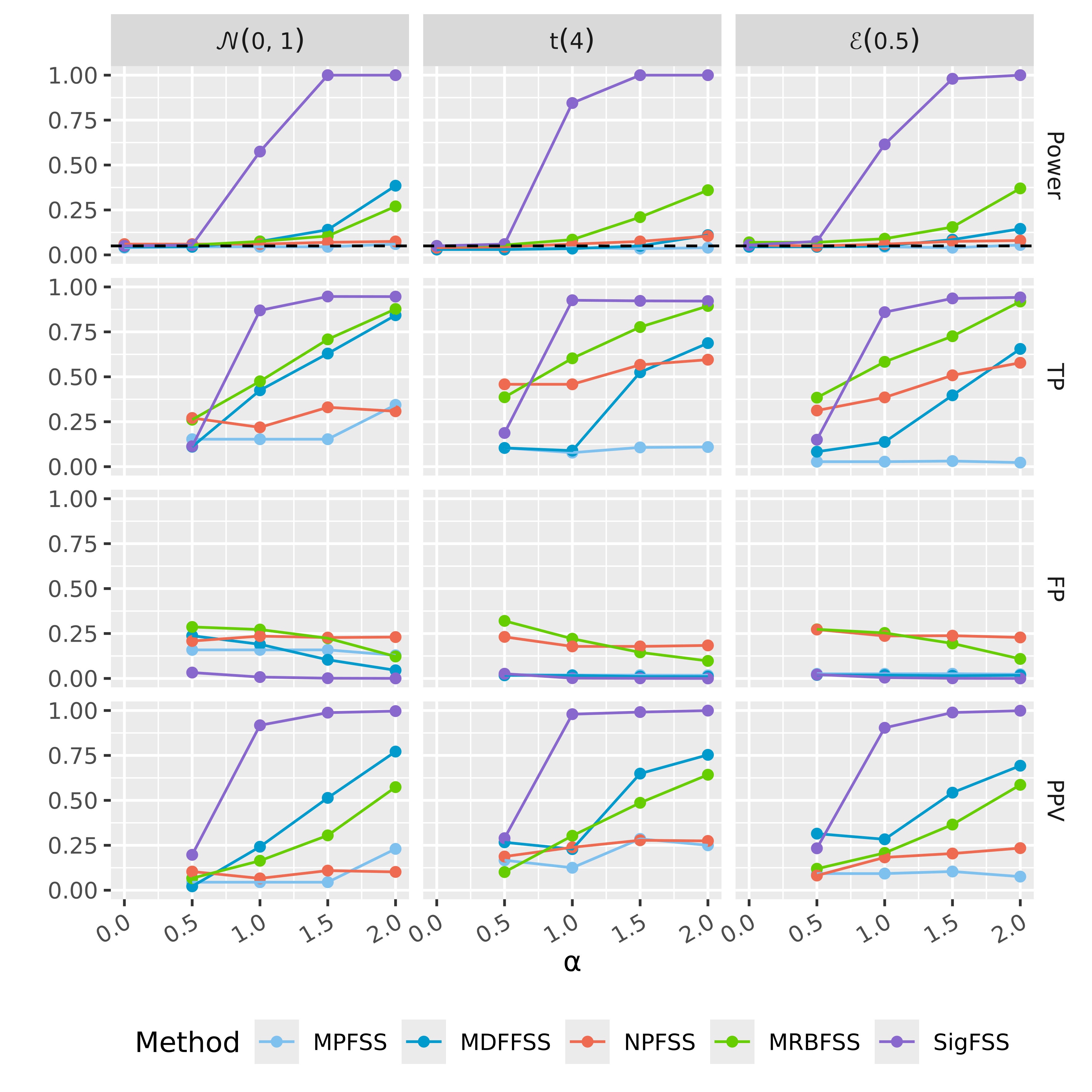}
\end{minipage}
\begin{minipage}{0.49\linewidth}
\centering $\Delta_4$
\includegraphics[width=\linewidth]{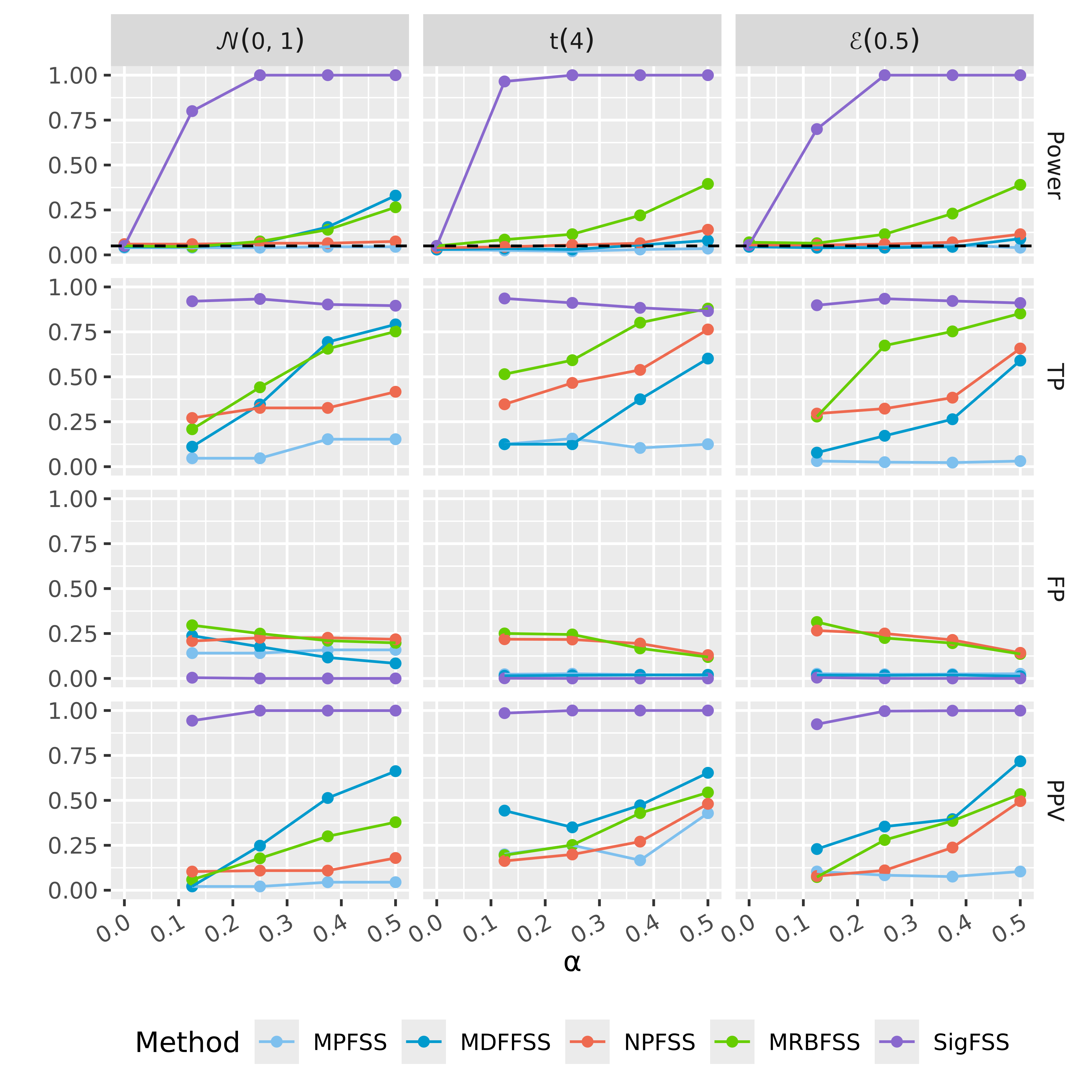}
\end{minipage}
\caption{The simulation study (multivariate case): a comparison of the SigFSS, NPFSS, MDFFSS, MRBFSS, and MPFSS methods for detection of the spatial cluster as the MLC, when $\rho = 0.2$ and considering a graphical approach for choosing $K$ in the SigFSS. $\alpha$ is the parameter that controls the cluster intensity.}
\label{fig:rho0.2nothre}
\end{figure}

\begin{figure}[h!]
\begin{minipage}{0.49\linewidth}
\centering $\Delta_1$
\includegraphics[width=\linewidth]{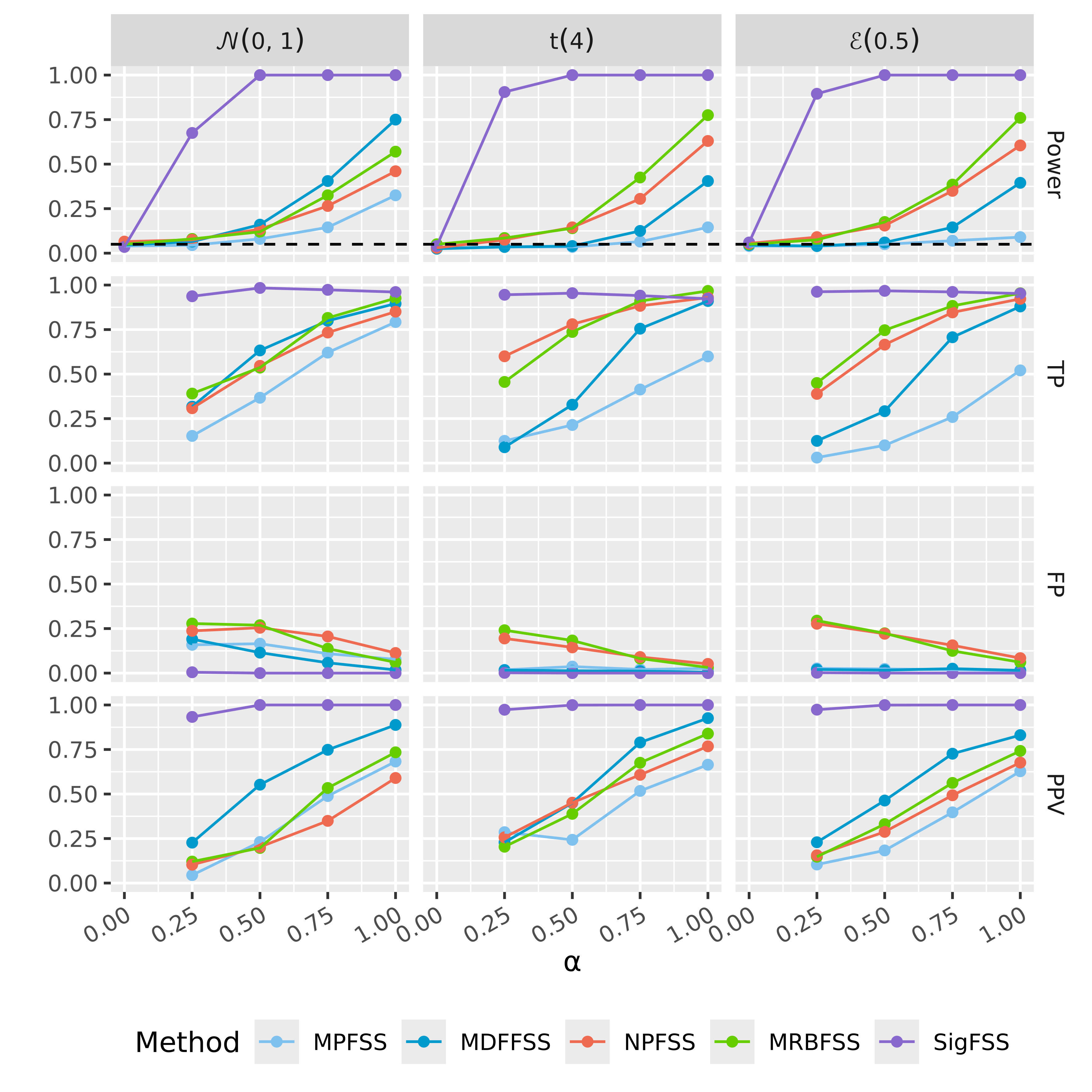}
\end{minipage}
\begin{minipage}{0.49\linewidth}
\centering $\Delta_2$
\includegraphics[width=\linewidth]{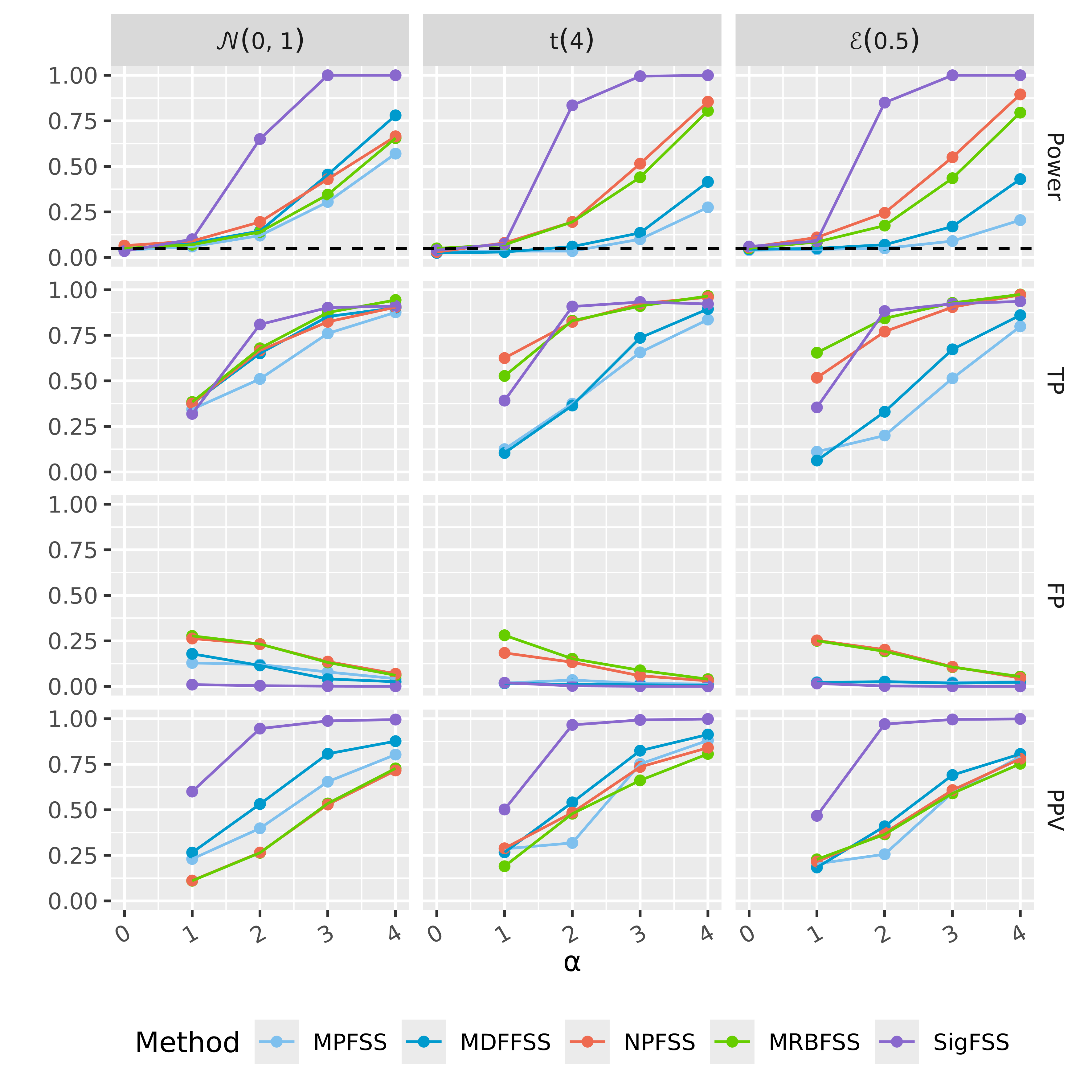}
\end{minipage}
\begin{minipage}{0.49\linewidth}
\centering $\Delta_3$
\includegraphics[width=\linewidth]{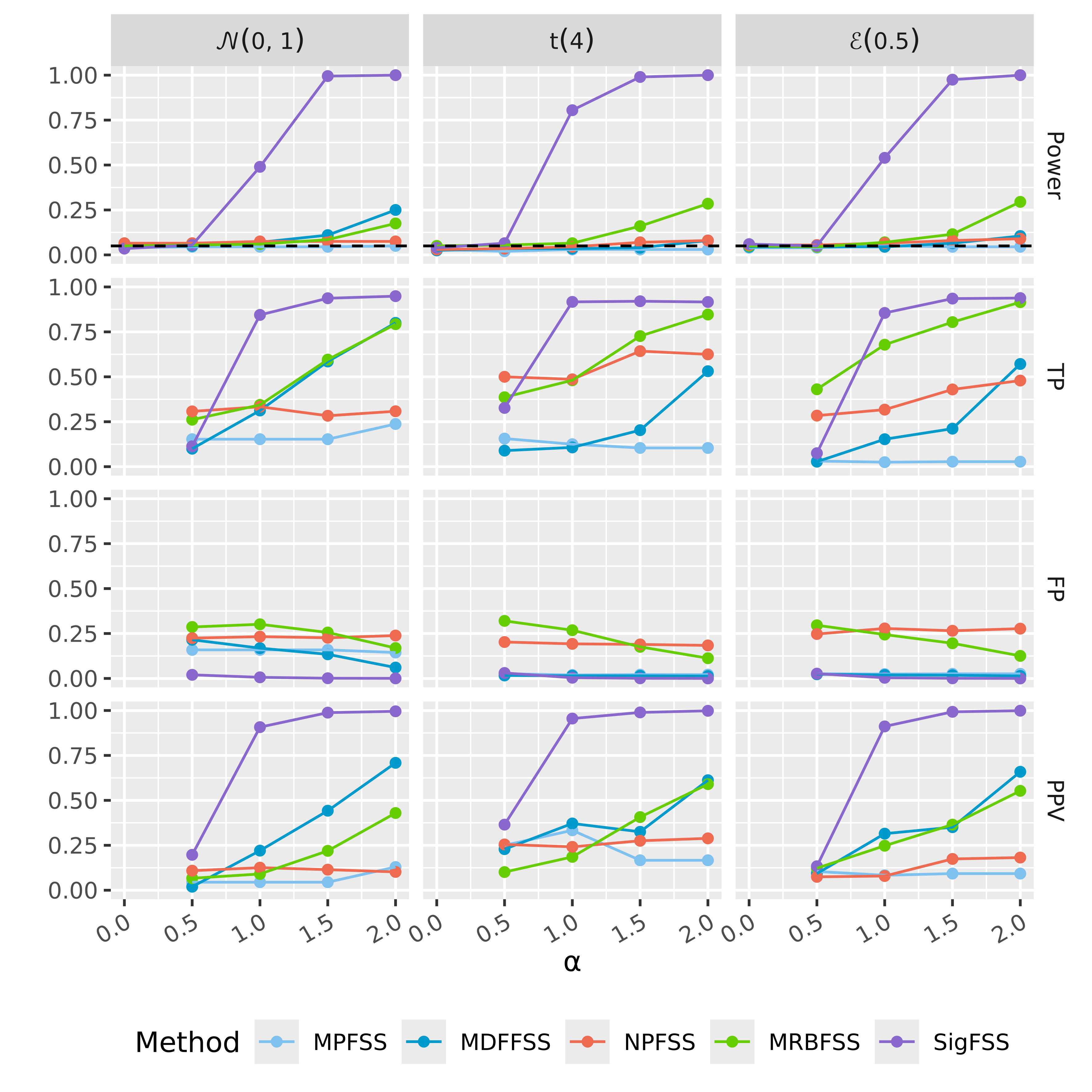}
\end{minipage}
\begin{minipage}{0.49\linewidth}
\centering $\Delta_4$
\includegraphics[width=\linewidth]{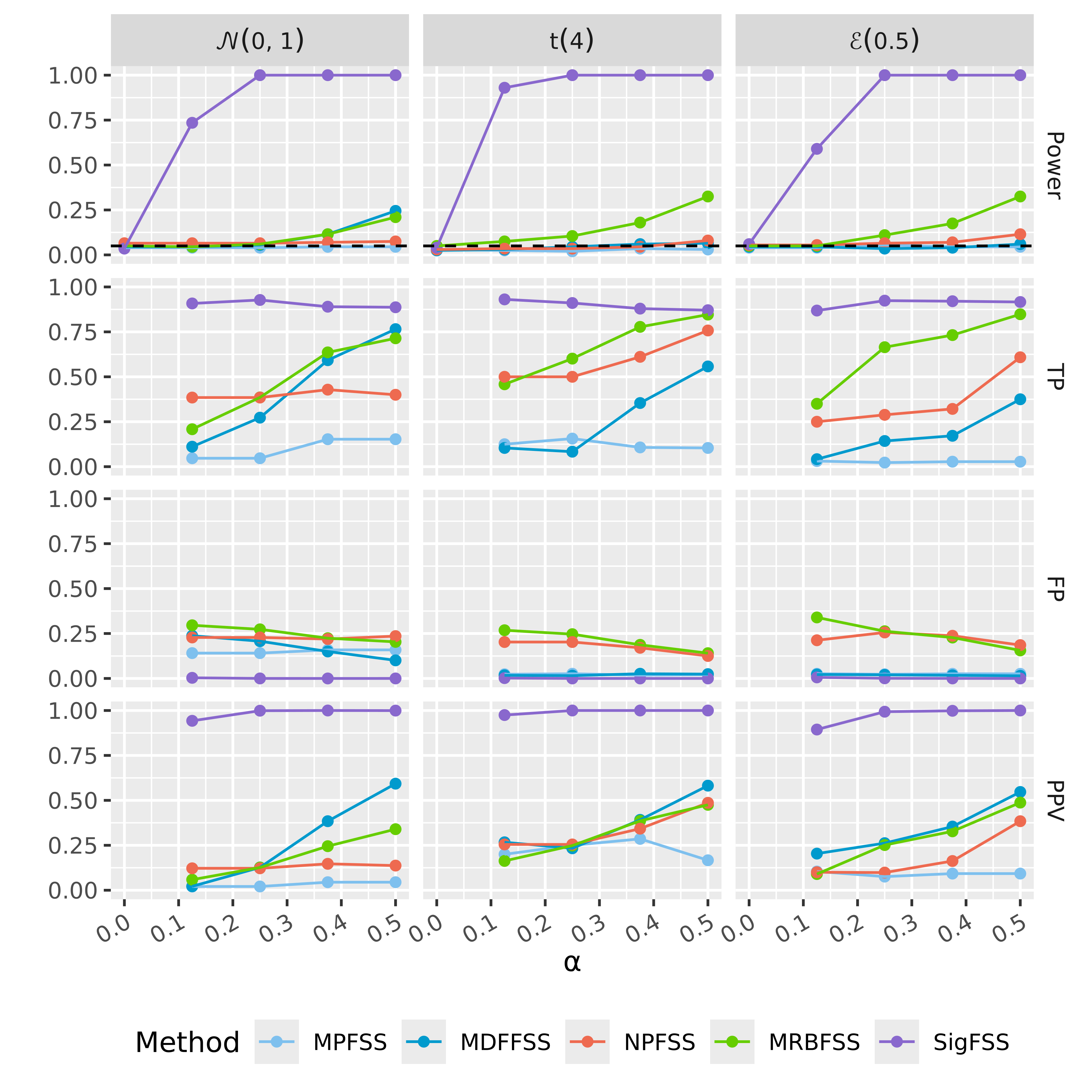}
\end{minipage}
\caption{The simulation study (multivariate case): a comparison of the SigFSS, NPFSS, MDFFSS, MRBFSS, and MPFSS methods for detection of the spatial cluster as the MLC, when $\rho = 0.5$ and considering a graphical approach for choosing $K$ in the SigFSS. $\alpha$ is the parameter that controls the cluster intensity.}
\label{fig:rho0.5nothre}
\end{figure}

\begin{figure}[h!]
\begin{minipage}{0.49\linewidth}
\centering $\Delta_1$
\includegraphics[width=\linewidth]{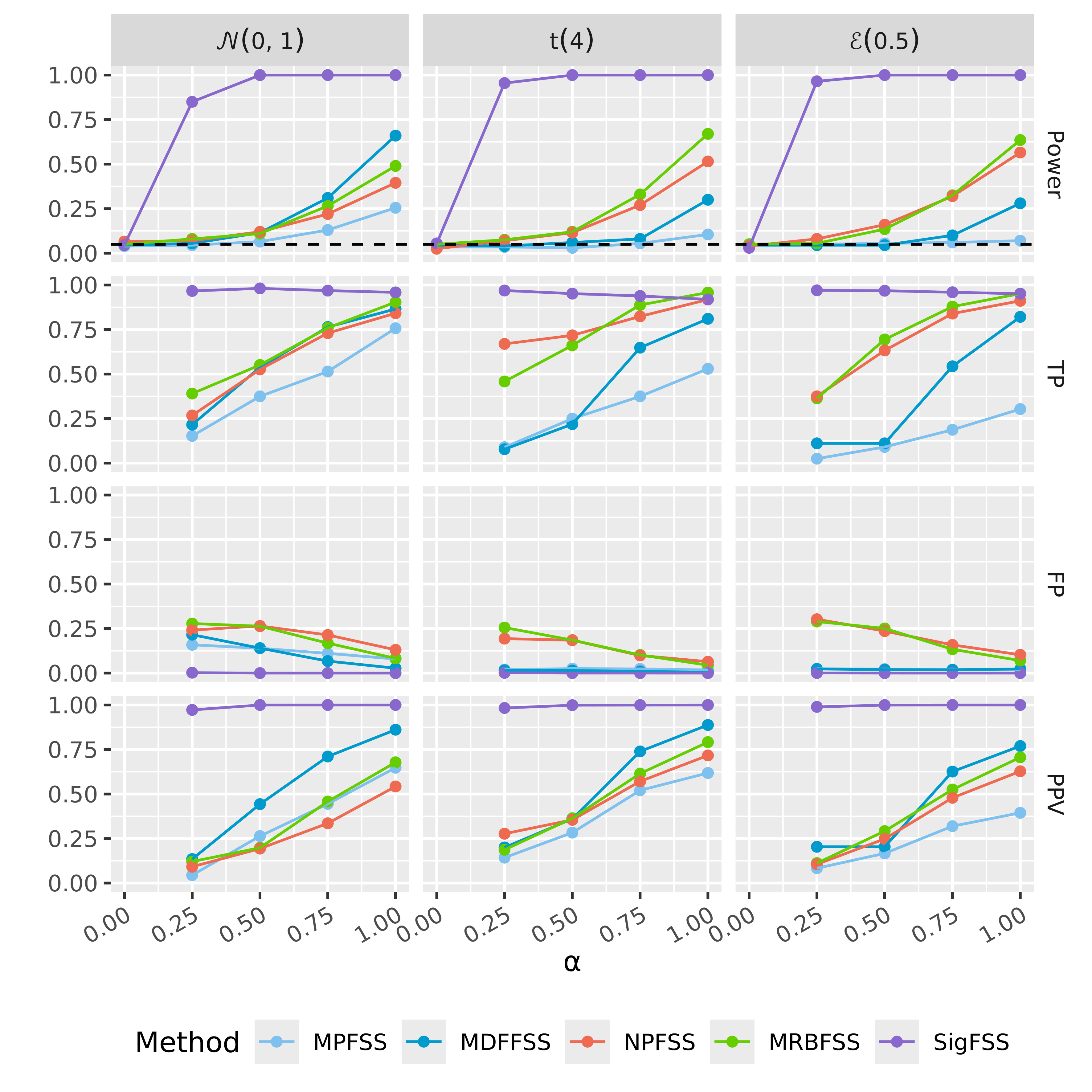}
\end{minipage}
\begin{minipage}{0.49\linewidth}
\centering $\Delta_2$
\includegraphics[width=\linewidth]{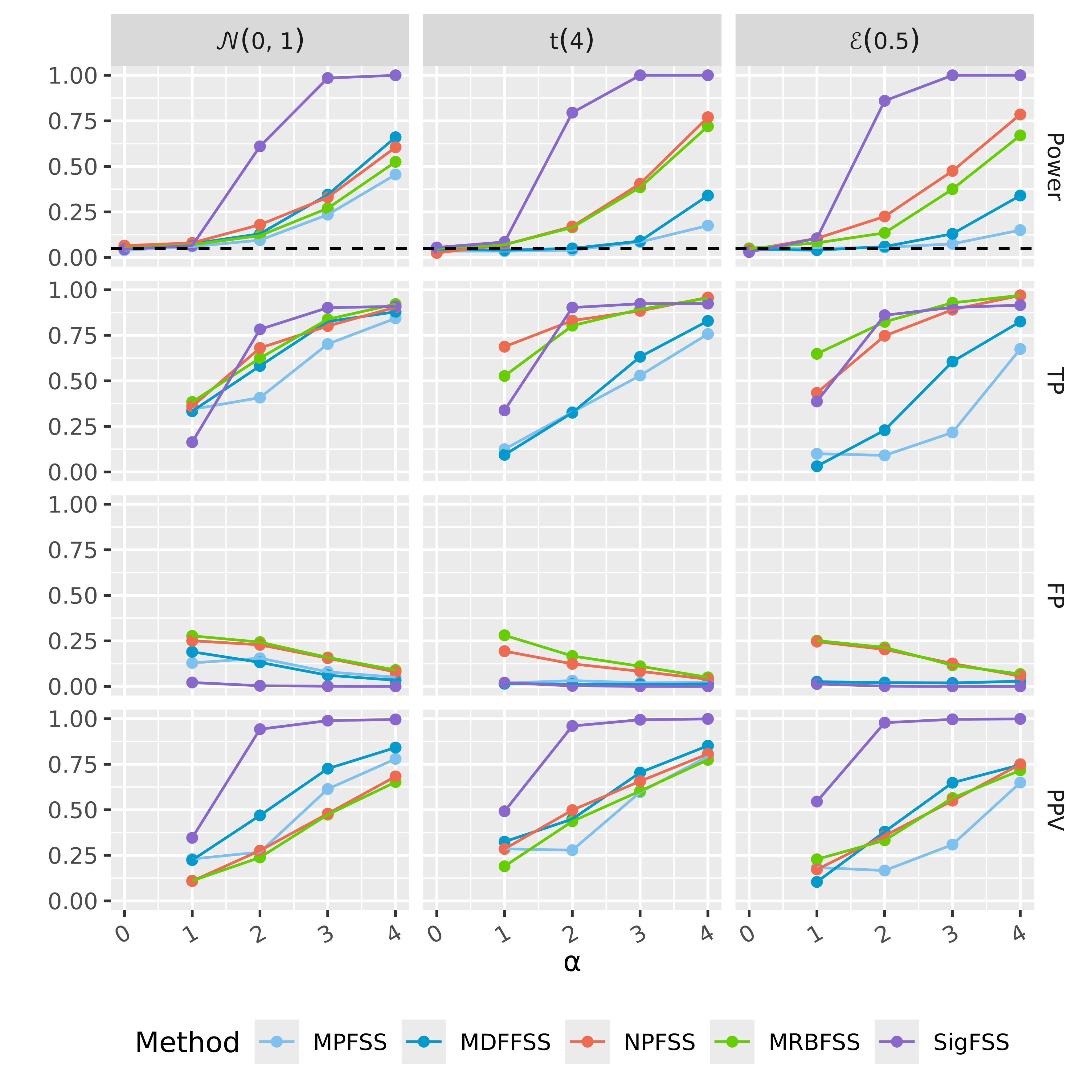}
\end{minipage}
\begin{minipage}{0.49\linewidth}
\centering $\Delta_3$
\includegraphics[width=\linewidth]{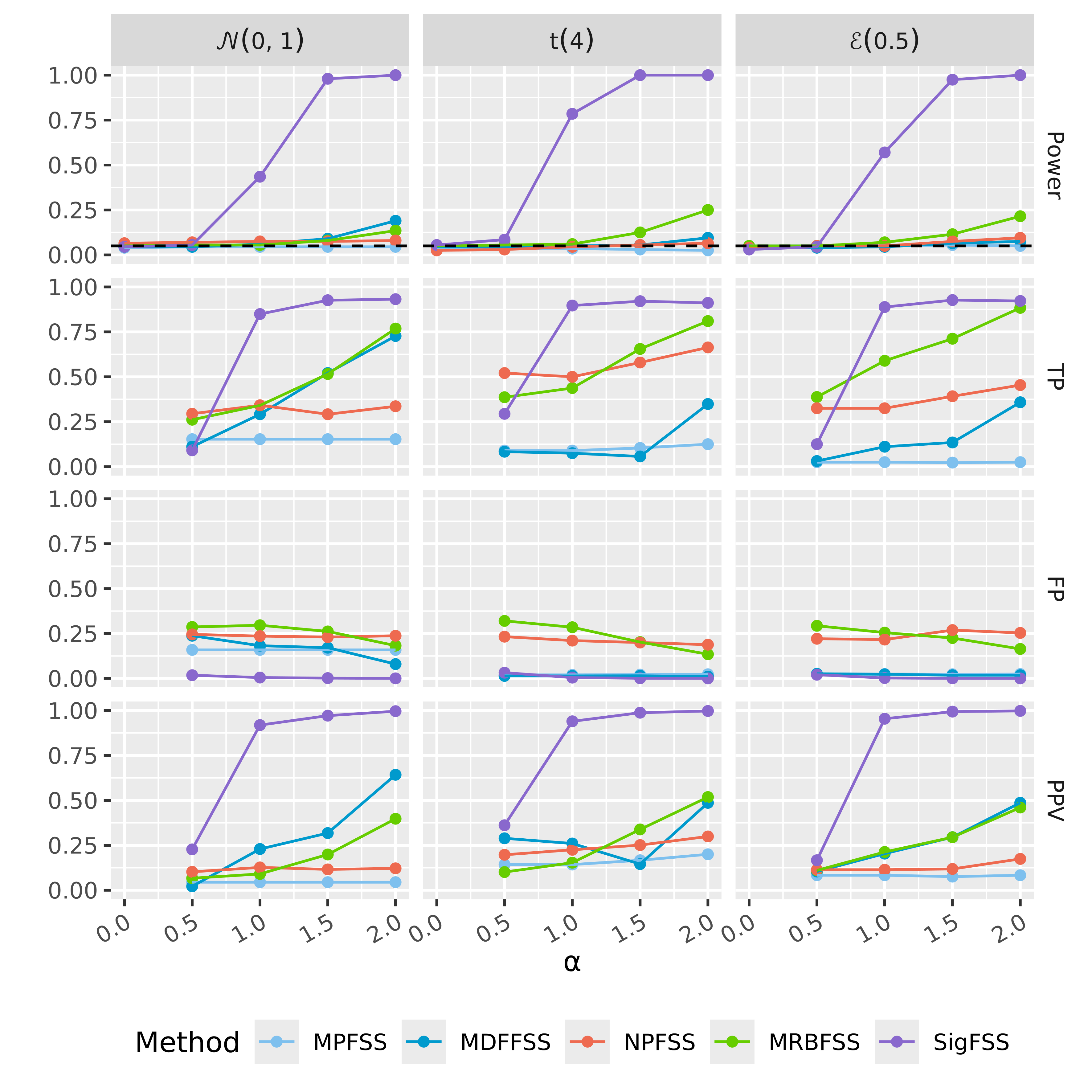}
\end{minipage}
\begin{minipage}{0.49\linewidth}
\centering $\Delta_4$
\includegraphics[width=\linewidth]{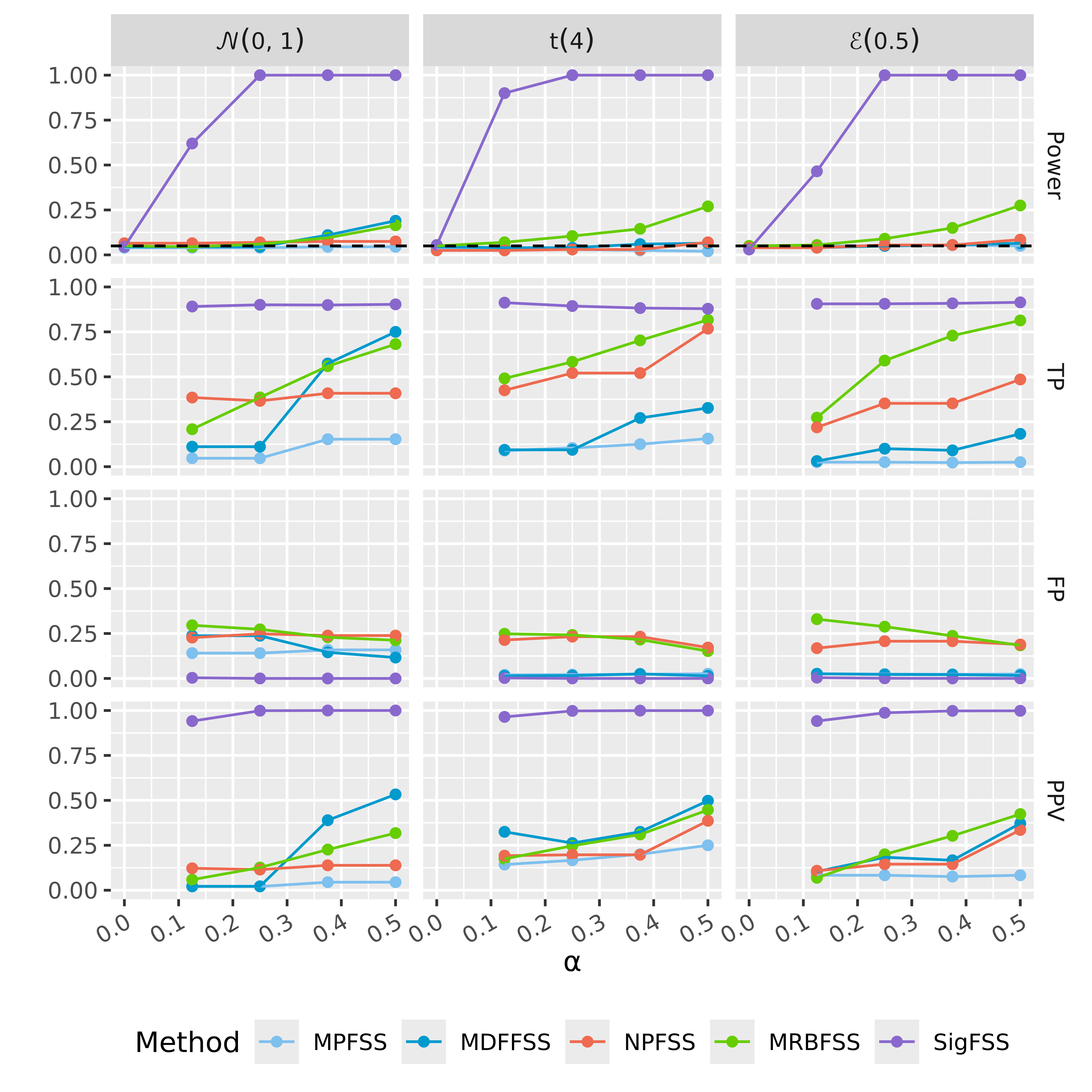}
\end{minipage}
\caption{The simulation study (multivariate case): a comparison of the SigFSS, NPFSS, MDFFSS, MRBFSS, and MPFSS methods for detection of the spatial cluster as the MLC, when $\rho = 0.8$ and considering a graphical approach for choosing $K$ in the SigFSS. $\alpha$ is the parameter that controls the cluster intensity.}
\label{fig:rho0.8nothre}
\end{figure}

The results of the simulation are presented in Figures \ref{fig:uninothre}, \ref{fig:rho0.2nothre}, \ref{fig:rho0.5nothre} and \ref{fig:rho0.8nothre} for when $K$ was chosen graphically (see also Figures \ref{fig:simuKHFSS}, \ref{fig:simuKSigFSS} and \ref{fig:simuKSigFSSmulti} in the Supplementary Materials), in Figures \ref{fig:uni95}, \ref{fig:rho0.2_95},  \ref{fig:rho0.5_95}, \ref{fig:rho0.8_95} for when a fixed threshold of 95\% for the cumulative inertia was applied, and in Figures \ref{fig:uni99}, \ref{fig:rho0.2_99}, \ref{fig:rho0.5_99}, \ref{fig:rho0.8_99} in the Supplementary Materials for when a fixed threshold of 99\% was applied.

For $\alpha = 0$, all methods yielded a type I error of 0.05, regardless of the type of process, the type of $\Delta$, and the level of correlation $\rho$ (for the multivariate processes). \\

In the univariate case, when $K$ is chosen graphically and $\Delta = \Delta_1$, the SigFSS yielded the highest power, true positive rate and positive predictive value and the lowest false positive rate. When $\Delta = \Delta_2$ and $\Delta = \Delta_3$, the SigFSS presented the lowest false positive rate and the highest positive predictive value. It also presented a high true positive rate and the greatest power when considering a non-Gaussian process. When $\Delta = \Delta_4$, the power of the SigFSS was higher than those of the PFSS, NPFSS, DFFSS and URBFSS but lower than that of the HFSS. However, the false positive rates and the positive predictive values were similar.

When considering a fixed threshold of 95\% or 99\% for the cumulative inertia, the SigFSS always presented the highest power, true positive rate and positive predictive value and the lowest false positive rate. \\

When considering a multivariate process ($p=2$), regardless of whether $K$ is chosen graphically or according to a fixed threshold of 95\% or 99\% for the cumulative inertia and regardless of the value of $\rho$, the SigFSS always performed best with regard to the power, true positive rate, false positive rate, and positive predictive value.

\section{Application to real data} \label{sec:appli}

\subsection{The mortality rate in France}

\begin{figure}[h!]
\centering
\includegraphics[width=\linewidth]{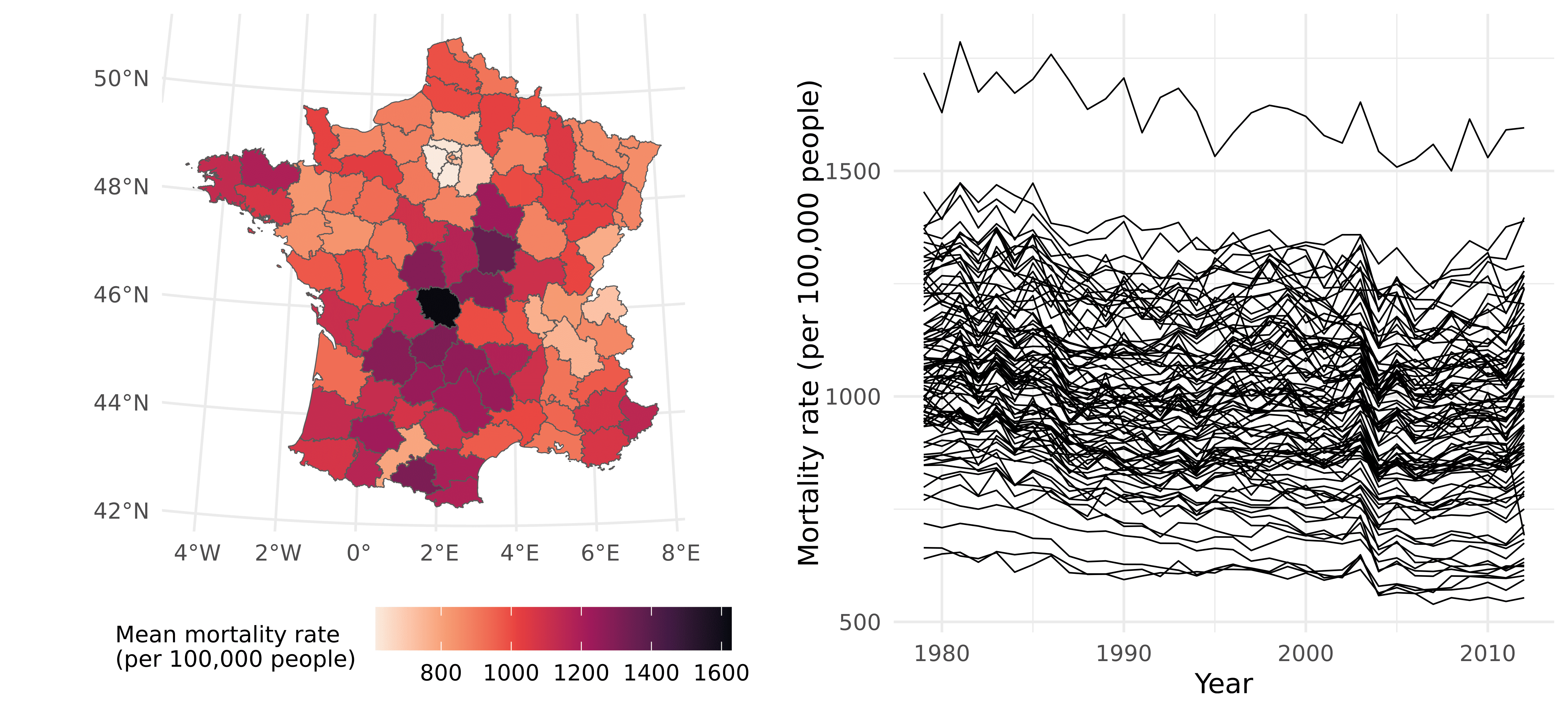}
\caption{Spatial distribution of the mean mortality rate per 100,000 people over the period 1979-2012 in each French \textit{département}, together with the associated mortality rate curves.}
\label{fig:presdatauni}
\end{figure}

\begin{figure}[h!]
\centering
\includegraphics[width=\linewidth]{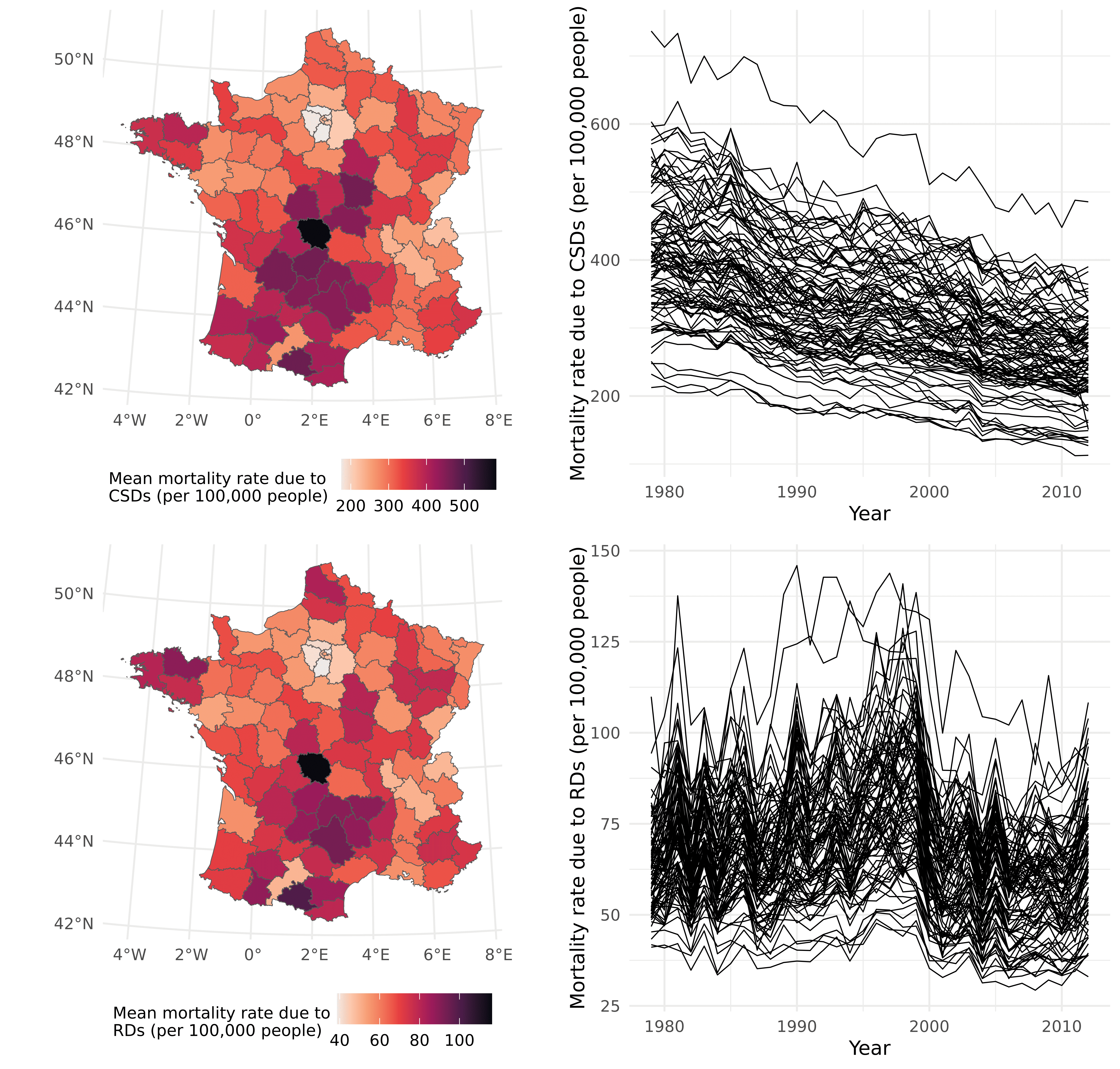}
\caption{Spatial distribution of the mean mortality rate per 100,000 people due to circulatory system diseases (CSDs) and respiratory diseases (RDs) over the period 1979-2012 in each French \textit{département}, together with the associated mortality rate curves.}
\label{fig:presdatamulti}
\end{figure}

We considered data provided by the \textit{Institut National de la Statistique et des Etudes Economiques} (Paris, France) on the annual mortality rate in each of the 94 French \textit{départements} between 1979 and 2012. The all-cause mortality rate was used for the univariate case, and the mortality rate due to circulatory system diseases (CSDs) and respiratory diseases (RDs) was used for the multivariate case (Figures \ref{fig:presdatauni} and \ref{fig:presdatamulti}).

\subsection{Spatial cluster detection}

We used the DFFSS, the PFSS, the NPFSS, the URBFSS, the HFSS and the SigFSS to detect spatial clusters of low or elevated mortality rate. For the multivariate case, we used the MDFFSS, the MPFSS, the NPFSS, the MRBFSS and the SigFSS to detect atypical areas with regard to the mortality rate for CSDs or RDs. \\
For the HFSS and the SigFSS we considered a graphical approach for the choice of $K$ (see Figure \ref{fig:choiceK} in Supplementary Materials) but we also considered fixed thresholds of 95\% and 99\% for the cumulative inertia (see Figures \ref{fig:clusterunisupp} and \ref{fig:clustermultisupp} and Tables \ref{tab:clusteruni_supp} and \ref{tab:clustermulti_supp} in Supplementary Materials). \\

For all methods, we considered the set of potential circular clusters $\mathcal{W}$ defined in Section \ref{sec:method}. The statistical significance of the MLC or the secondary clusters was evaluated in 999 Monte-Carlo permutations. A spatial cluster was considered to be statistically significant when the associated p-value was below 0.05.

\subsection{Results for the univariate case}

\begin{figure}[h!]
\centering
\includegraphics[width=\linewidth]{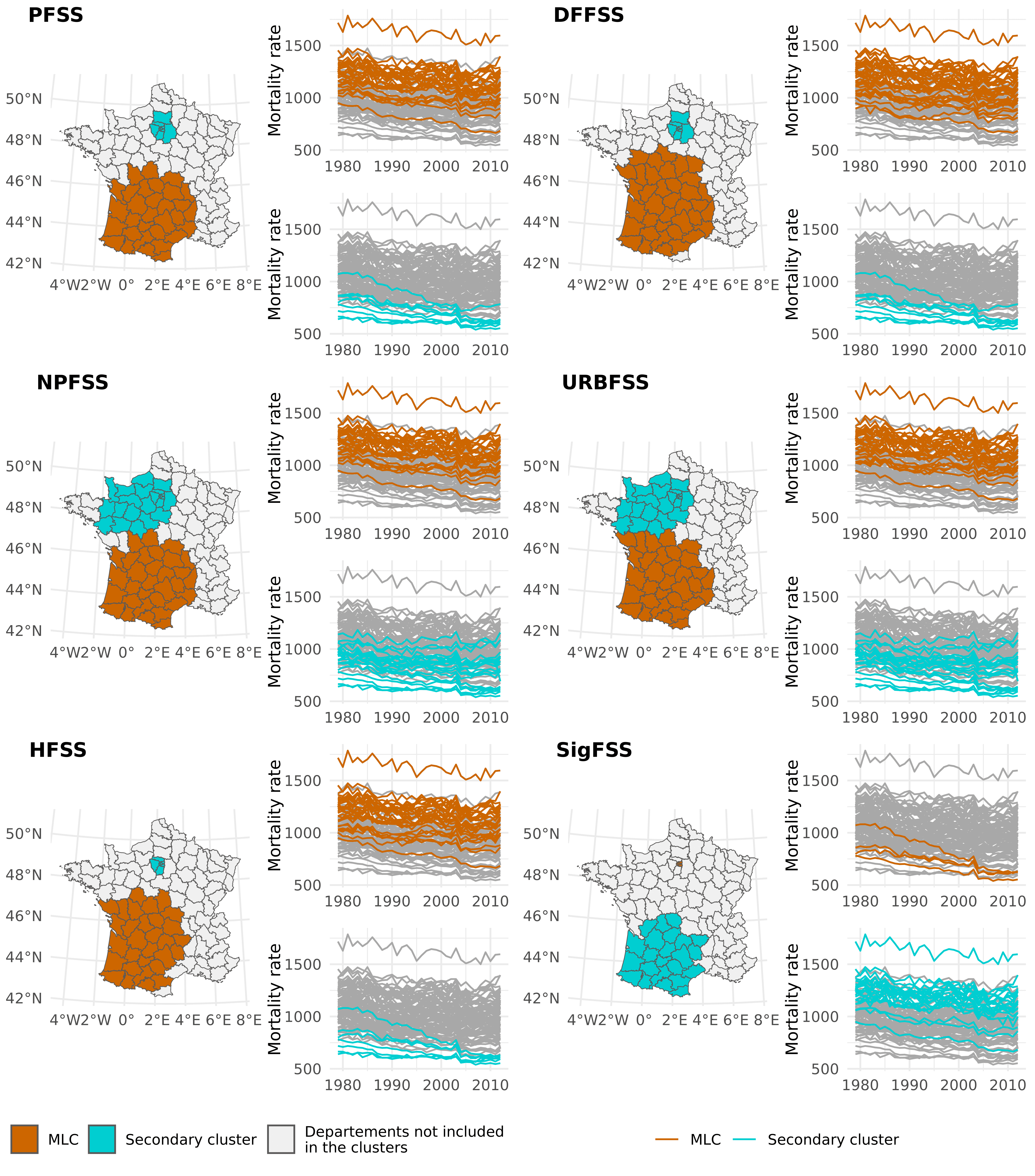}
\caption{Statistically significant spatial clusters with a higher or lower mortality rate (per 100,000 people).}
\label{fig:clusteruni}
\end{figure}

All the methods detected two statistically significant spatial clusters (see Figure \ref{fig:clusteruni} and Table \ref{tab:clusteruni}). One corresponded to a spatial aggregation of lower mortality rates and comprised \textit{départements} from the Île-de-France region; depending on the method, some other \textit{départements} were added to those in the Île-de-France. The other statistically significant spatial cluster was characterized by an elevated mortality rate relative to the rest of the study area and covered a large number of \textit{départements} in southwestern France. 

It is noteworthy that the SigFSS method yielded more precise (i.e., smaller) clusters. \\

When considering a threshold of 95\% or 99\% for the cumulative inertia, the clusters detected with the HFSS and the SigFSS were fairly similar.

\begin{table}[h!]
\caption{Statistically significant spatial clusters with a higher or lower mortality rate.}
\label{tab:clusteruni}
\centering
\begin{tabular}{cccc}
\hline
Method & Cluster & Number of \textit{départements} & p-value \\ \hline
DFFSS & Most likely cluster & 37 & 0.001 \\
      & Secondary cluster & 9 & 0.001 \\ \hline
PFSS & Most likely cluster & 31 & 0.001 \\
     & Secondary cluster & 9 & 0.001 \\ \hline
NPFSS & Most likely cluster & 31 & 0.001 \\
      & Secondary cluster & 23 & 0.001 \\ \hline
URBFSS & Most likely cluster & 32 & 0.001 \\
       & Secondary cluster & 23 & 0.001 \\ \hline
HFSS & Most likely cluster & 30 & 0.001 \\
     & Secondary cluster & 7 & 0.002 \\ \hline
SigFSS & Most likely cluster & 4 & 0.001 \\
       & Secondary cluster & 23 & 0.002 \\ \hline
\end{tabular}
\end{table}

\subsection{Spatial clusters detection in the multivariate case}

\begin{figure}[h!]
 \centering
 \includegraphics[width=\linewidth]{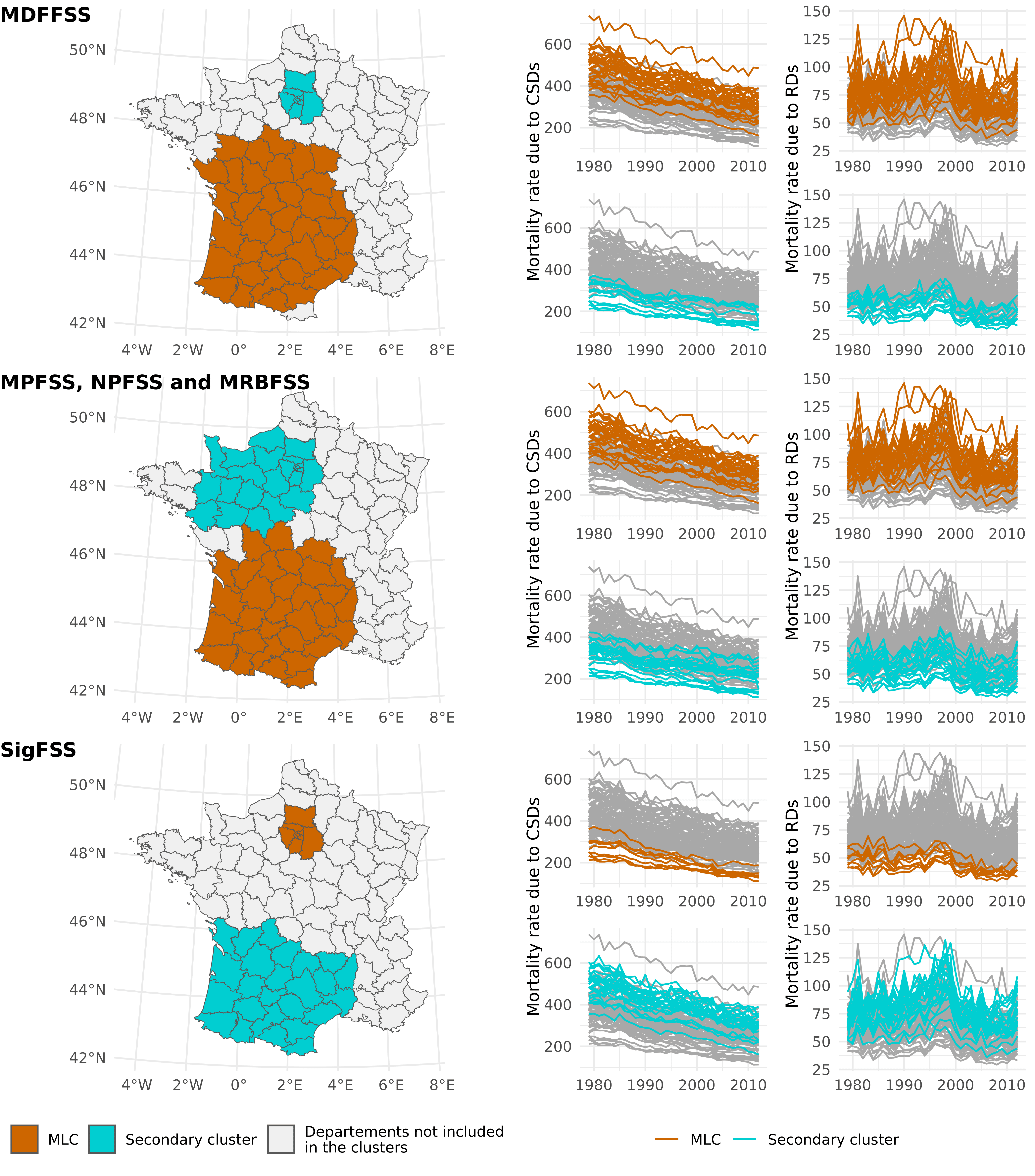}
 \caption{Statistically significant spatial clusters of abnormal mortality rate (per 100,000 people) due to circulatory system diseases (CSDs) and respiratory diseases (RDs).}
 \label{fig:clustermulti}
 \end{figure}

As in the univariate case, all the methods detected two statistically significant spatial clusters in the multivariate case (see Figure \ref{fig:clustermulti} and Table \ref{tab:clustermulti}). Furthermore, the clusters detected in the multivariate case were very similar to those detected in the univariate case. One cluster was located around the Île-de-France region and was associated with a lower CSD and RD mortality rate, compared with the rest of the spatial region. The other cluster was located in southwestern France and was associated with higher CSD and RD mortality rate than elsewhere. \\

Once again, the SigFSS method detected more precise clusters than the literature methods did. The SigFSS results were notably stable; there were few differences between the detected clusters when considering cumulative inertia thresholds of 95\% or 99\%.

\begin{table}[h!]
\caption{Statistically significant spatial clusters with an abnormal mortality rate due to circulatory system diseases and respiratory diseases.}
\label{tab:clustermulti}
\centering
\begin{tabular}{cccc}
\hline
Method & Cluster & Number of \textit{départements} & p-value \\ \hline
MDFFSS & Most likely cluster & 31 & 0.001 \\
       & Secondary cluster & 13 & 0.026 \\ \hline
MPFSS, NPFSS, & Most likely cluster & 31 & 0.001 \\
MRBFSS        & Secondary cluster & 23 & 0.001 \\ \hline
SigFSS & Most likely cluster & 9 & 0.001 \\
       & Secondary cluster & 25 & 0.005 \\ \hline
\end{tabular}
\end{table}

\section{Discussion} \label{sec:discussion}

We developed a SigFSS to detect spatial clusters in functional data. The method is based on signatures and presents the advantage of being applicable to both univariate and multivariate functional data. \\

In a simulation study of a univariate case, we compared the SigFSS with the PFSS \citep{frevent2021detecting}, the DFFSS \citep{frevent2021detecting}, the URBFSS \citep{frevent2022r}, the NPFSS \citep{smida2022wilcoxon} and the HFSS \citep{smida2025hotelling}. We found that the SigFSS (i) almost always yielded the greatest power and the highest true positive rate (ii) always yielded the lowest false positive rate and the highest positive predictive value; this resulted in more precise clusters.

In a simulation study of a multivariate case, we compared the SigFSS with the MPFSS, the MDFFSS, the MRBFSS \citep{frevent2023investigating} and the NPFSS \citep{smida2022wilcoxon}. Again, the results showed that the SigFSS always outperformed the literature methods. \\

Next, we used the methods to detect spatial clusters of abnormal mortality rate in France. Two statistically significant clusters were detected by all methods: one near the Île-de-France region (characterized by a lower mortality rate, compared with the rest of France) and one in southwestern France (characterized by higher mortality rate). However, the clusters detected with the SigFSS were smaller, which was consistent with the results of the simulation study.

We also applied the methods to a multivariate functional dataset corresponding to the mortality rate due to CSDs and RDs; as in the univariate case, the methods detected two similar statistically significant clusters. One cluster was located near the Île-de-France region and was characterized by a lower mortality rate due to CSDs and RDs (compared with the rest of France), and the other was located in southwestern France and was characterized by a higher mortality rate due to CSDs and RDs. The clusters detected with the SigFSS were smaller than those detected by the MPFSS, the NPFSS, the MRBFSS and the MDFFSS. \\

It should be noted that we considered circular spatial clusters only. In fact, other cluster shapes might be relevant. For instance, \cite{tango2005flexibly} considered arbitrarily shaped clusters with a predefined maximum size by considering all sets of connected sites. However, it is important to note that this approach generates a significantly larger number of potential clusters (thereby involving greater computational time) than the method developed by \cite{spatialscanstat}. The elliptical cluster approach developed by \cite{elliptic} is also associated with greater computational time. \cite{lin2016spatial} developed a potential alternative by suggesting the aggregation of circular clusters to form larger clusters of arbitrary shapes. \\

In practice, the detection of secondary clusters is often relevant. Here, we followed the approach developed by \cite{spatialscanstat}, in which statistical inference is performed on other potential clusters as it is for the MLC. \cite{zhang2010spatial} suggests an alternative, sequential approach: once a statistically significant MLC has been detected, data are removed from the MLC (as if there were no spatial units there) and the MLC in the remaining data is then determined. If it is statistically significant, it is considered to be a secondary cluster. This approach is repeated until none of the clusters is statistically significant.
A key limitation of these approaches is their failure to maintain the nominal type I error. The reliable inference of secondary clusters is challenging and requires further research. \\

It is noteworthy that in both the simulation study and the application to real data, the observation times were identical for all the spatial locations. While this represents an ideal scenario for functional data analysis, it is often not representative of real-world datasets. Nevertheless, our new method is applicable to spatial sites with different observation times because the signatures at each of the locations are computed independently. \\

The spatial scan statistic models presented here rely on the assumption whereby observations at different spatial locations are independent. This is a classical but strong assumption in the field of spatial scan statistics and is frequently violated in real-world applications - particularly in environmental datasets such as pollution measurements, where spatially close locations tend to exhibit similar values. The presence of spatial autocorrelation is known to lead to an inflation of the type I error in the random permutation procedure \citep{10.1214/07-AOAS129, lin2014generalized, lee2020spatial, ahmed2021spatial}. Although several recent works have explored the incorporation of spatial dependence into scan statistics \citep{lee2020spatial,ahmed2021spatial}, they focused on relatively simple models. The development of a method that account for spatial dependence within functional spatial scan statistics remains a challenge.

\bibliography{bibliographie.bib}
\bibliographystyle{chicago}

\appendix

\setcounter{figure}{0}
\setcounter{table}{0}
\renewcommand{\thefigure}{S\arabic{figure}}
\renewcommand{\thetable}{S\arabic{table}}

\section{Supplementary materials for the simulation study}

\begin{figure}[H]
\centering
\includegraphics[width=0.5\linewidth]{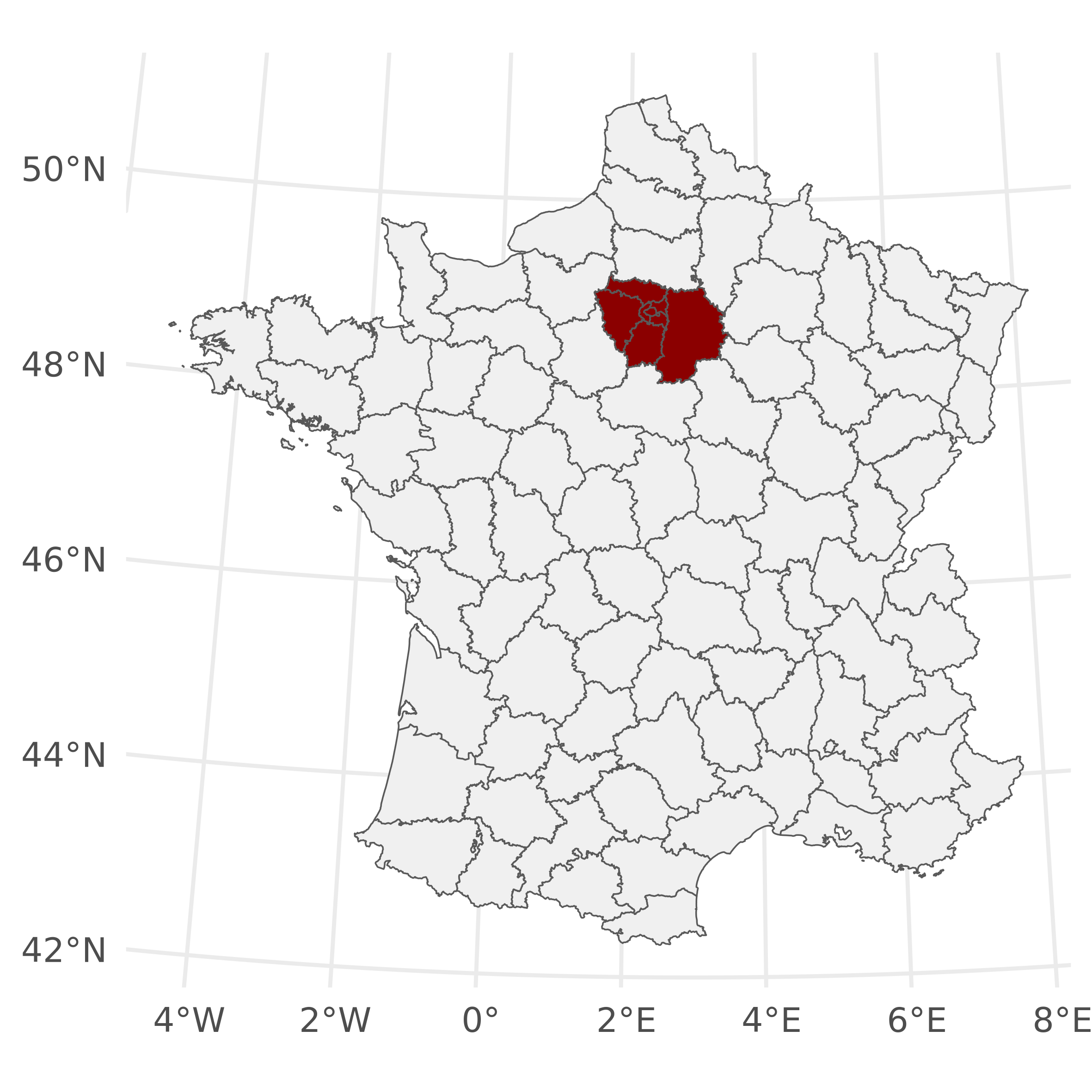}
\caption{The simulation study: The 94 French \textit{départements} and the spatial cluster (in red) simulated for each artificial dataset.}
\label{fig:simulatedcluster}
\end{figure}

\begin{figure}[H]
\centering
\includegraphics[width=\linewidth]{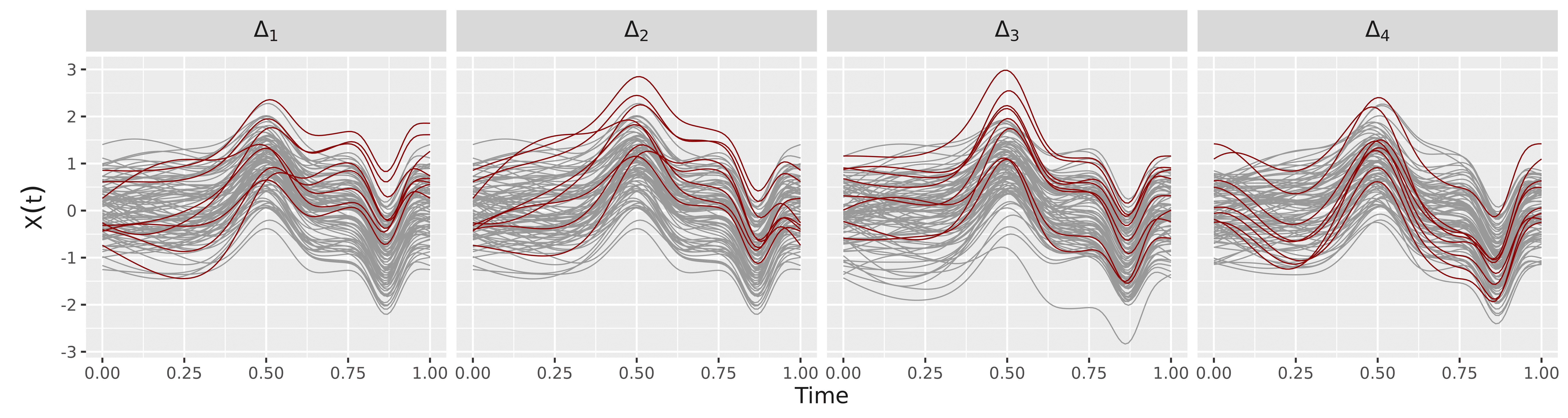}
\caption{The simulation study: an example of the data generated for the Gaussian process in the univariate case, with $\Delta(t) = \Delta_1(t) = t, \Delta(t) = \Delta_2(t) = 4t(1-t), \Delta(t) = \Delta_3(t) = 2\exp{(-100(t-0.5)^2)}/3$ and $\Delta(t) = \Delta_4(t) = 0.5\cos{(4\pi(t-0.5))}$. The red curves correspond to the observations in the cluster.}
\label{fig:examplesimu_uni}
\end{figure}

\begin{figure}[H]
\centering
\includegraphics[width=\linewidth]{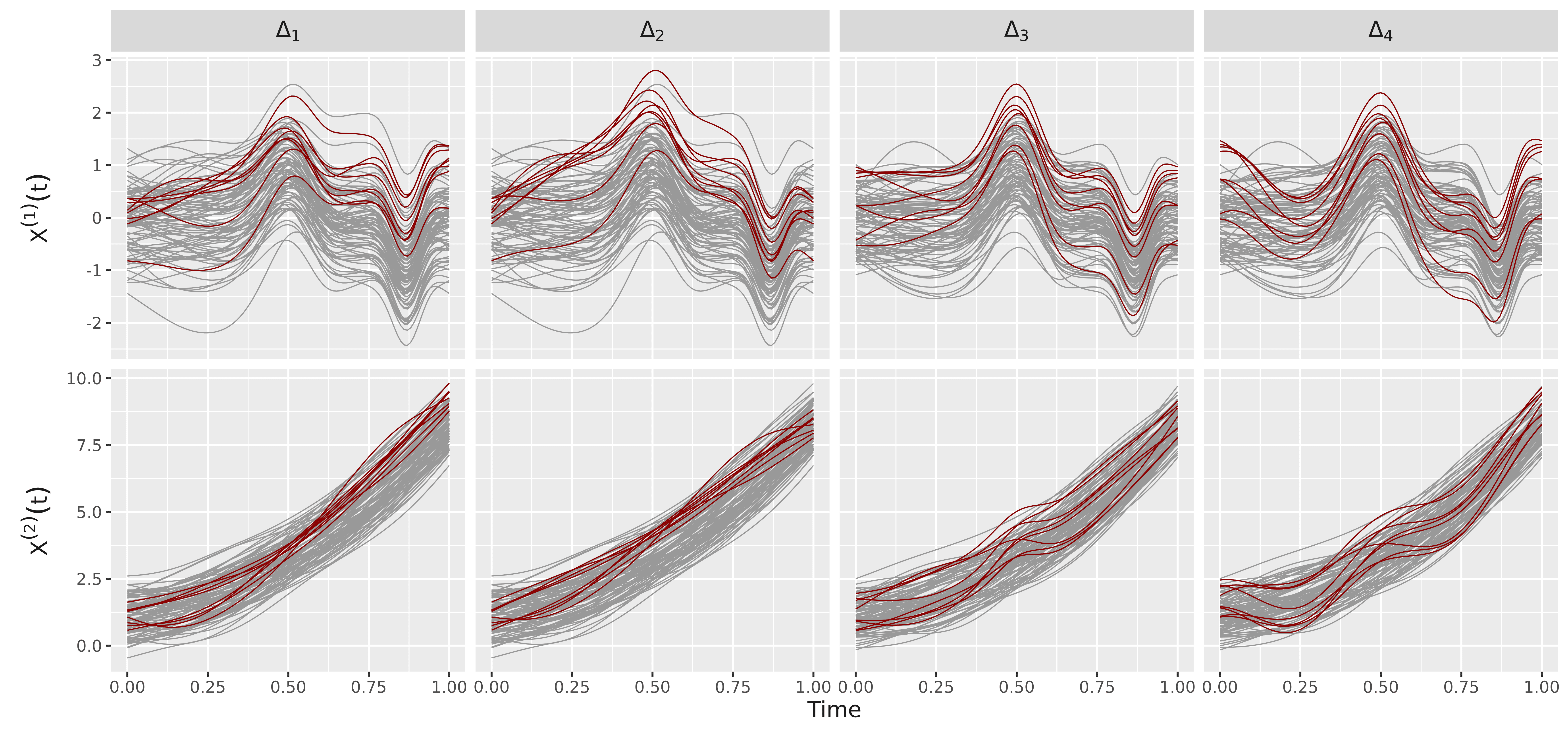}
\caption{The simulation study: an example of the data generated for the Gaussian process in the multivariate case ($\rho = 0.2$), with $\Delta(t) = \Delta_1(t) = (t,t)^\top, \Delta(t) = \Delta_2(t) = 4(t(1-t),t(1-t))^\top, \Delta(t) = \Delta_3(t) = 2(\exp{(-100(t-0.5)^2)}/3,\exp{(-100(t-0.5)^2)}/3)^\top$ and $\Delta(t) = \Delta_4(t) = 0.5(\cos{(4\pi(t-0.5))},\cos{(4\pi(t-0.5))})^\top$. The red curves correspond to the observations in the cluster.}
\label{fig:examplesimu_multi}
\end{figure}

\begin{figure}[H]
\begin{minipage}{0.49\linewidth}
\centering $\Delta_1$
\includegraphics[width=\linewidth]{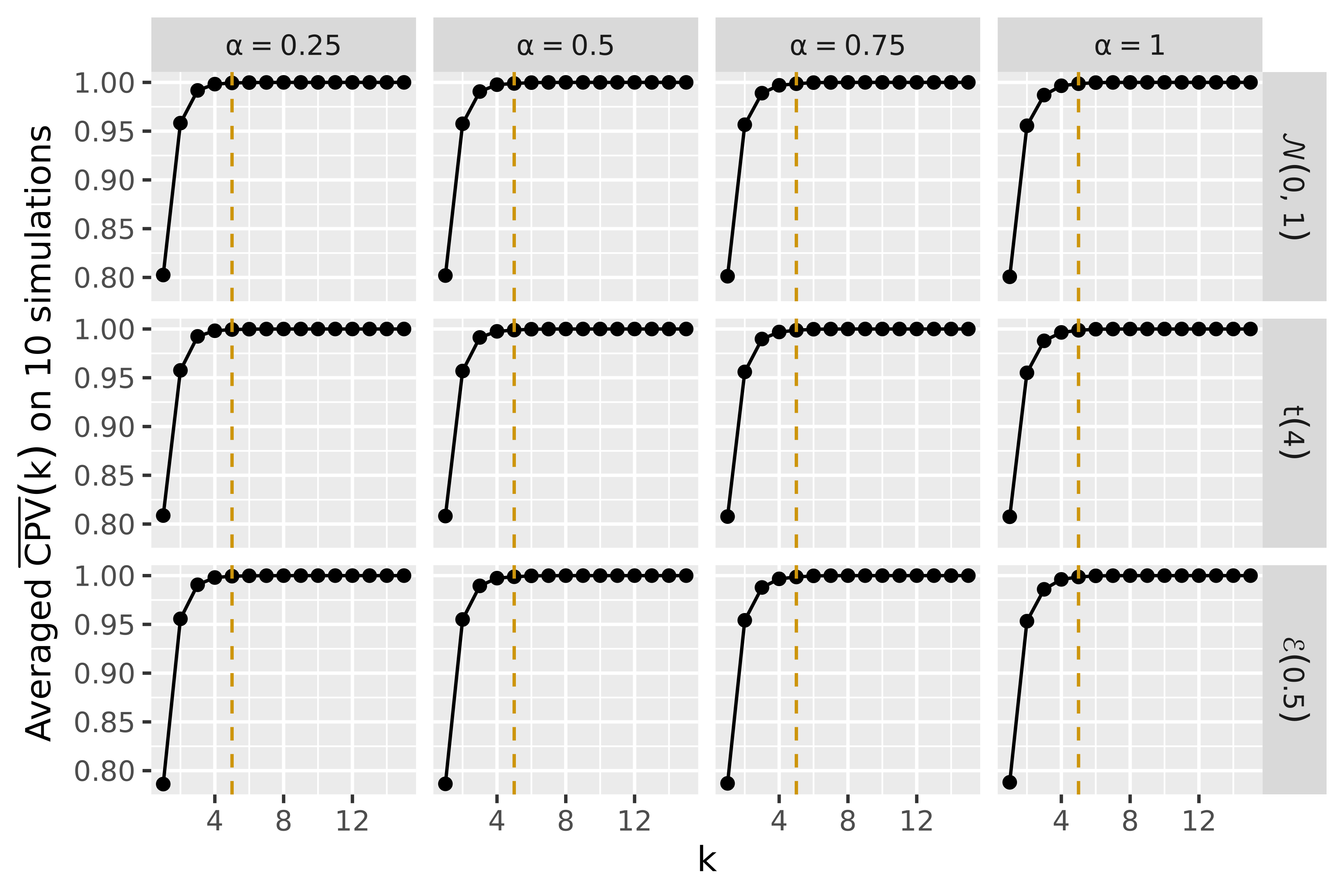}
\end{minipage}
\begin{minipage}{0.49\linewidth}
\centering $\Delta_2$
\includegraphics[width=\linewidth]{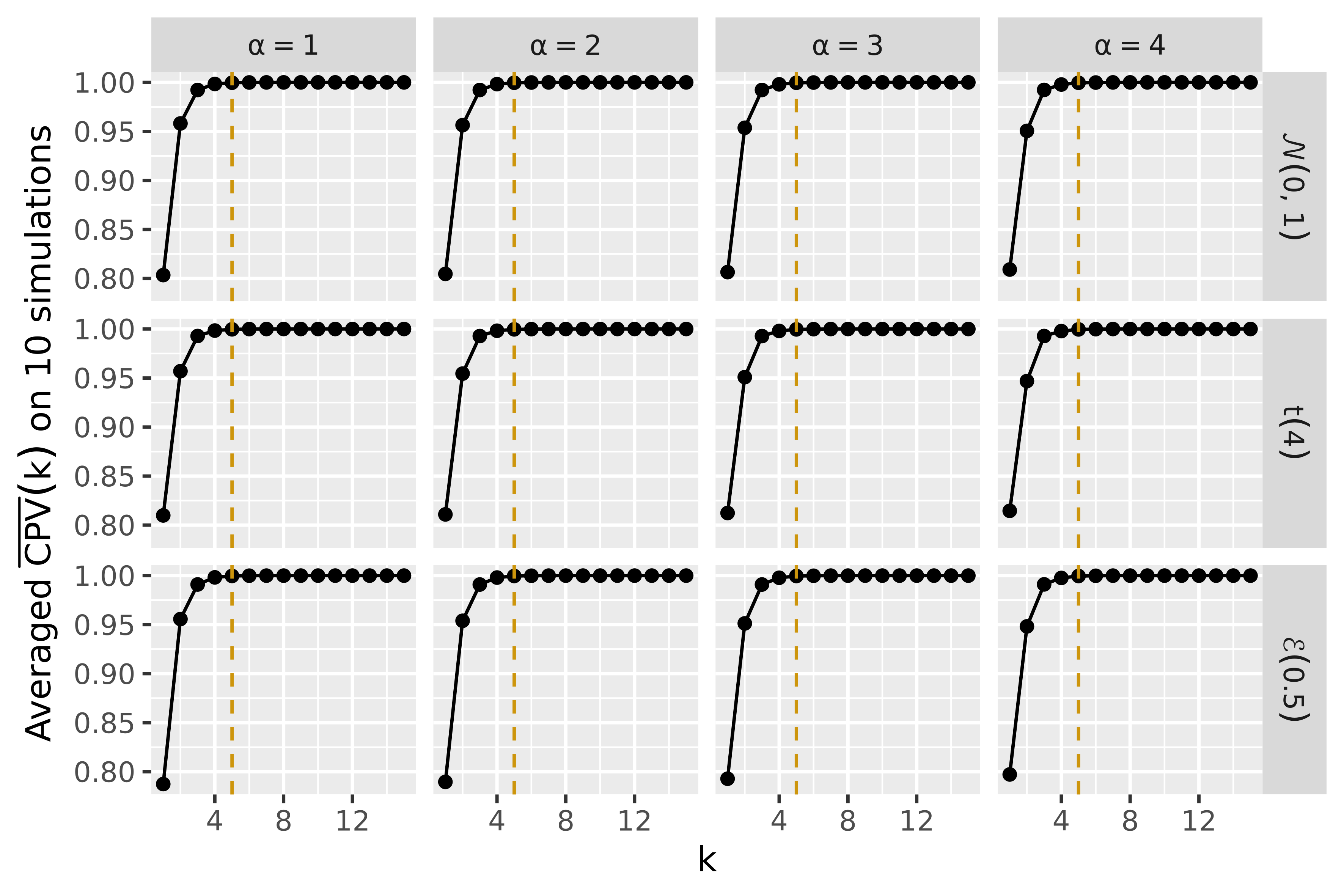}
\end{minipage}
\begin{minipage}{0.49\linewidth}
\centering $\Delta_3$
\includegraphics[width=\linewidth]{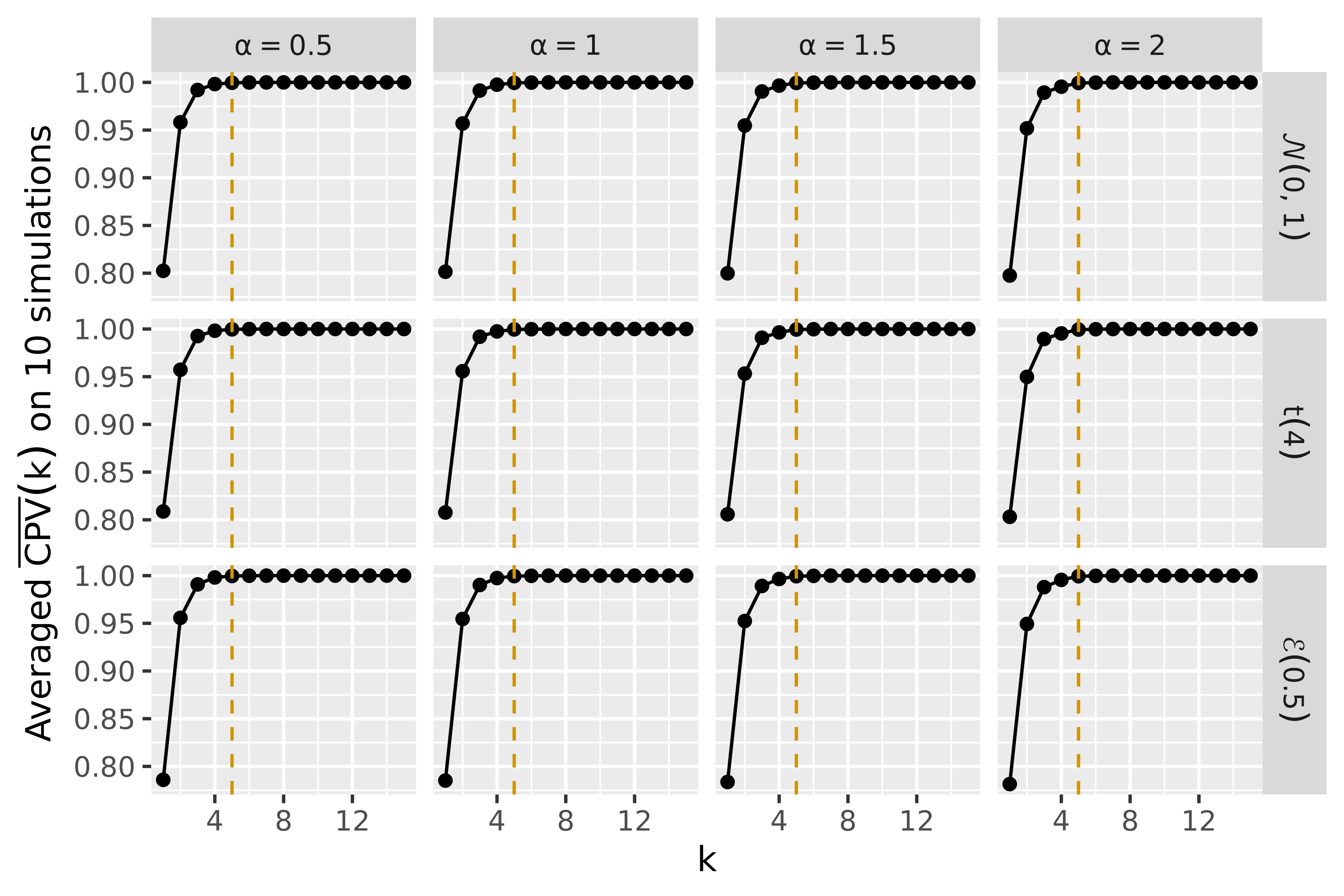}
\end{minipage}
\begin{minipage}{0.49\linewidth}
\centering $\Delta_4$
\includegraphics[width=\linewidth]{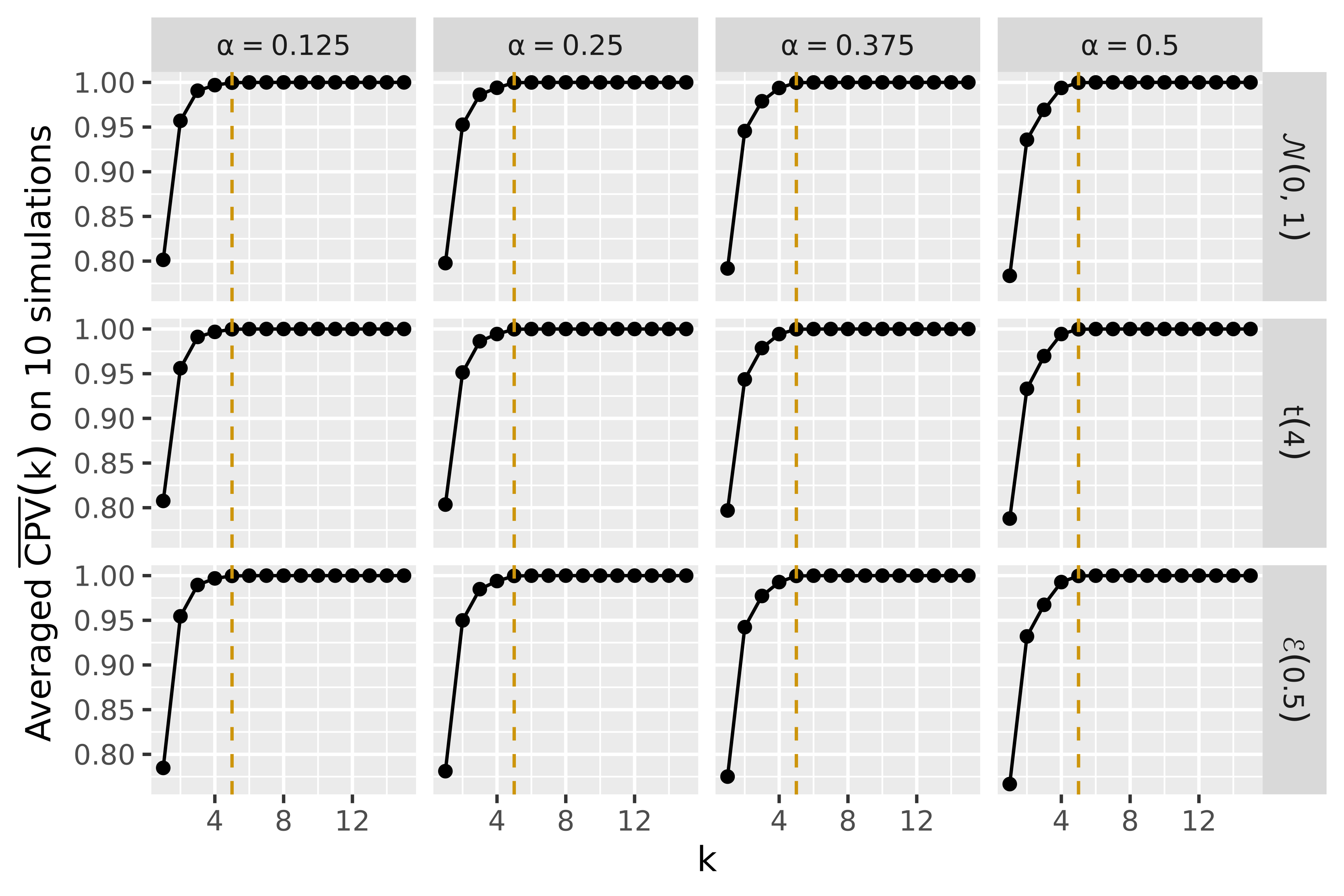}
\end{minipage}
\caption{The simulation study (univariate case): the cumulative inertia for the choice of $K$ in the HFSS. The selected value is highlighted by a vertical line.}
\label{fig:simuKHFSS}
\end{figure}

\begin{figure}[H]
\begin{minipage}{0.49\linewidth}
\centering $\Delta_1$
\includegraphics[width=\linewidth]{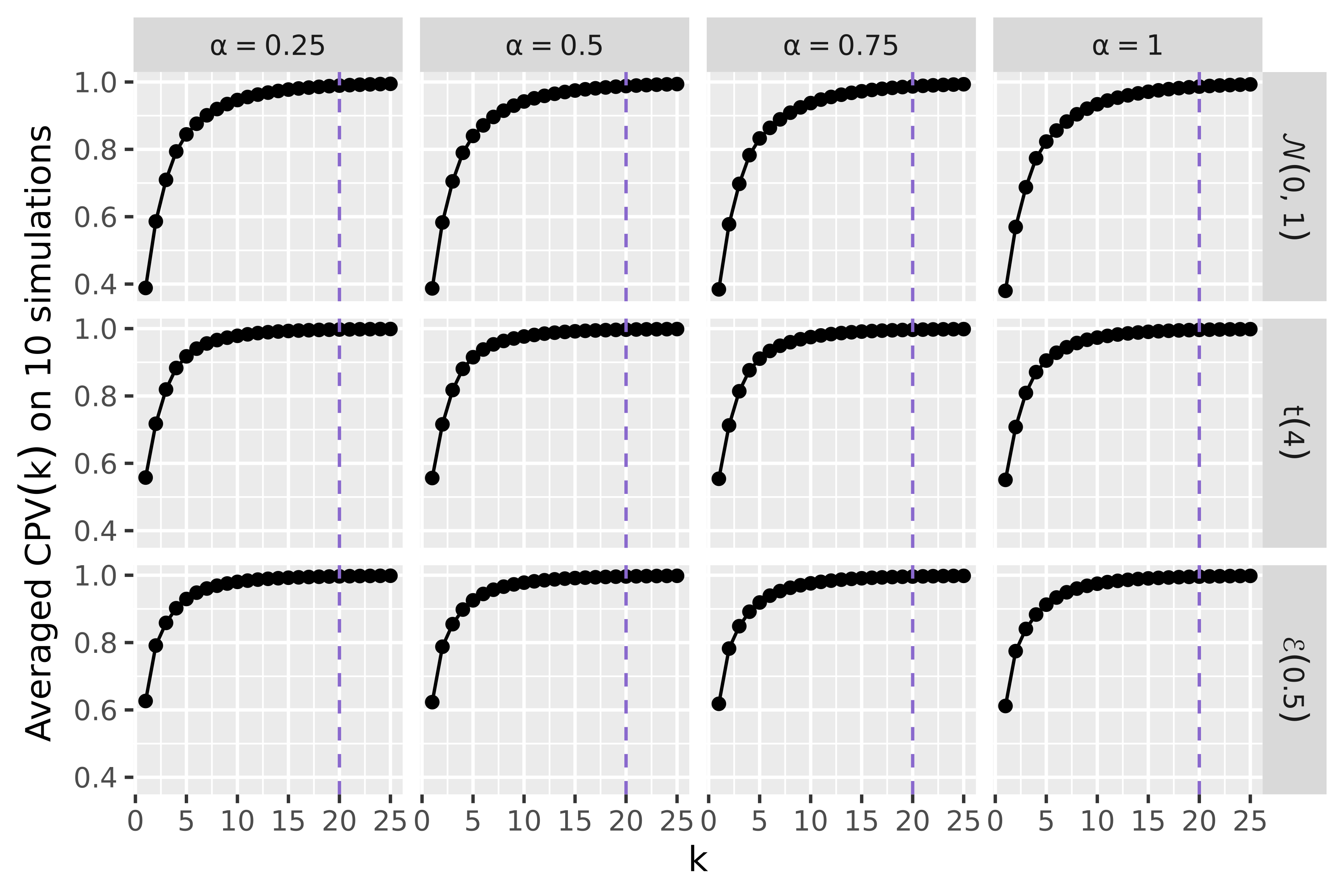}
\end{minipage}
\begin{minipage}{0.49\linewidth}
\centering $\Delta_2$
\includegraphics[width=\linewidth]{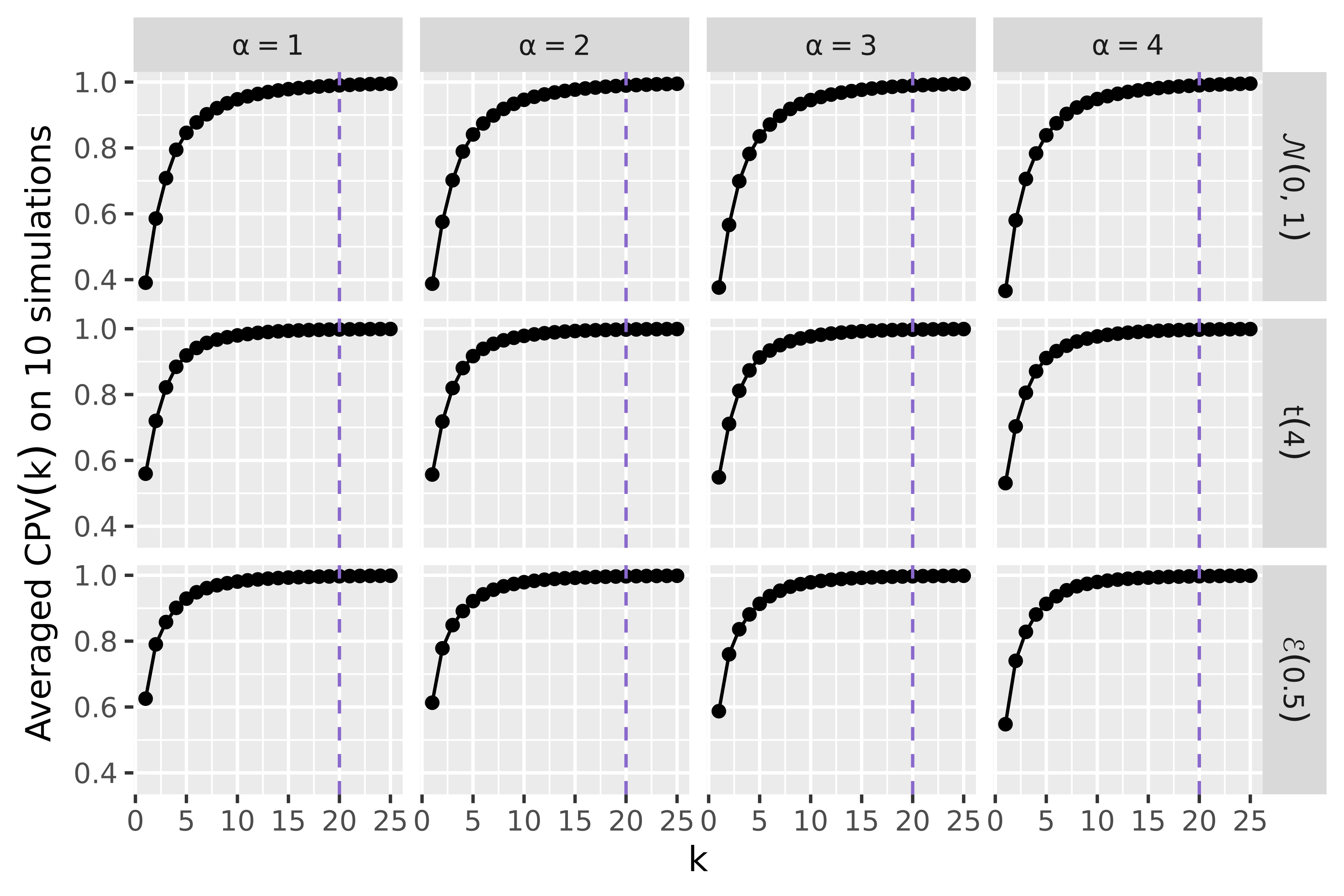}
\end{minipage}
\begin{minipage}{0.49\linewidth}
\centering $\Delta_3$
\includegraphics[width=\linewidth]{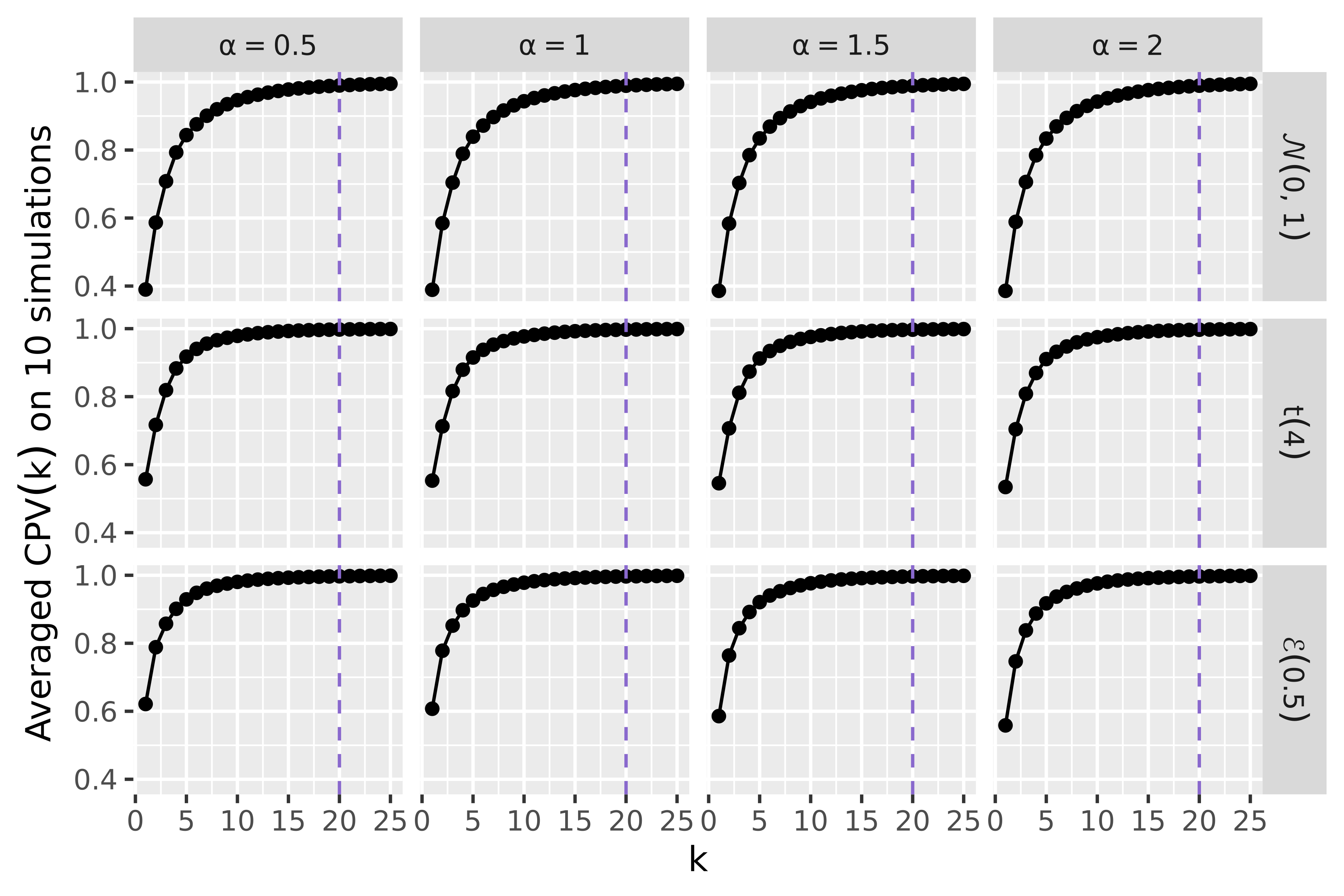}
\end{minipage}
\begin{minipage}{0.49\linewidth}
\centering $\Delta_4$
\includegraphics[width=\linewidth]{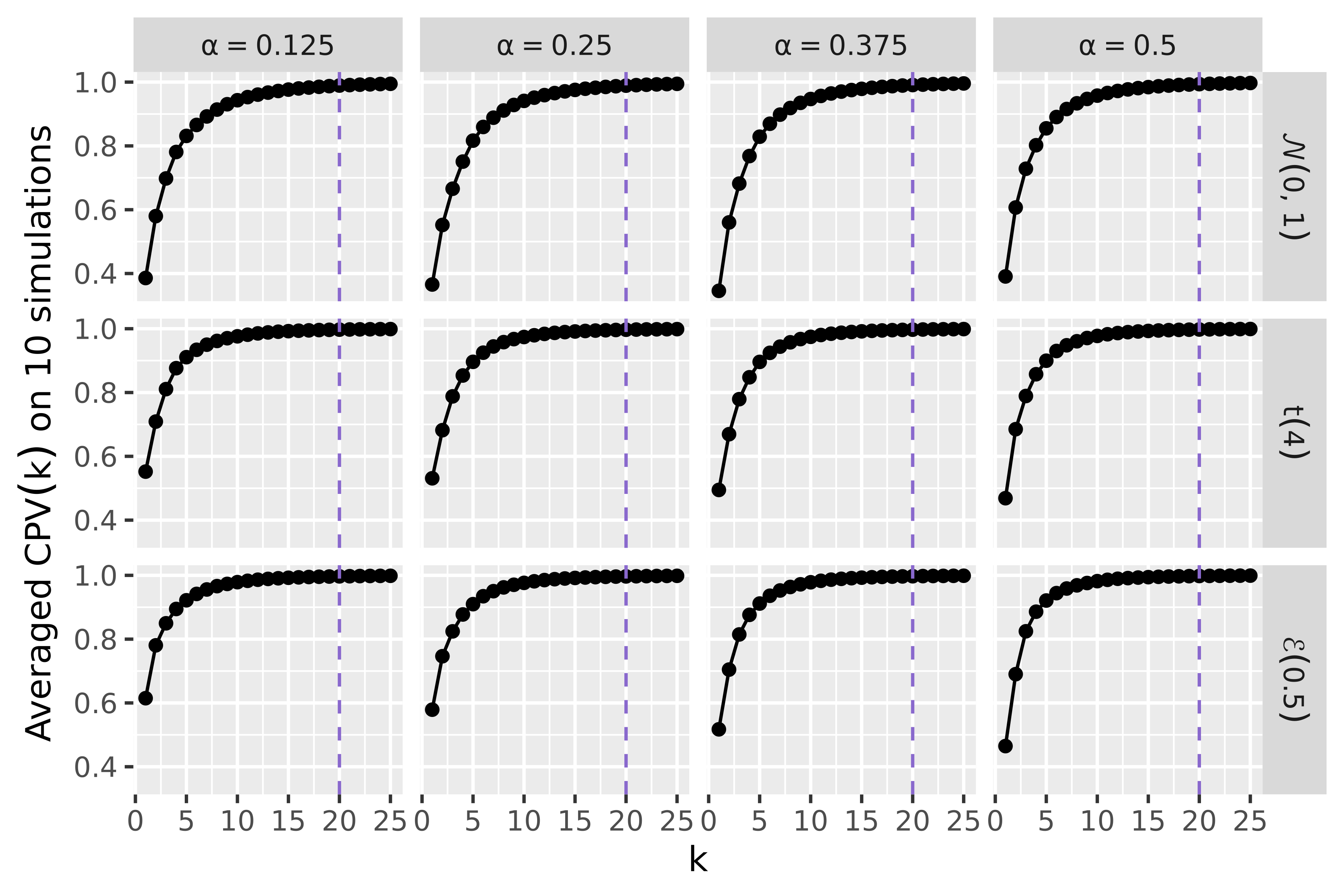}
\end{minipage}
\caption{The simulation study (univariate case): the cumulative inertia for the choice of $K$ in the SigFSS. The selected value is highlighted by a vertical line.}
\label{fig:simuKSigFSS}
\end{figure}

\begin{figure}[H]
\begin{minipage}{0.49\linewidth}
\centering $\Delta_1$
\includegraphics[width=\linewidth]{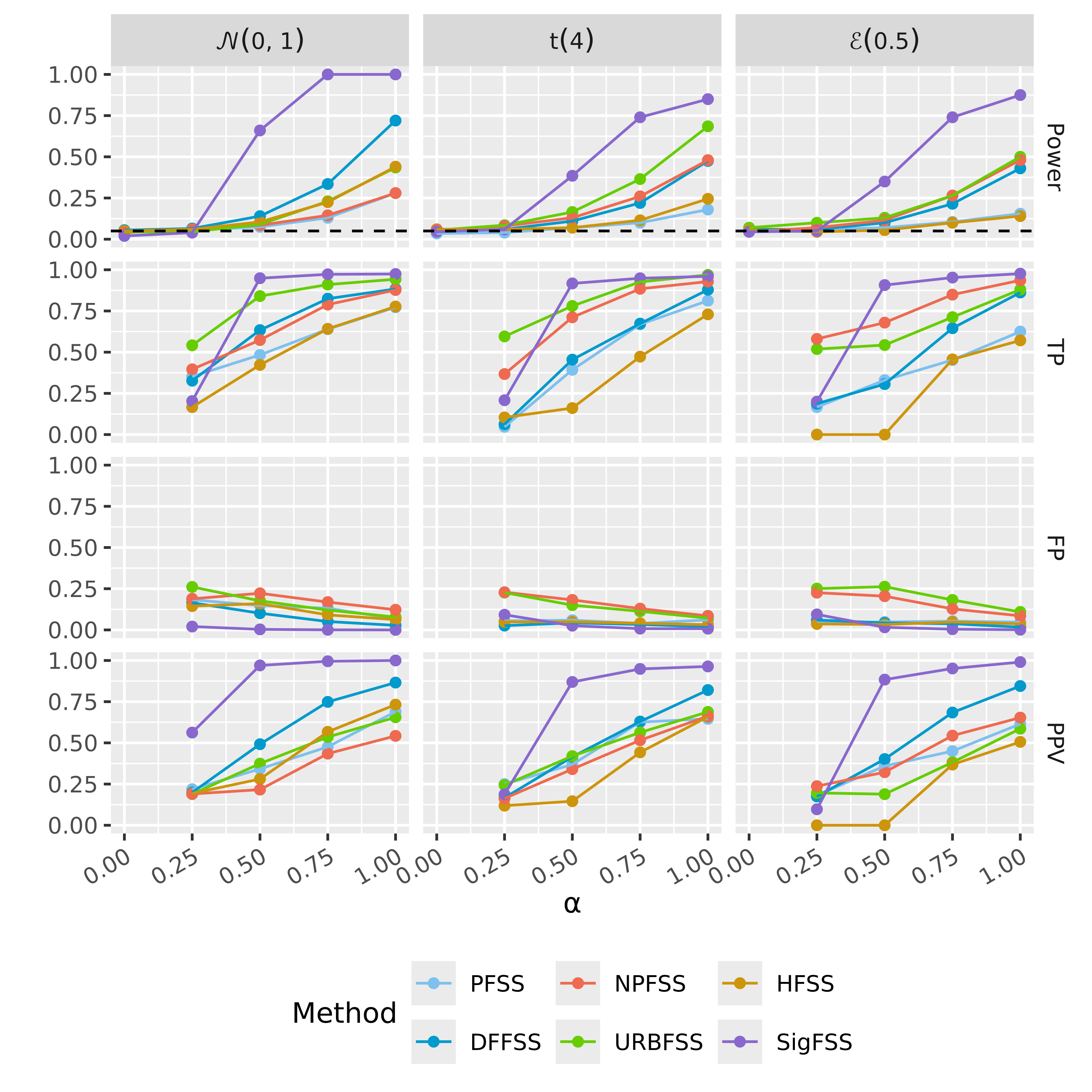}
\end{minipage}
\begin{minipage}{0.49\linewidth}
\centering $\Delta_2$
\includegraphics[width=\linewidth]{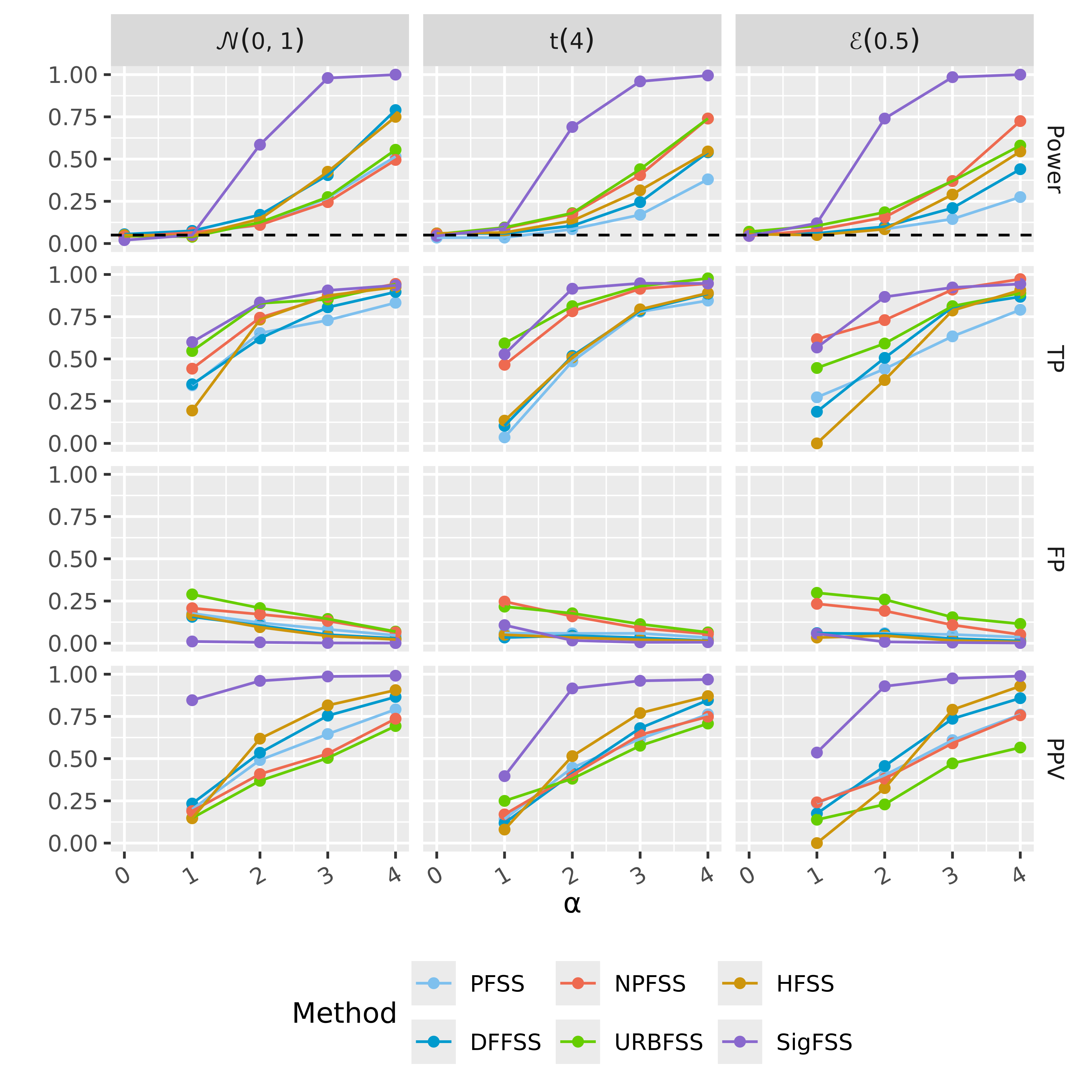}
\end{minipage}
\begin{minipage}{0.49\linewidth}
\centering $\Delta_3$
\includegraphics[width=\linewidth]{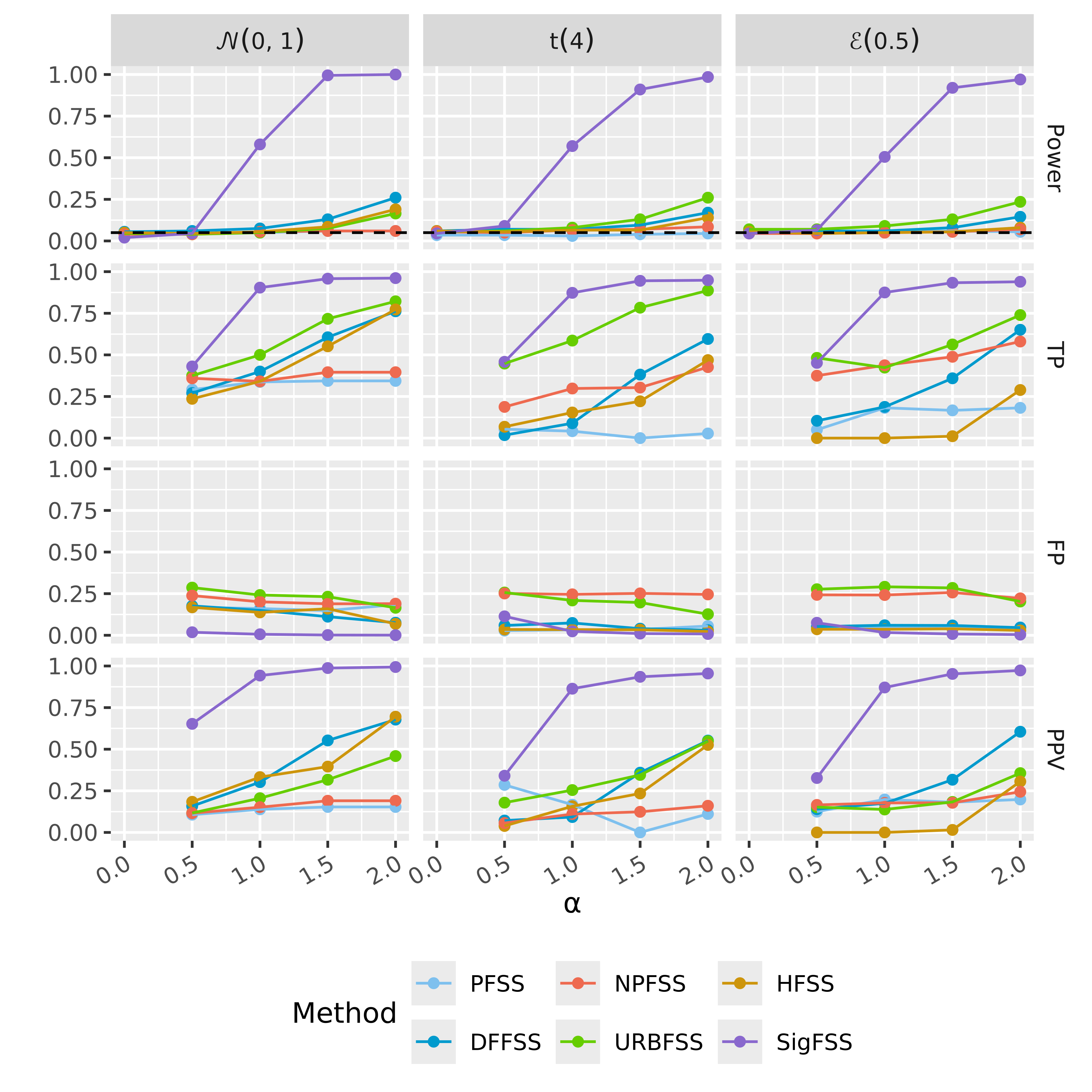}
\end{minipage}
\begin{minipage}{0.49\linewidth}
\centering $\Delta_4$
\includegraphics[width=\linewidth]{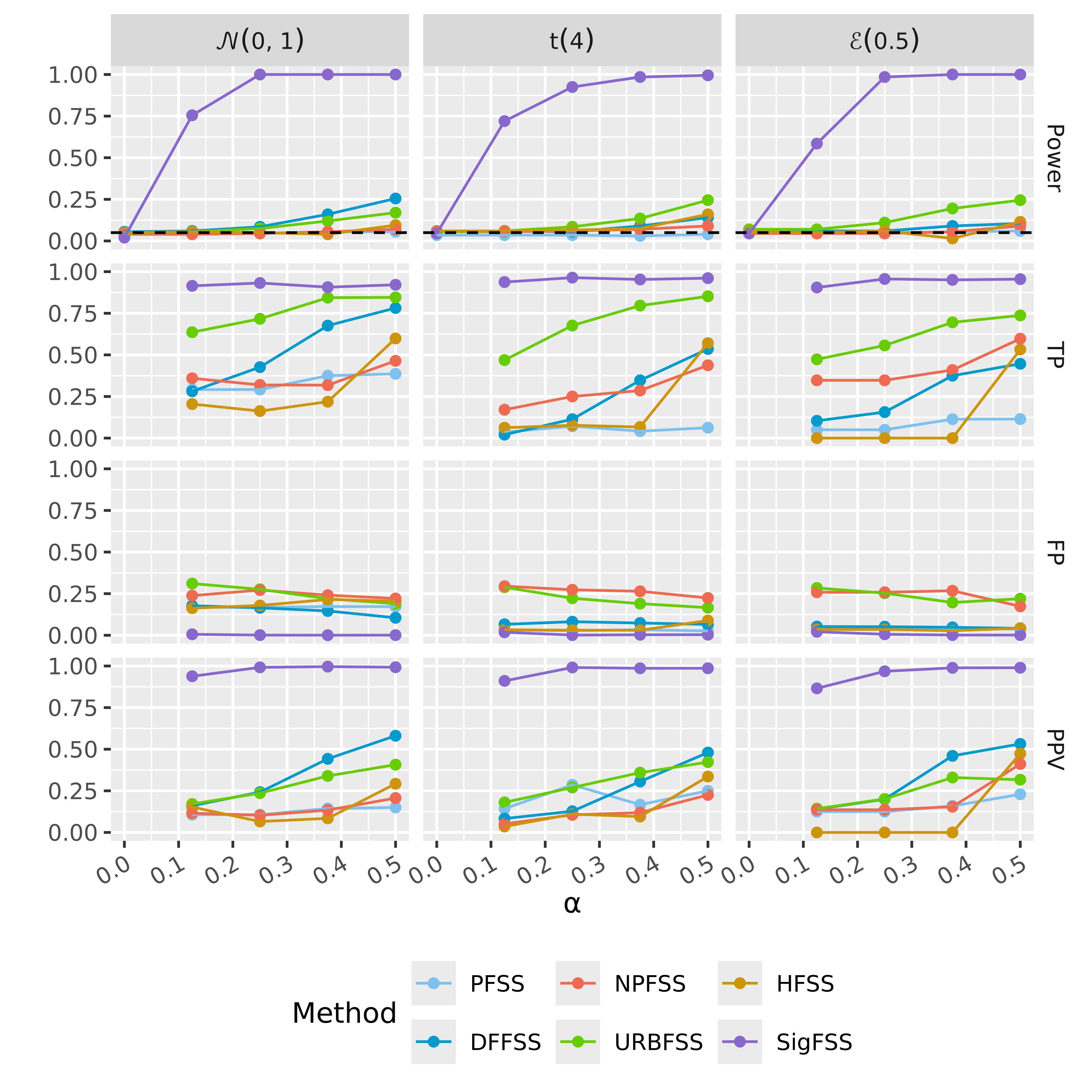}
\end{minipage}
\caption{The simulation study (univariate case): a comparison of the SigFSS, HFSS, NPFSS, DFFSS, URBFSS, and PFSS methods for detection of the spatial cluster as the MLC, when considering a fixed threshold of 95\% for the cumulative inertia in the SigFSS and the HFSS. $\alpha$ is the parameter that controls the cluster intensity.}
\label{fig:uni95}
\end{figure}

\begin{figure}[H]
\begin{minipage}{0.49\linewidth}
\centering $\Delta_1$
\includegraphics[width=\linewidth]{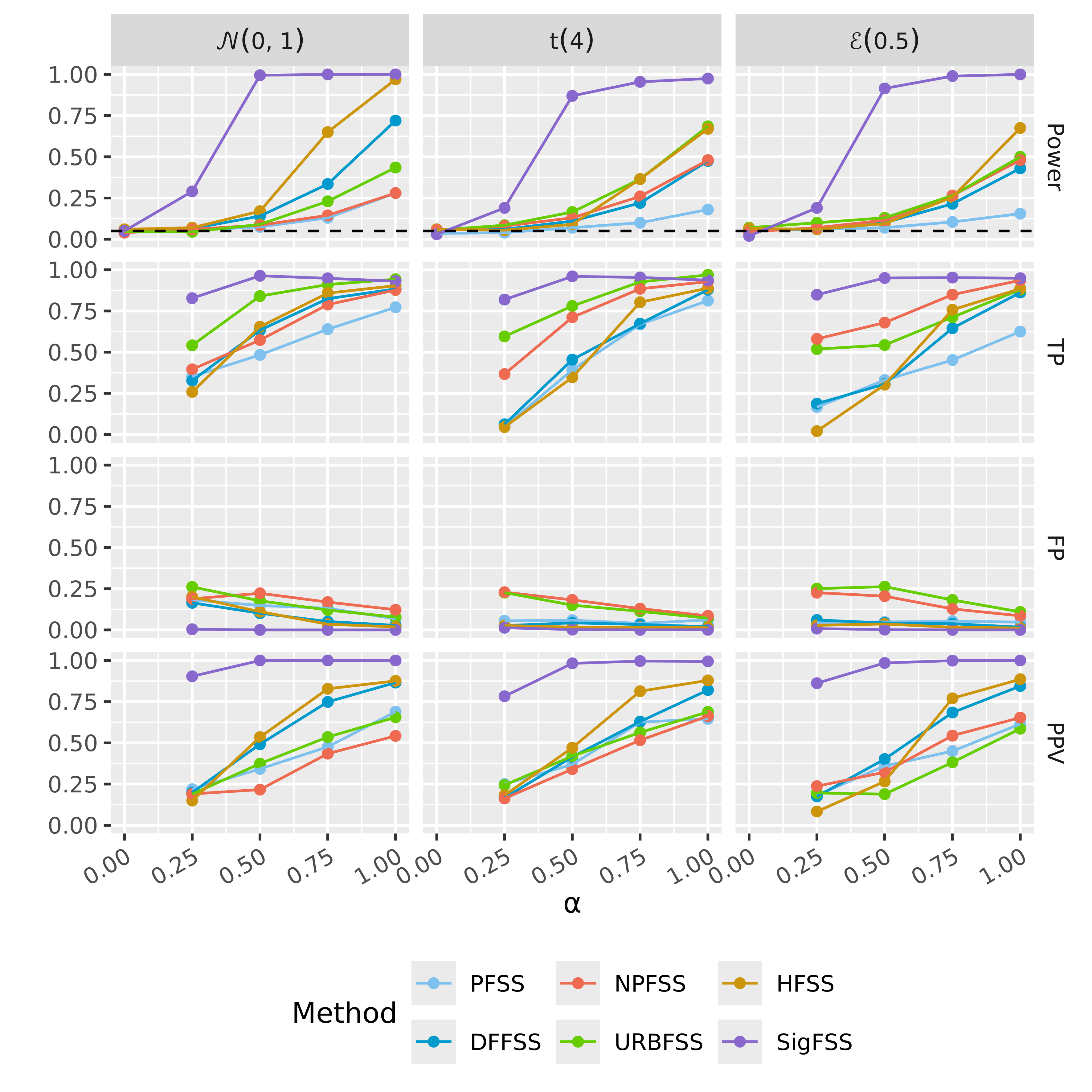}
\end{minipage}
\begin{minipage}{0.49\linewidth}
\centering $\Delta_2$
\includegraphics[width=\linewidth]{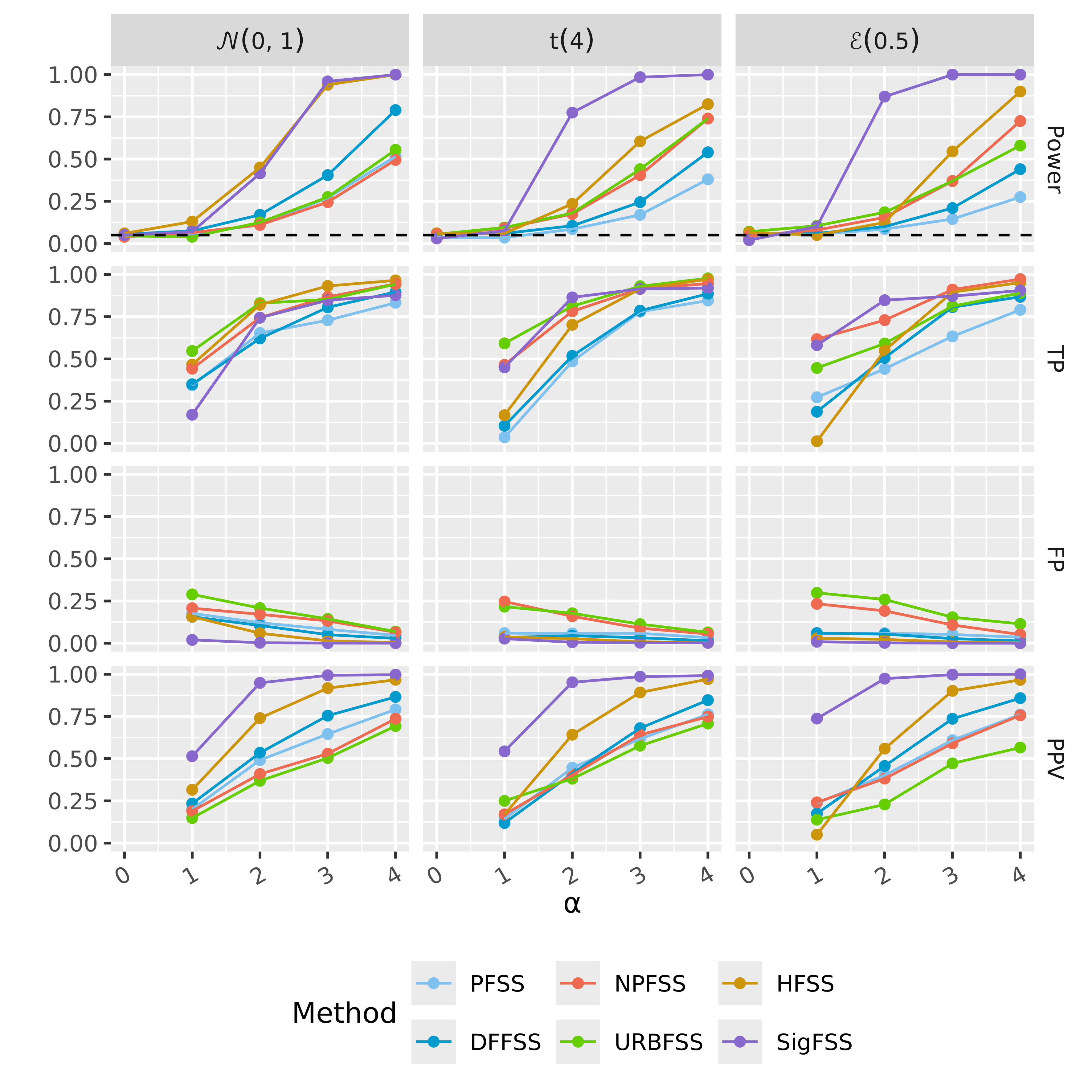}
\end{minipage}
\begin{minipage}{0.49\linewidth}
\centering $\Delta_3$
\includegraphics[width=\linewidth]{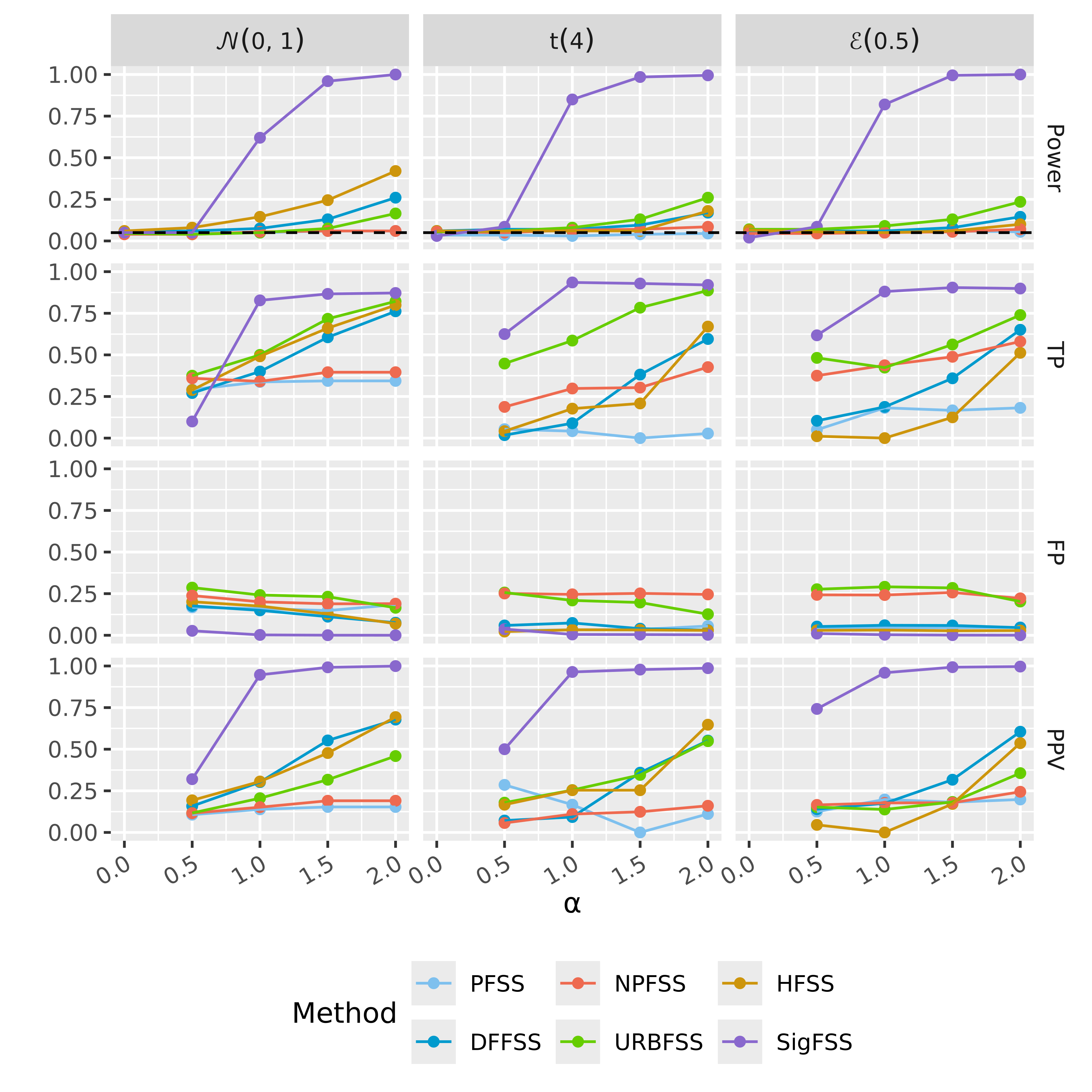}
\end{minipage}
\begin{minipage}{0.49\linewidth}
\centering $\Delta_4$
\includegraphics[width=\linewidth]{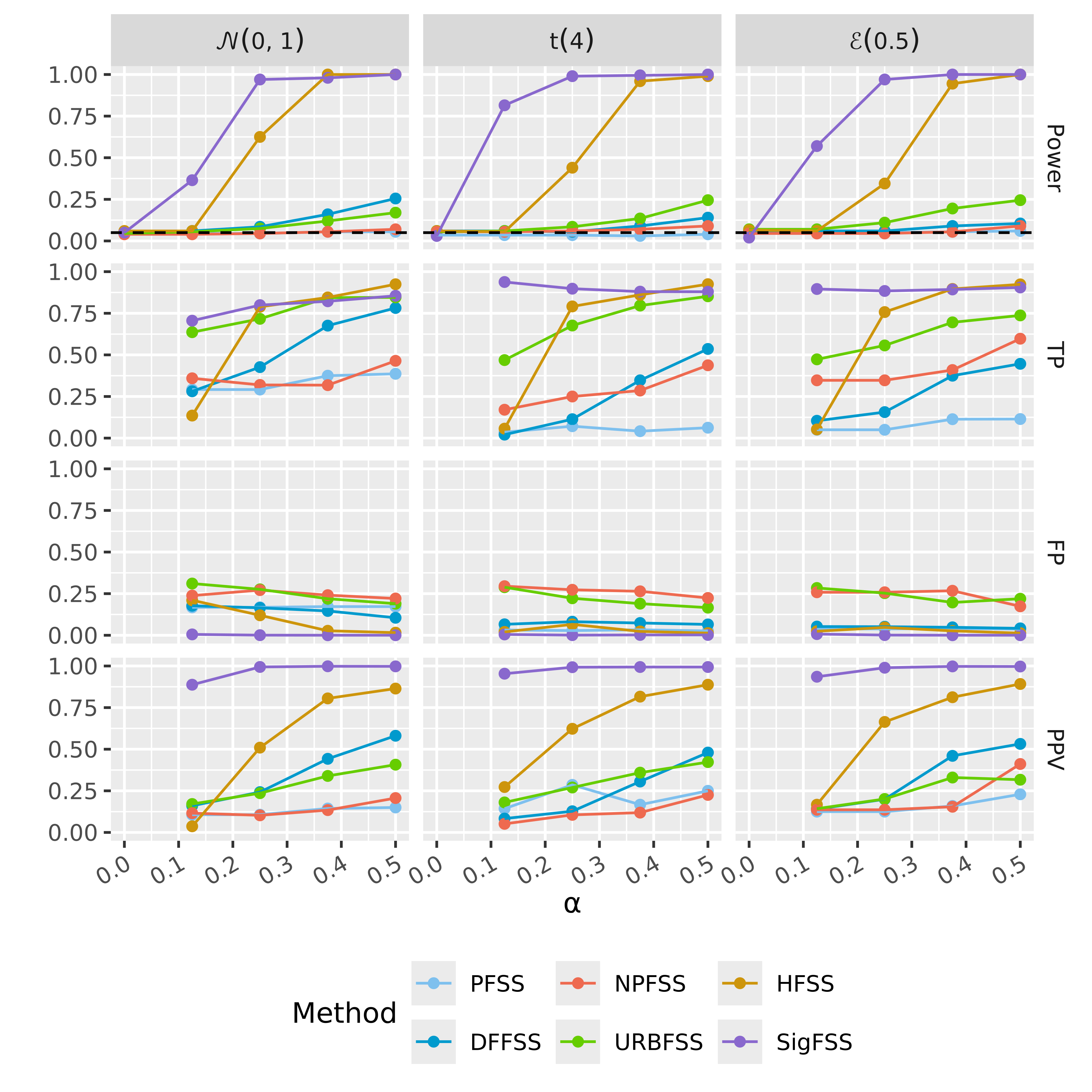}
\end{minipage}
\caption{The simulation study (univariate case): a comparison of the SigFSS, HFSS, NPFSS, DFFSS, URBFSS, and PFSS methods for detection of the spatial cluster as the MLC, when considering a fixed threshold of 99\% for the cumulative inertia in the SigFSS and the HFSS. $\alpha$ is the parameter that controls the cluster intensity.}
\label{fig:uni99}
\end{figure}

\begin{figure}[H]
\begin{minipage}{0.49\linewidth}
\centering $\Delta_1$
\includegraphics[width=\linewidth]{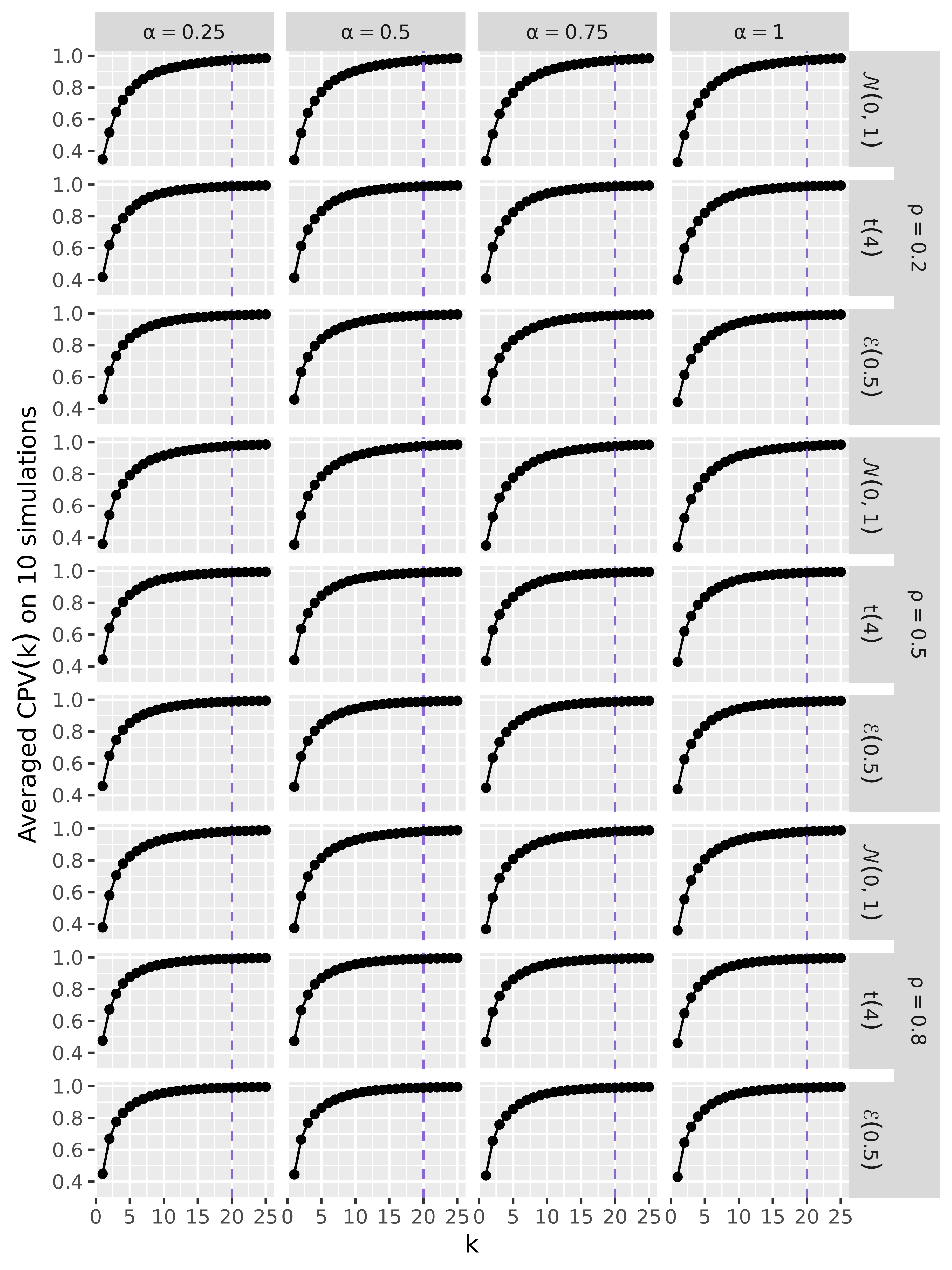}
\end{minipage}
\begin{minipage}{0.49\linewidth}
\centering $\Delta_2$
\includegraphics[width=\linewidth]{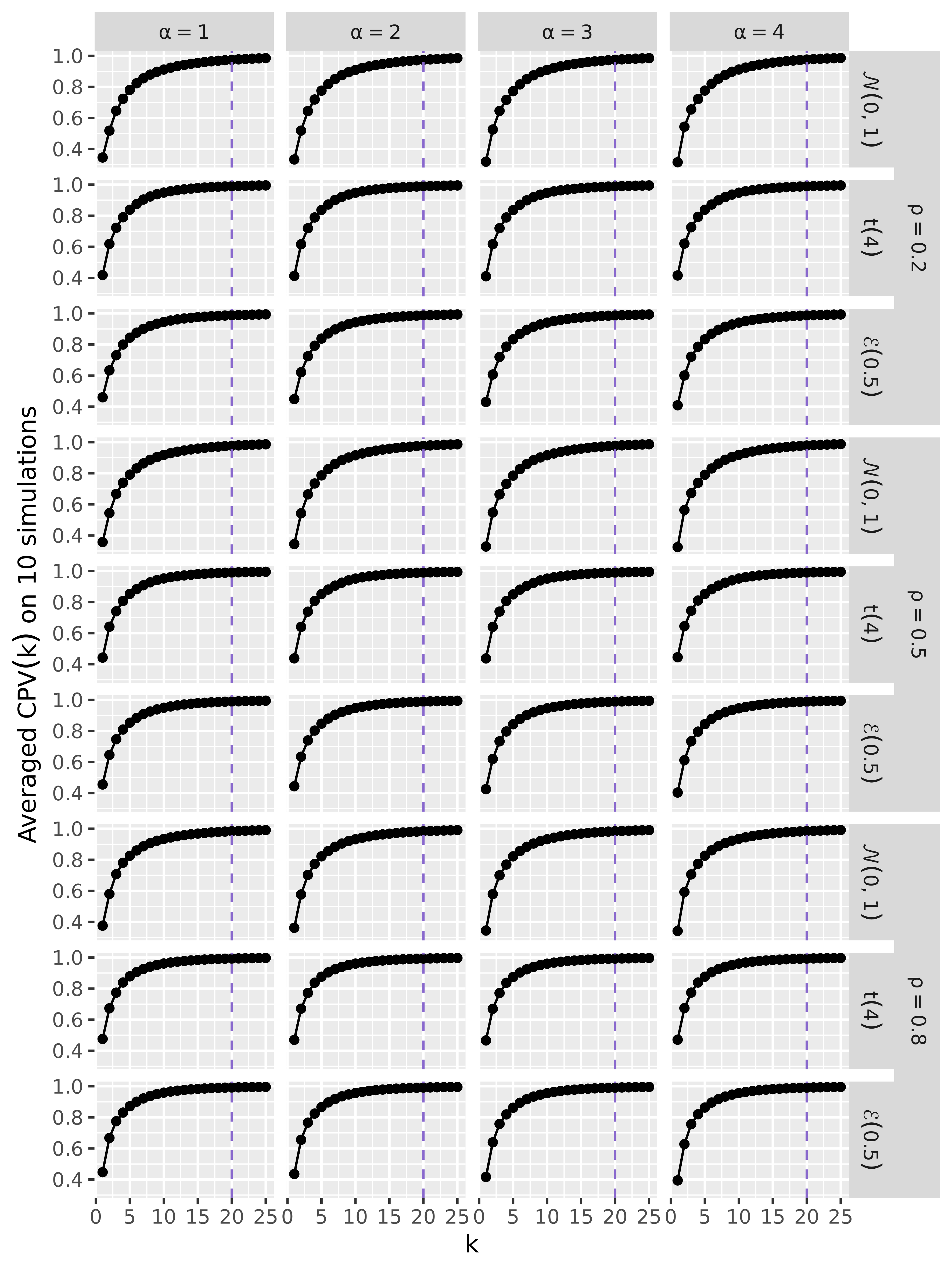}
\end{minipage}
\begin{minipage}{0.49\linewidth}
\centering $\Delta_3$
\includegraphics[width=\linewidth]{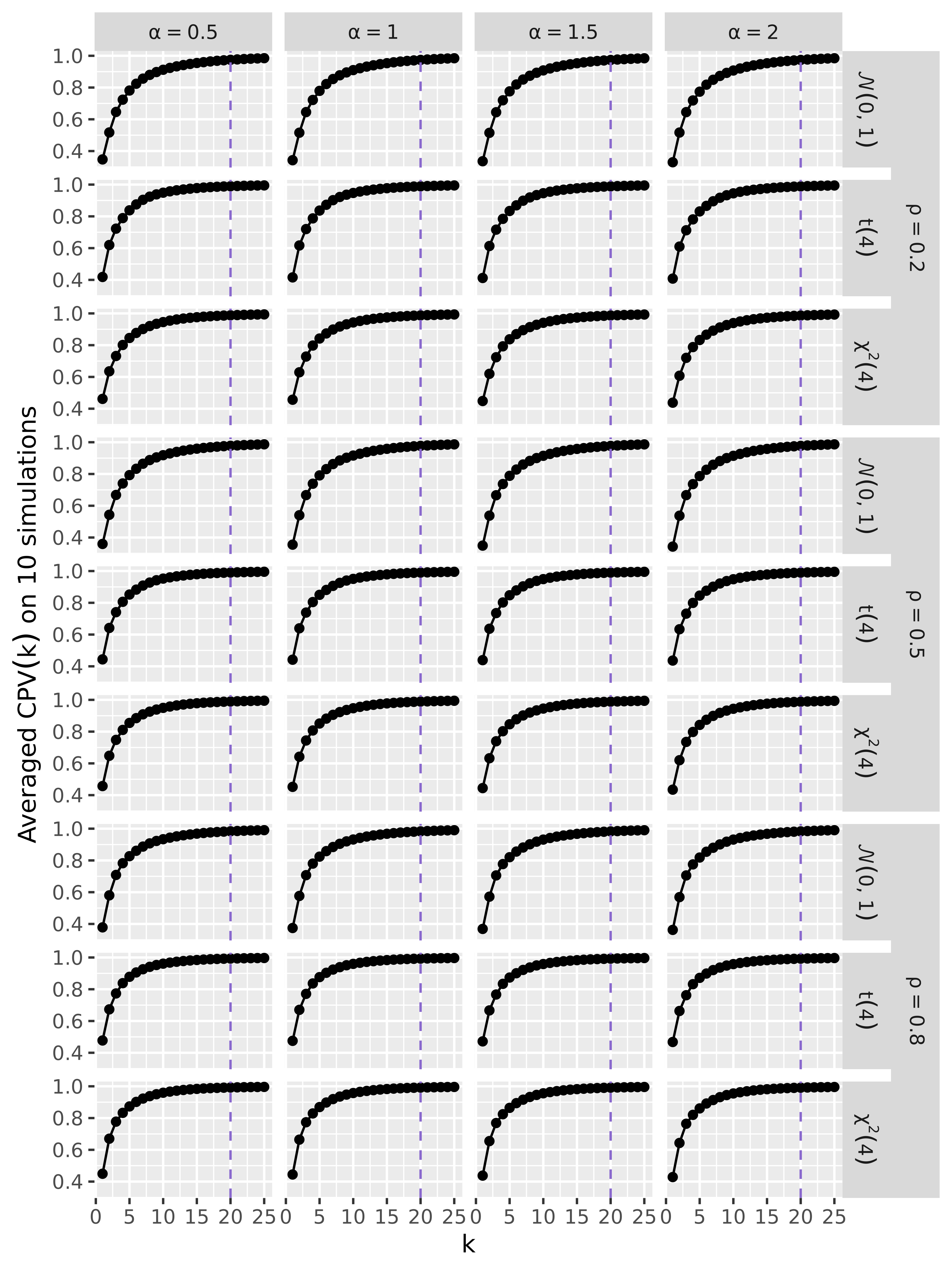}
\end{minipage}
\begin{minipage}{0.49\linewidth}
\centering $\Delta_4$
\includegraphics[width=\linewidth]{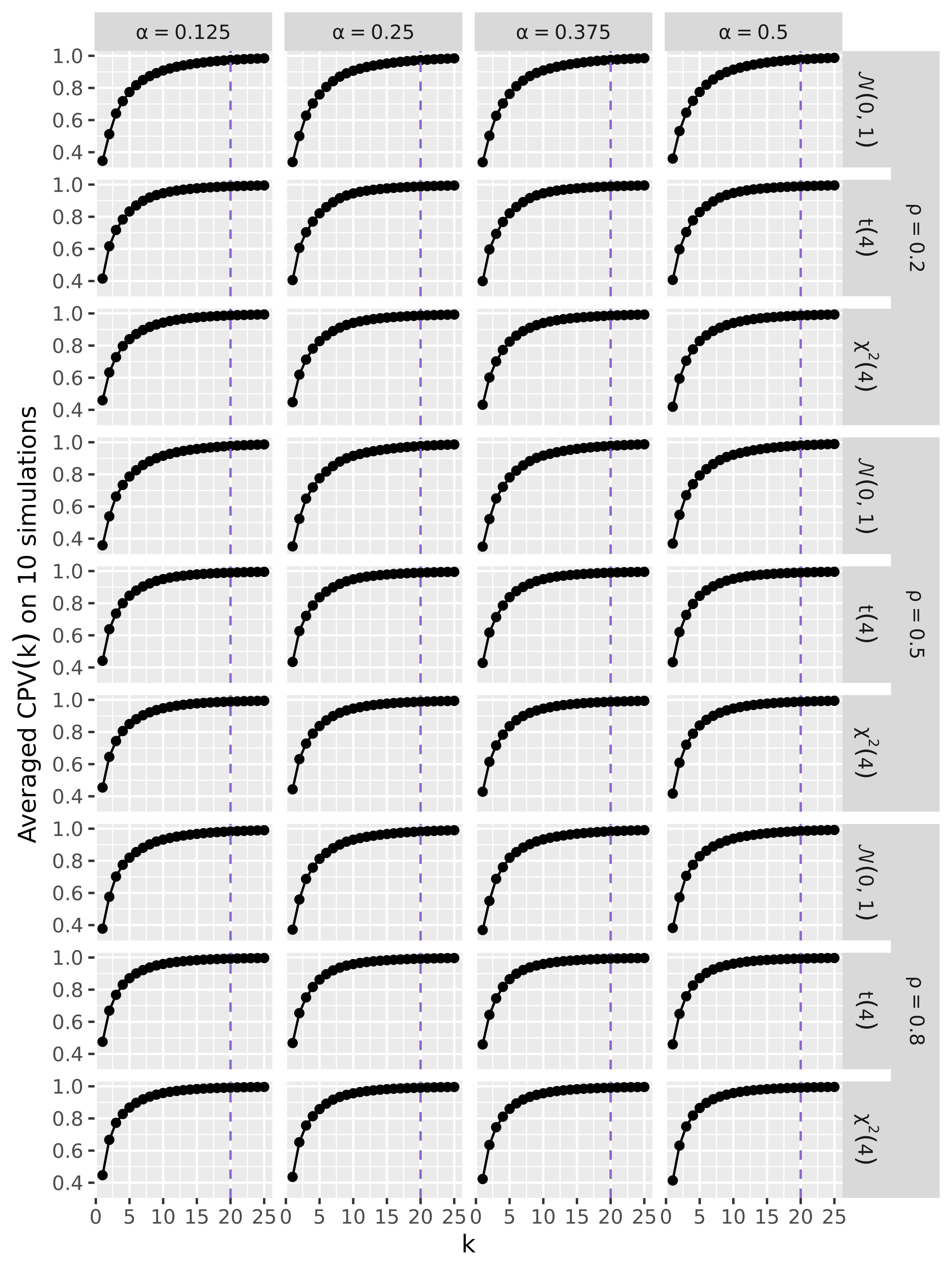}
\end{minipage}
\caption{The simulation study (multivariate case): the cumulative inertia for the choice of $K$ in the SigFSS. The selected value is highlighted by a vertical line.}
\label{fig:simuKSigFSSmulti}
\end{figure}

\begin{figure}[H]
\begin{minipage}{0.49\linewidth}
\centering $\Delta_1$
\includegraphics[width=\linewidth]{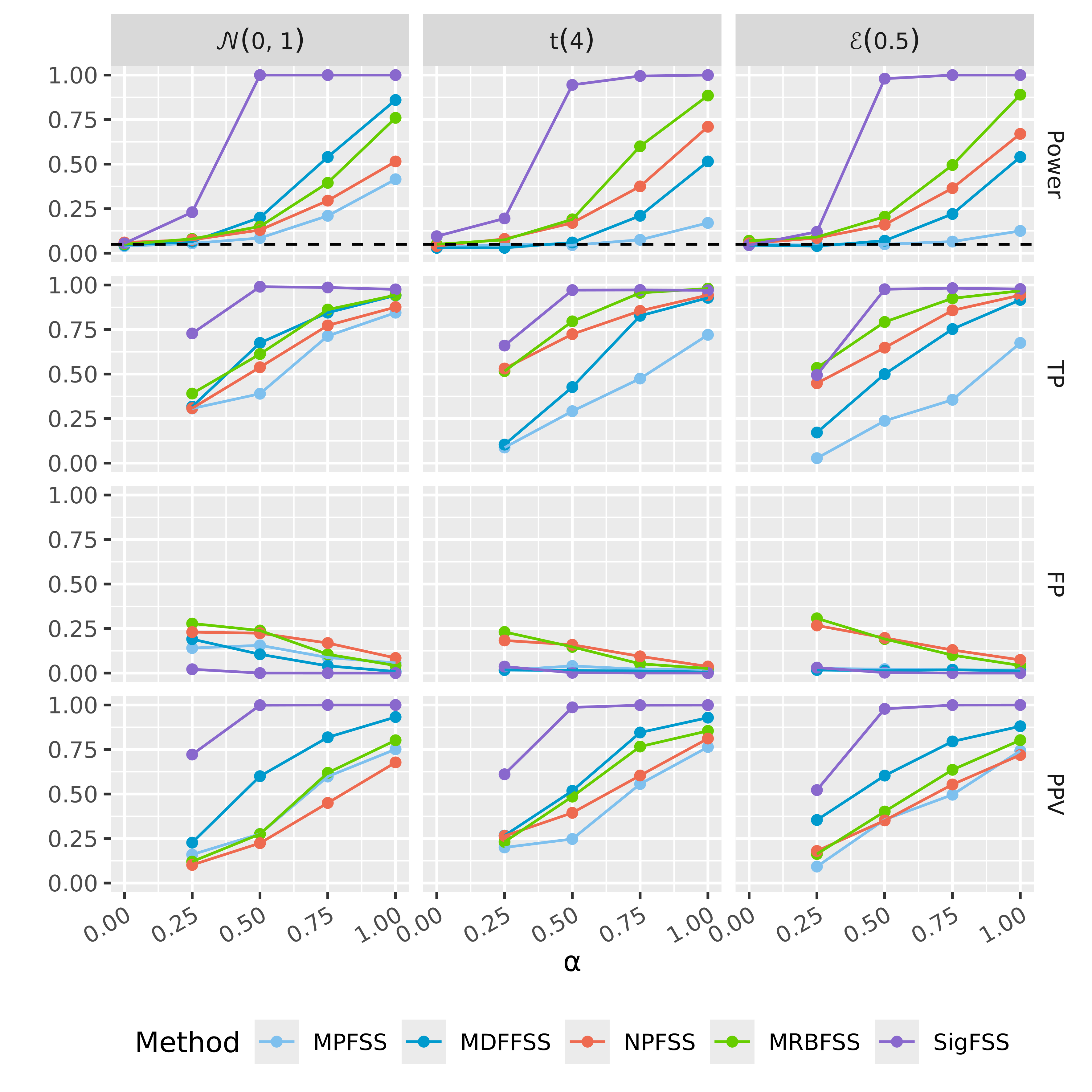}
\end{minipage}
\begin{minipage}{0.49\linewidth}
\centering $\Delta_2$
\includegraphics[width=\linewidth]{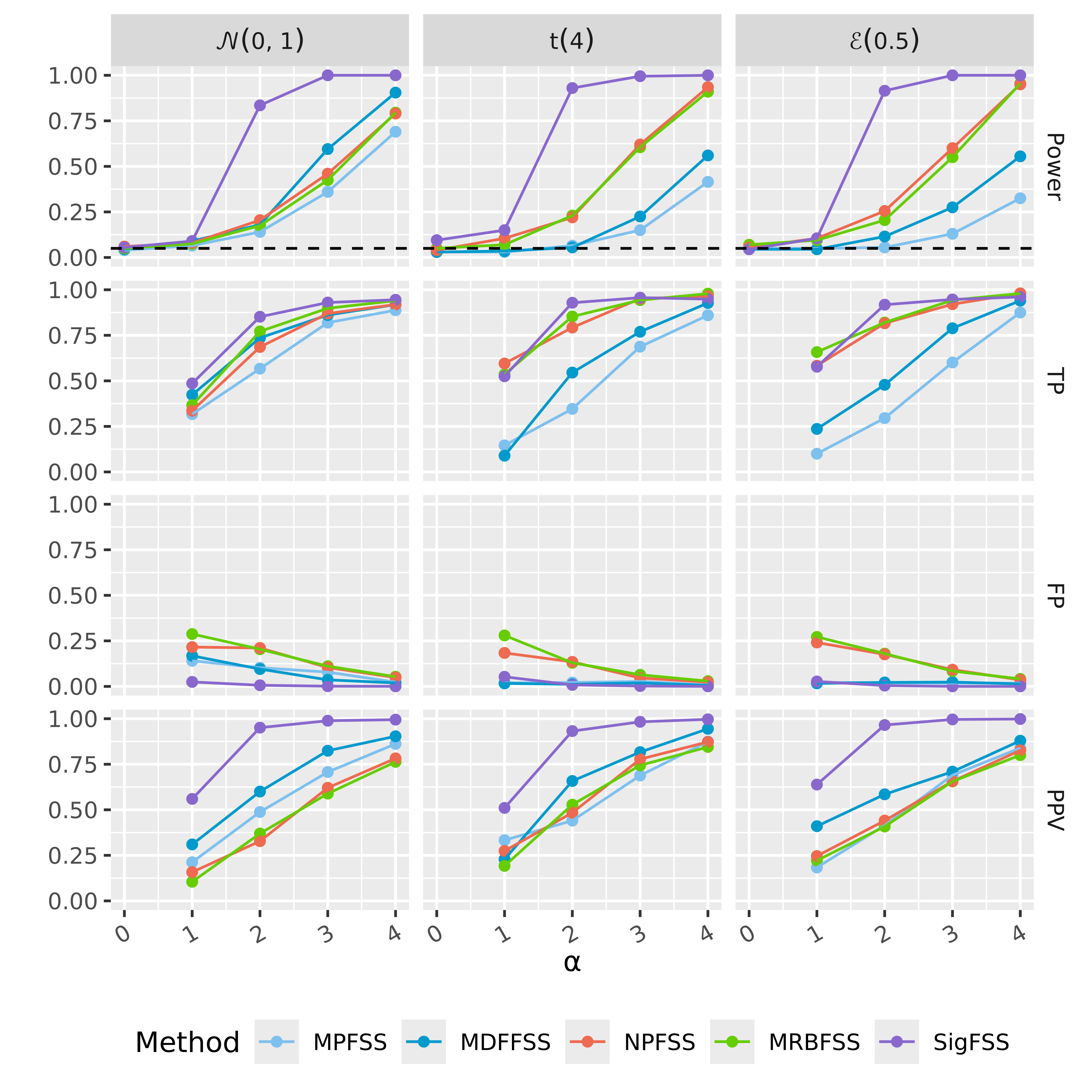}
\end{minipage}
\begin{minipage}{0.49\linewidth}
\centering $\Delta_3$
\includegraphics[width=\linewidth]{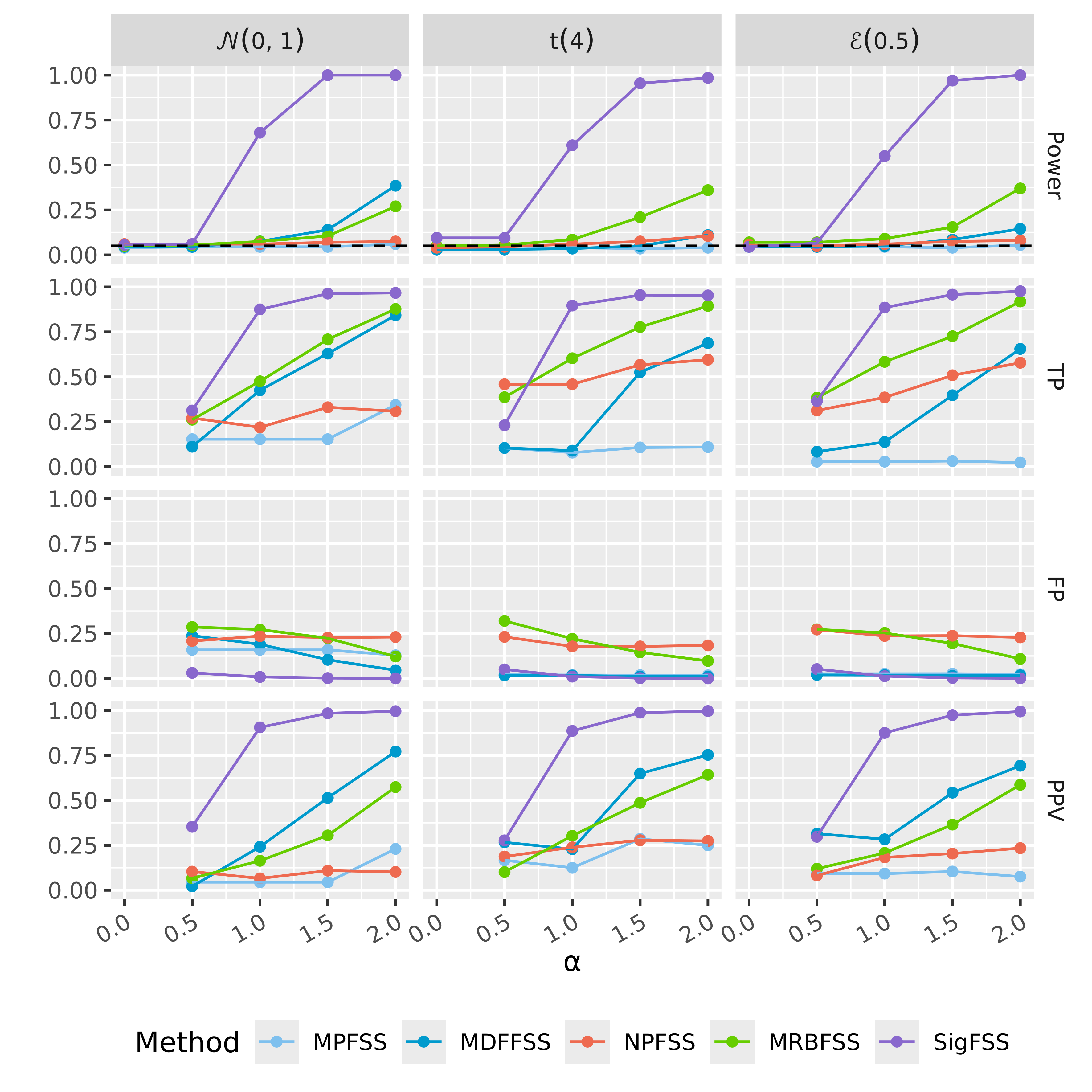}
\end{minipage}
\begin{minipage}{0.49\linewidth}
\centering $\Delta_4$
\includegraphics[width=\linewidth]{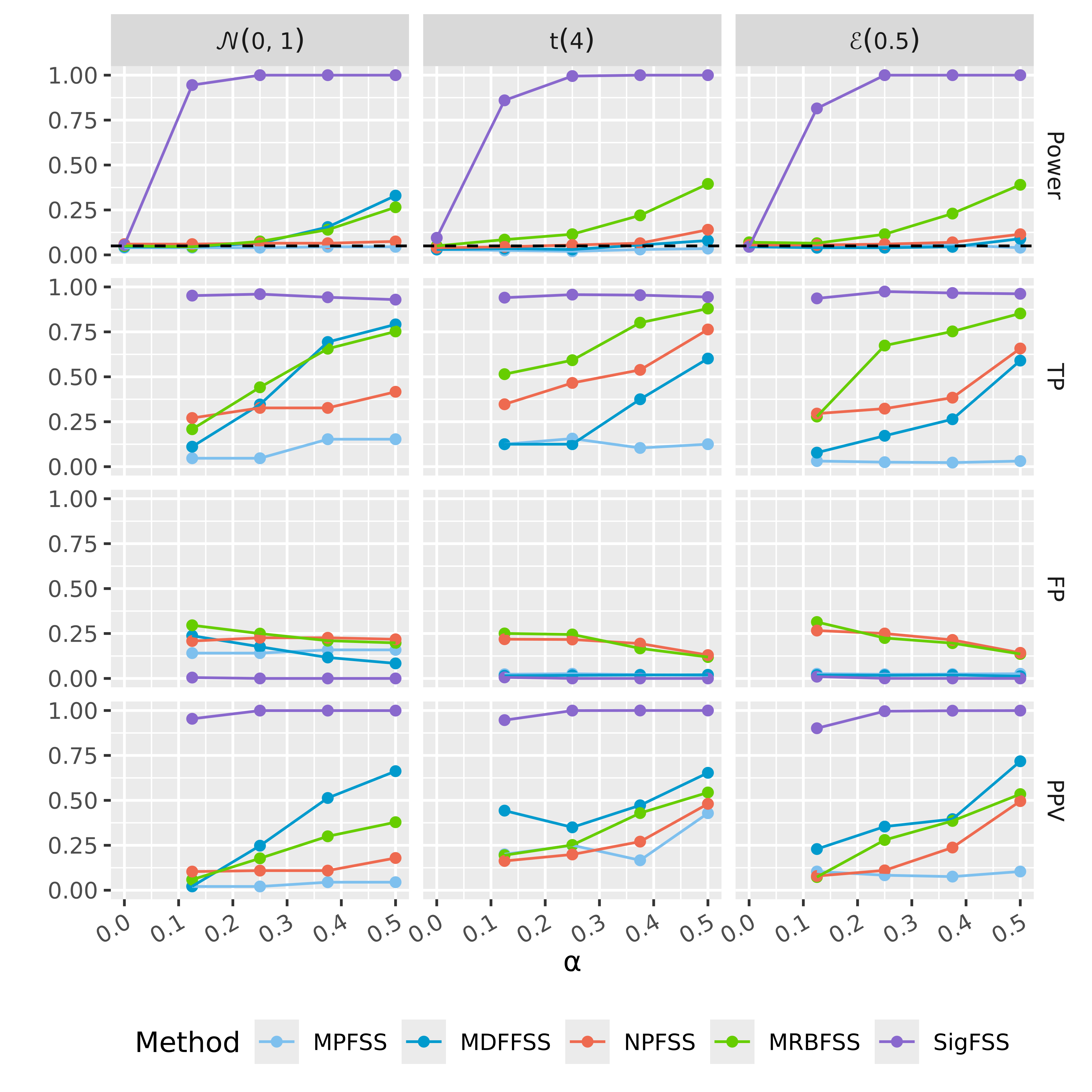}
\end{minipage}
\caption{The simulation study (multivariate case): a comparison of the SigFSS, NPFSS, MDFFSS, MRBFSS, and MPFSS methods for detection of the spatial cluster as the MLC, when $\rho = 0.2$ and considering a fixed threshold of 95\% for the cumulative inertia in the SigFSS. $\alpha$ is the parameter that controls the cluster intensity.}
\label{fig:rho0.2_95}
\end{figure}

\begin{figure}[H]
\begin{minipage}{0.49\linewidth}
\centering $\Delta_1$
\includegraphics[width=\linewidth]{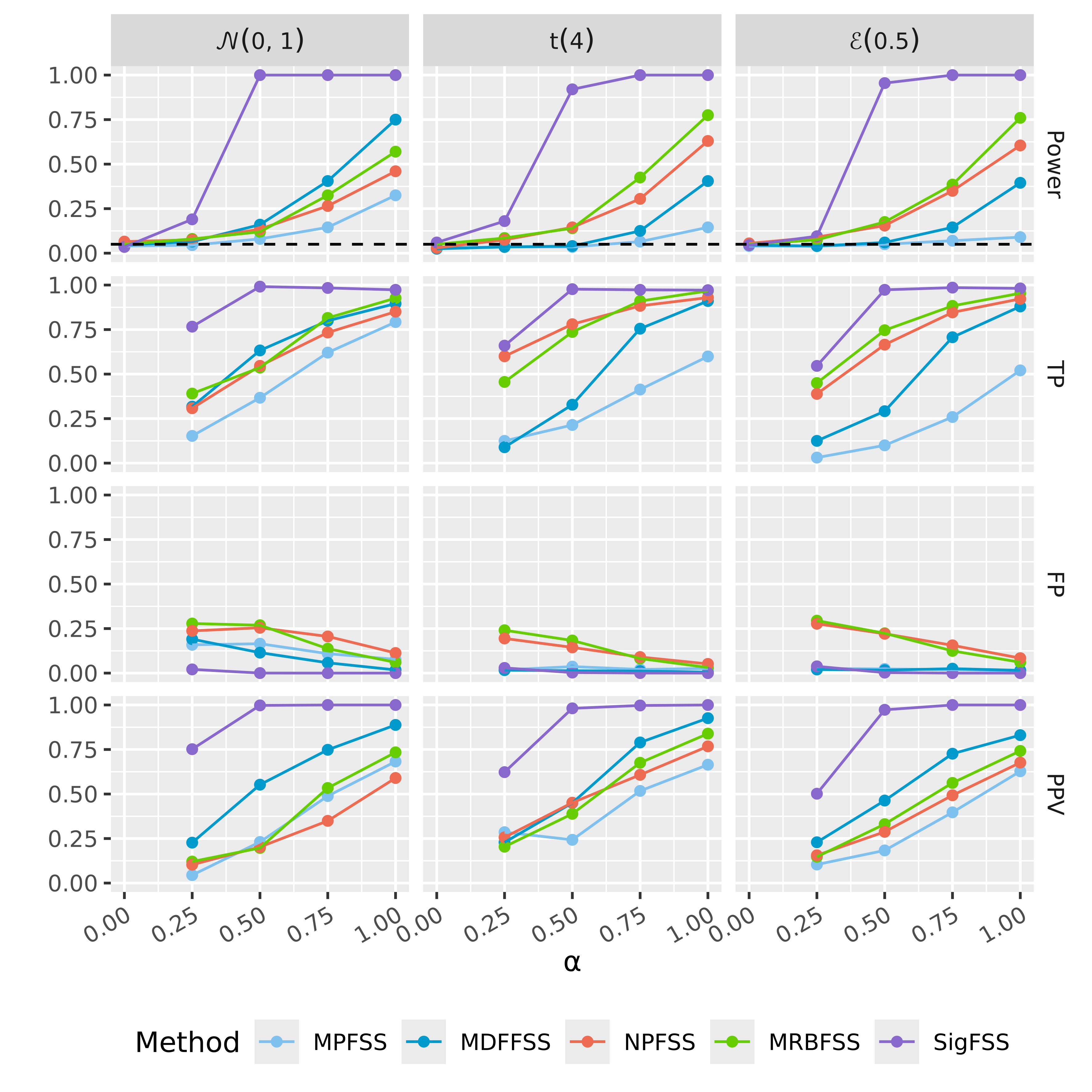}
\end{minipage}
\begin{minipage}{0.49\linewidth}
\centering $\Delta_2$
\includegraphics[width=\linewidth]{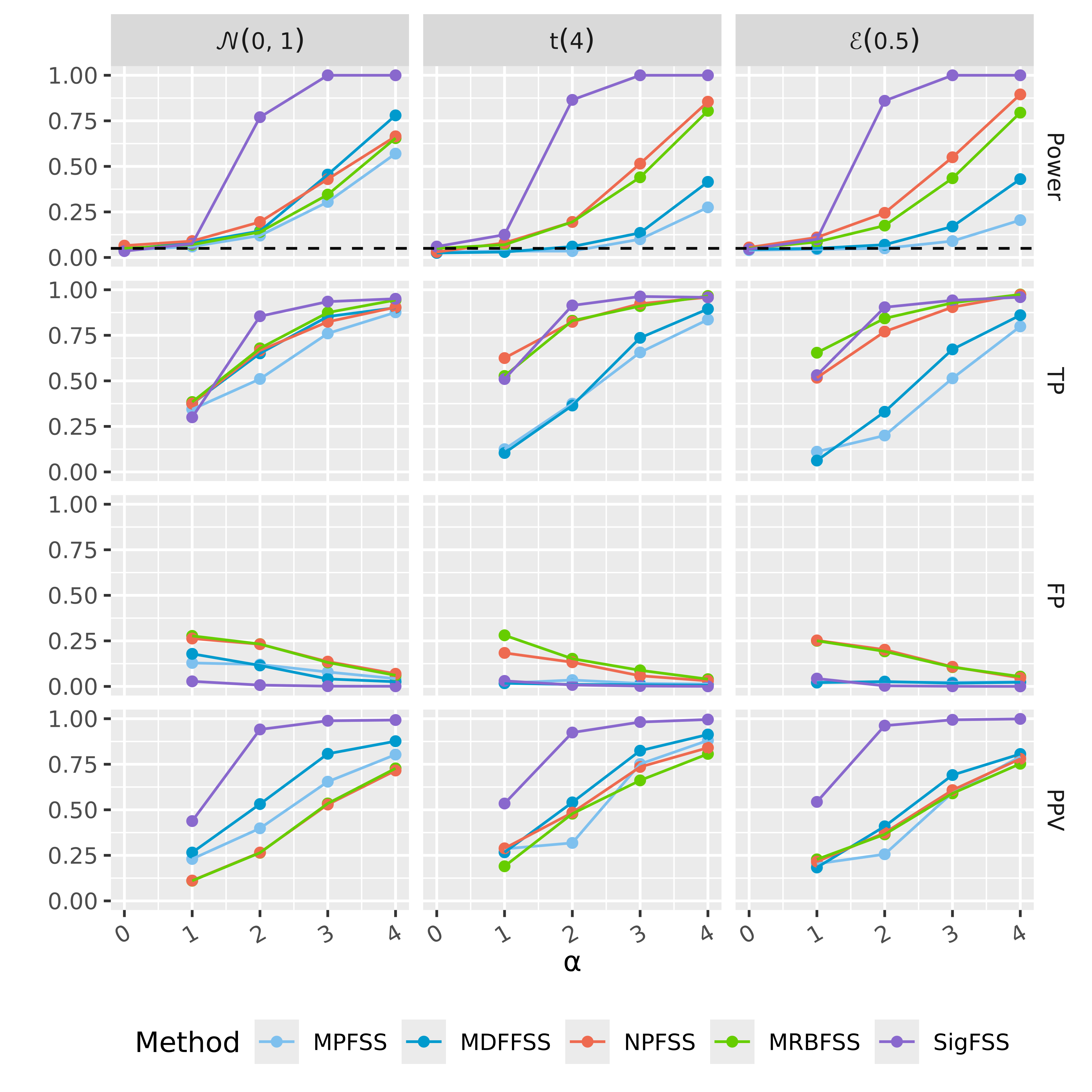}
\end{minipage}
\begin{minipage}{0.49\linewidth}
\centering $\Delta_3$
\includegraphics[width=\linewidth]{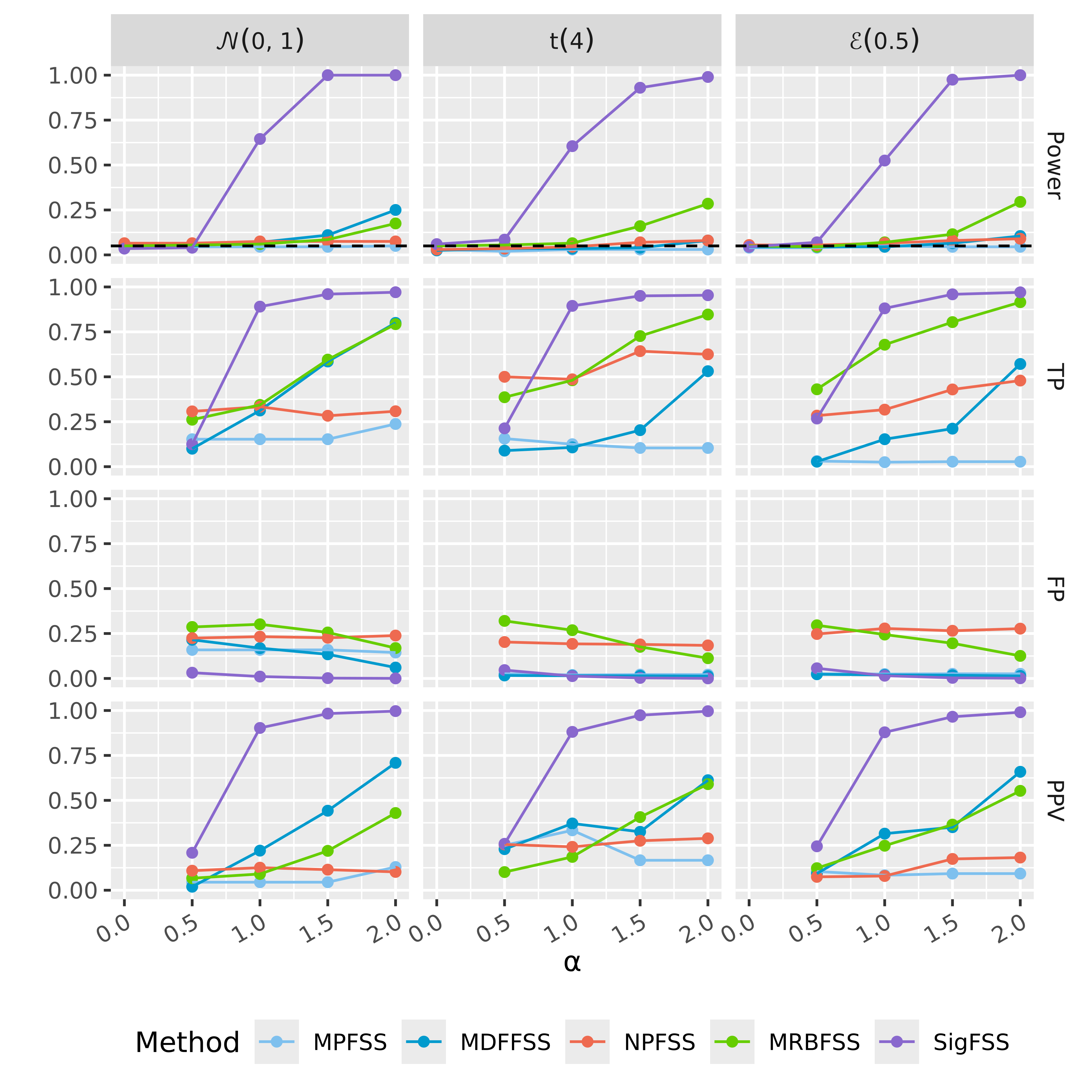}
\end{minipage}
\begin{minipage}{0.49\linewidth}
\centering $\Delta_4$
\includegraphics[width=\linewidth]{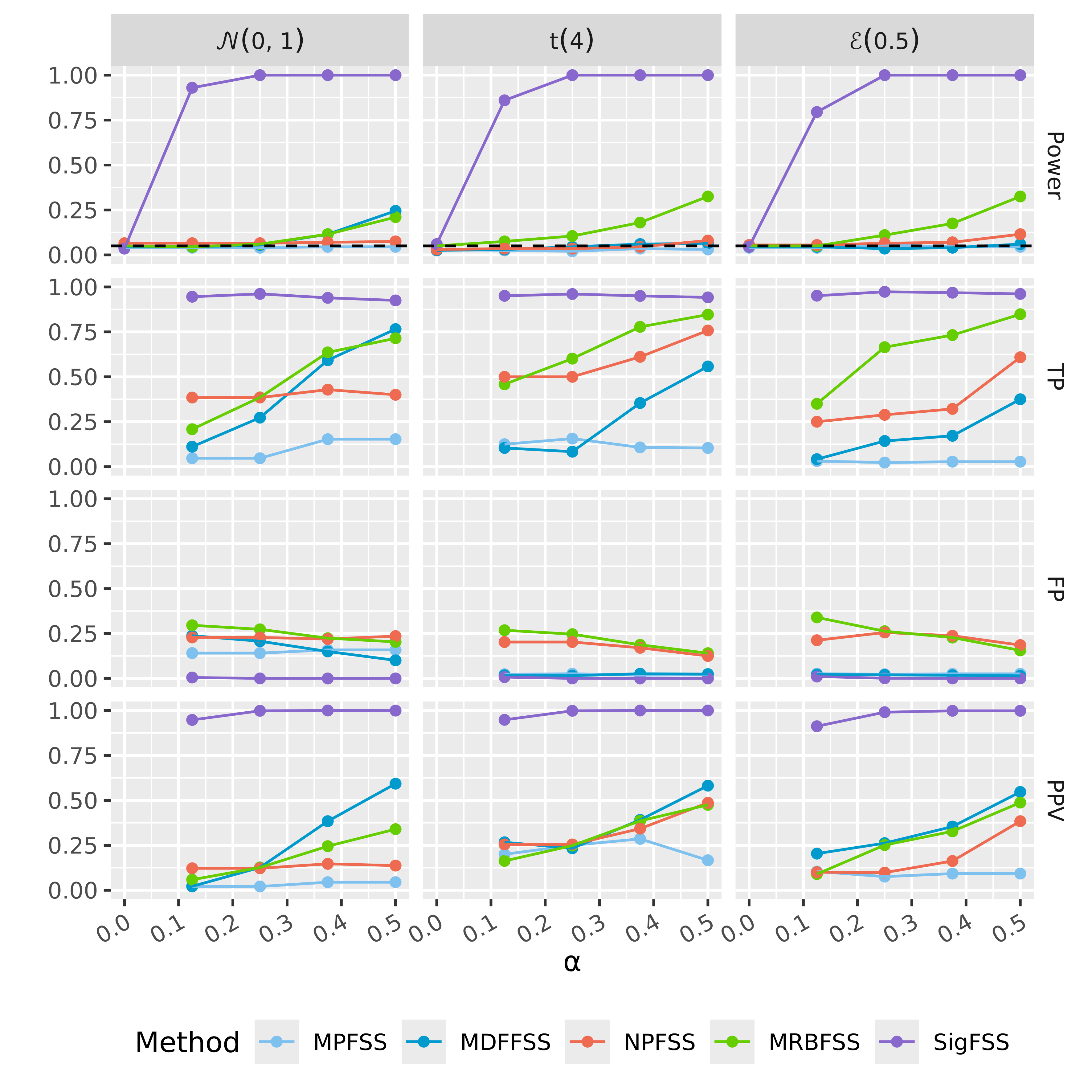}
\end{minipage}
\caption{The simulation study (multivariate case): a comparison of the SigFSS, NPFSS, MDFFSS, MRBFSS, and MPFSS methods for detection of the spatial cluster as the MLC, when $\rho = 0.5$ and considering a fixed threshold of 95\% for the cumulative inertia in the SigFSS. $\alpha$ is the parameter that controls the cluster intensity.}
\label{fig:rho0.5_95}
\end{figure}

\begin{figure}[H]
\begin{minipage}{0.49\linewidth}
\centering $\Delta_1$
\includegraphics[width=\linewidth]{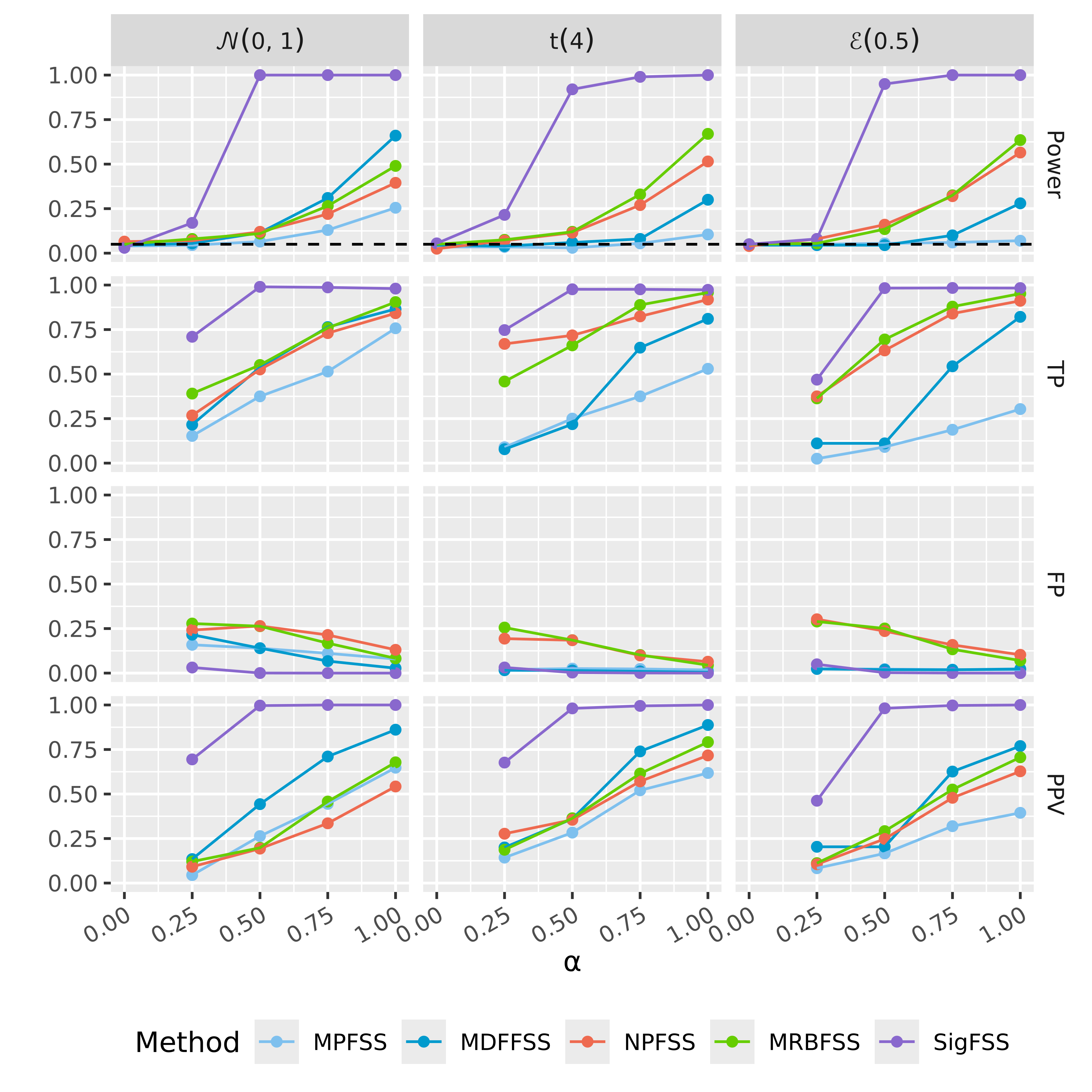}
\end{minipage}
\begin{minipage}{0.49\linewidth}
\centering $\Delta_2$
\includegraphics[width=\linewidth]{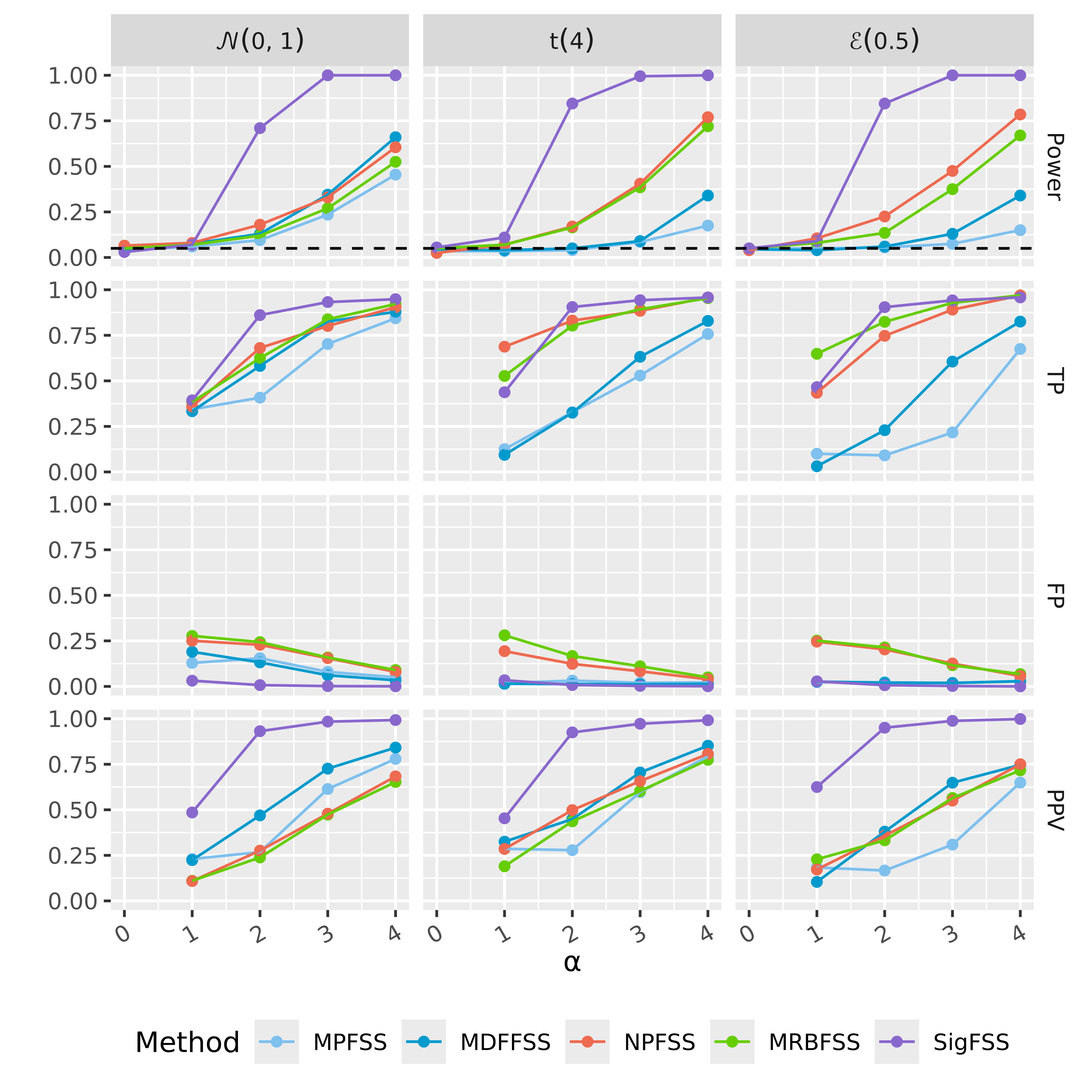}
\end{minipage}
\begin{minipage}{0.49\linewidth}
\centering $\Delta_3$
\includegraphics[width=\linewidth]{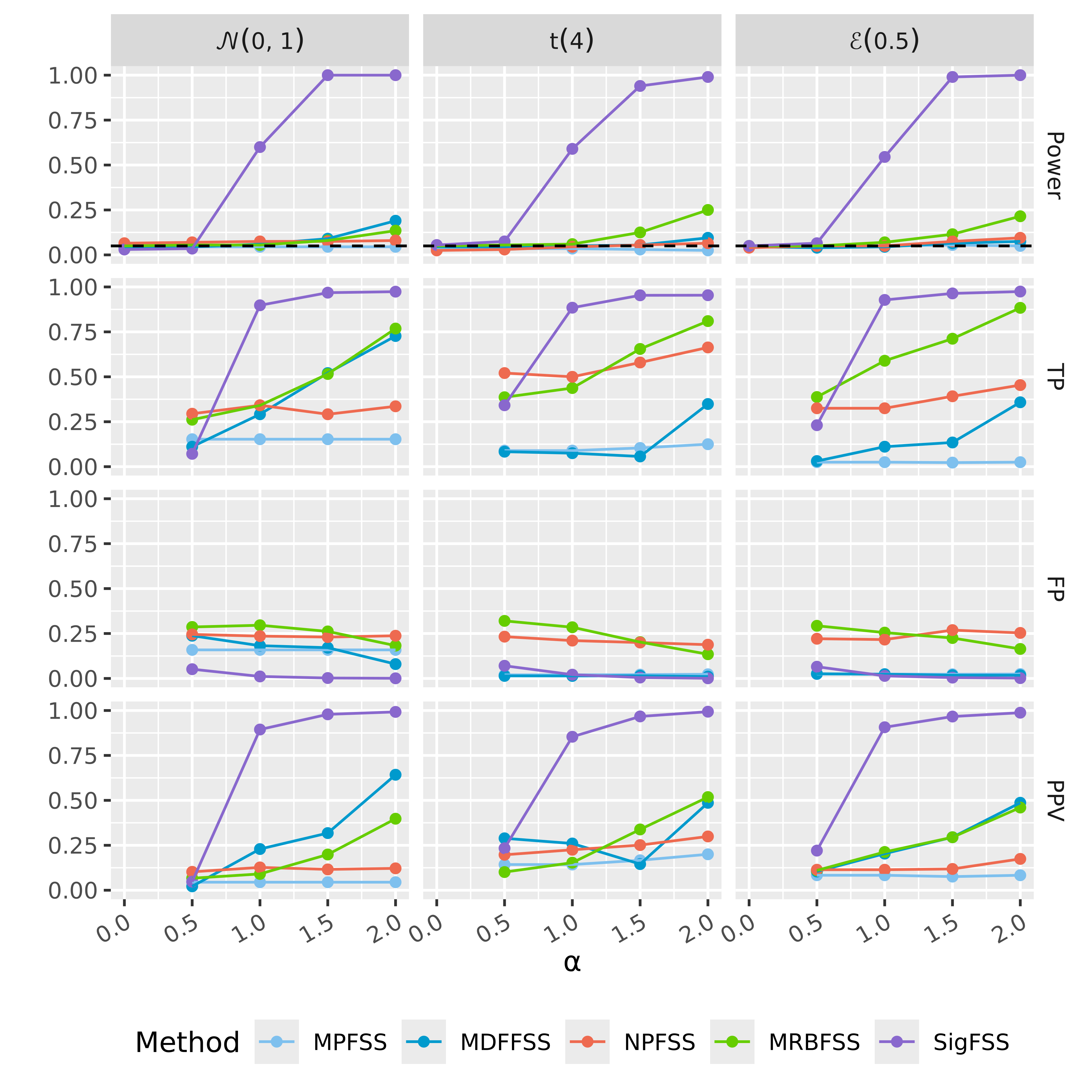}
\end{minipage}
\begin{minipage}{0.49\linewidth}
\centering $\Delta_4$
\includegraphics[width=\linewidth]{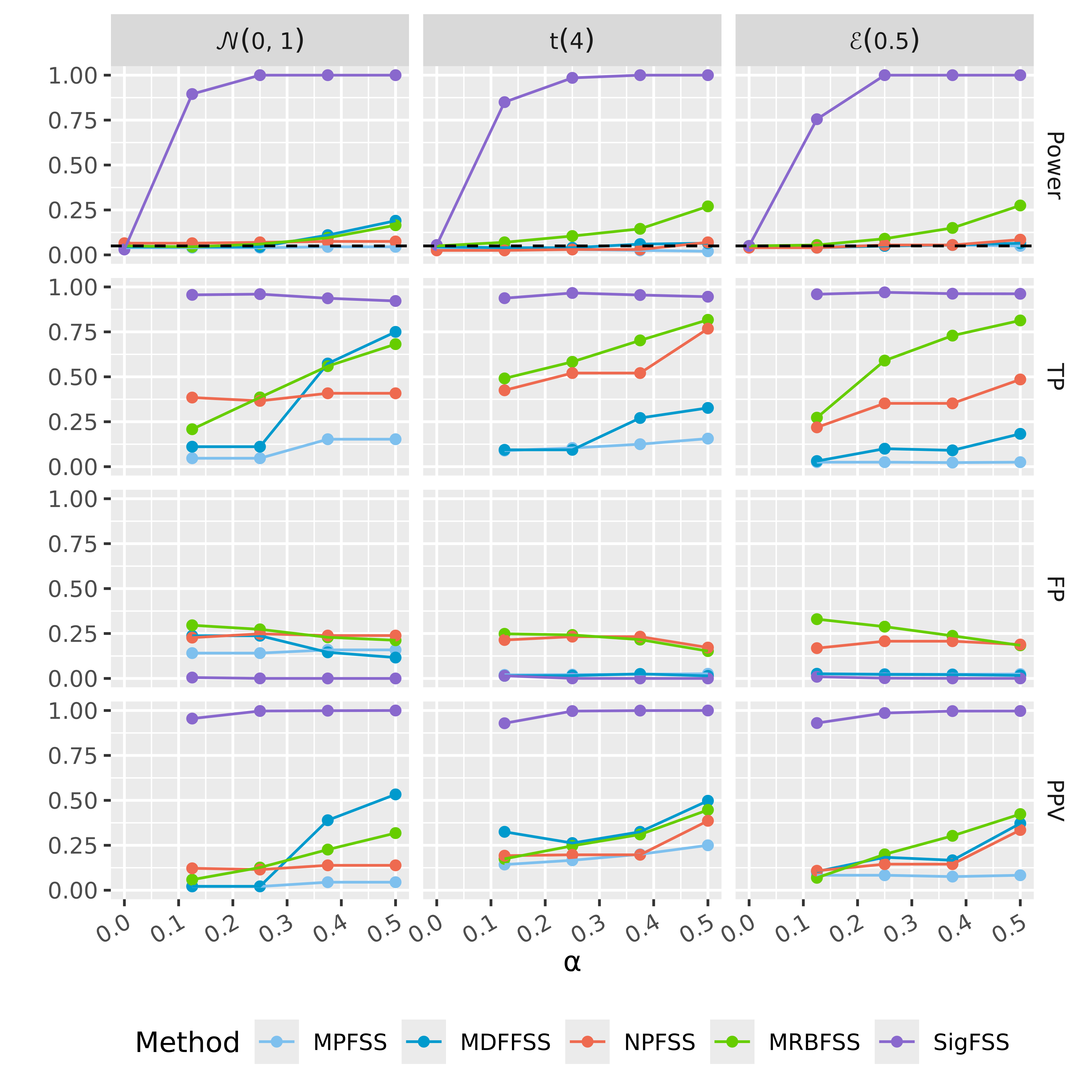}
\end{minipage}
\caption{The simulation study (multivariate case): a comparison of the SigFSS, NPFSS, MDFFSS, MRBFSS, and MPFSS methods for detection of the spatial cluster as the MLC, when $\rho = 0.8$ and considering a fixed threshold of 95\% for the cumulative inertia in the SigFSS. $\alpha$ is the parameter that controls the cluster intensity.}
\label{fig:rho0.8_95}
\end{figure}

\begin{figure}[H]
\begin{minipage}{0.49\linewidth}
\centering $\Delta_1$
\includegraphics[width=\linewidth]{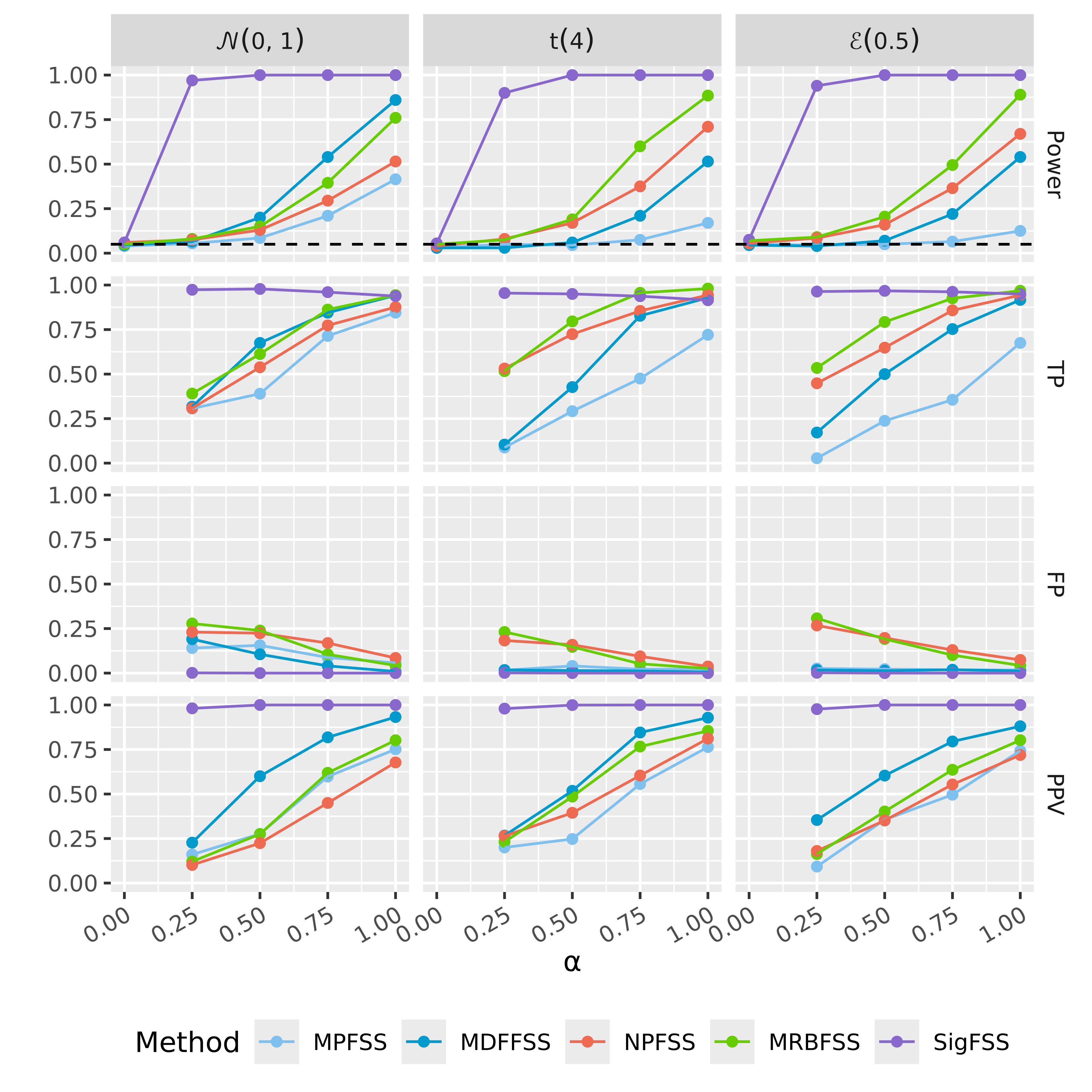}
\end{minipage}
\begin{minipage}{0.49\linewidth}
\centering $\Delta_2$
\includegraphics[width=\linewidth]{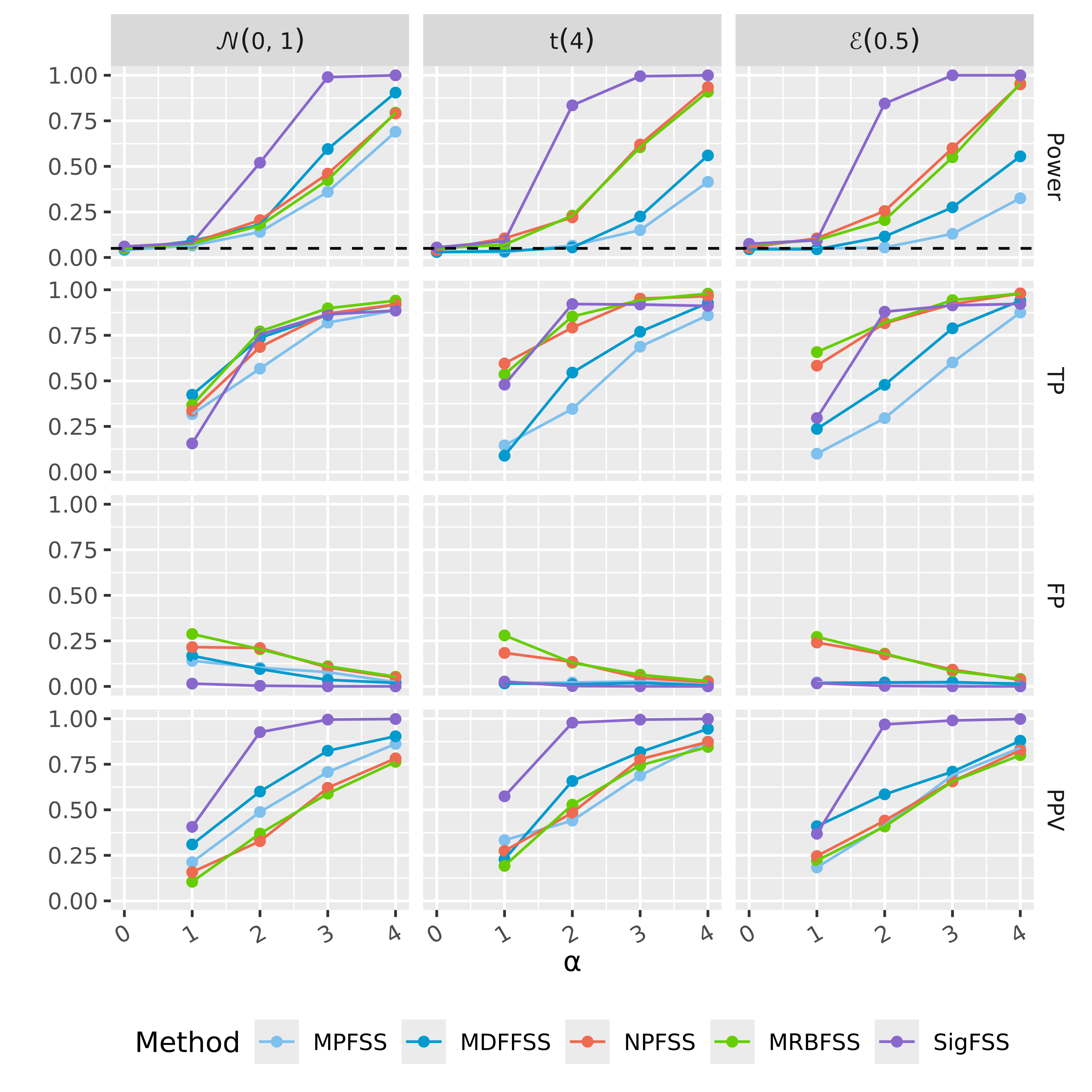}
\end{minipage}
\begin{minipage}{0.49\linewidth}
\centering $\Delta_3$
\includegraphics[width=\linewidth]{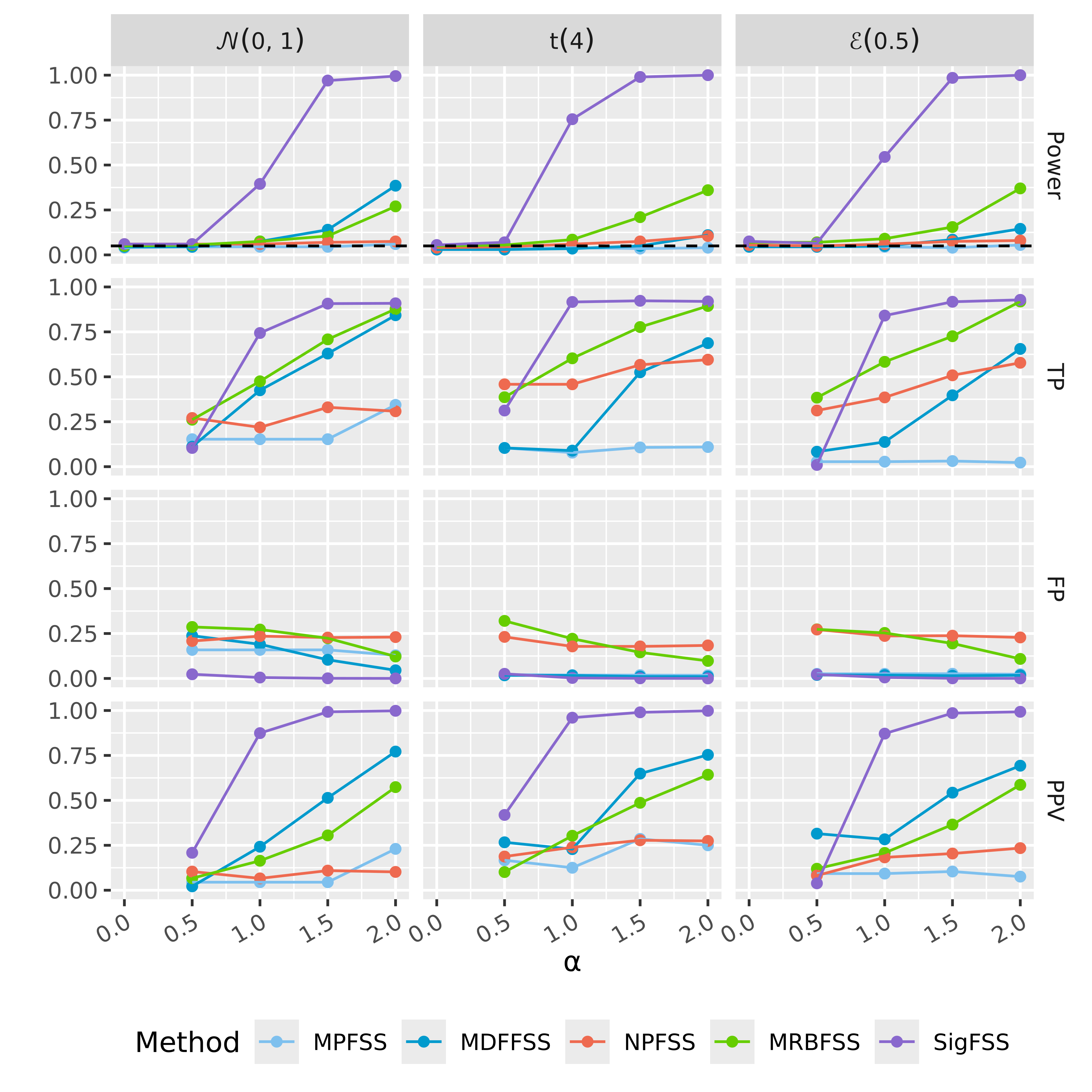}
\end{minipage}
\begin{minipage}{0.49\linewidth}
\centering $\Delta_4$
\includegraphics[width=\linewidth]{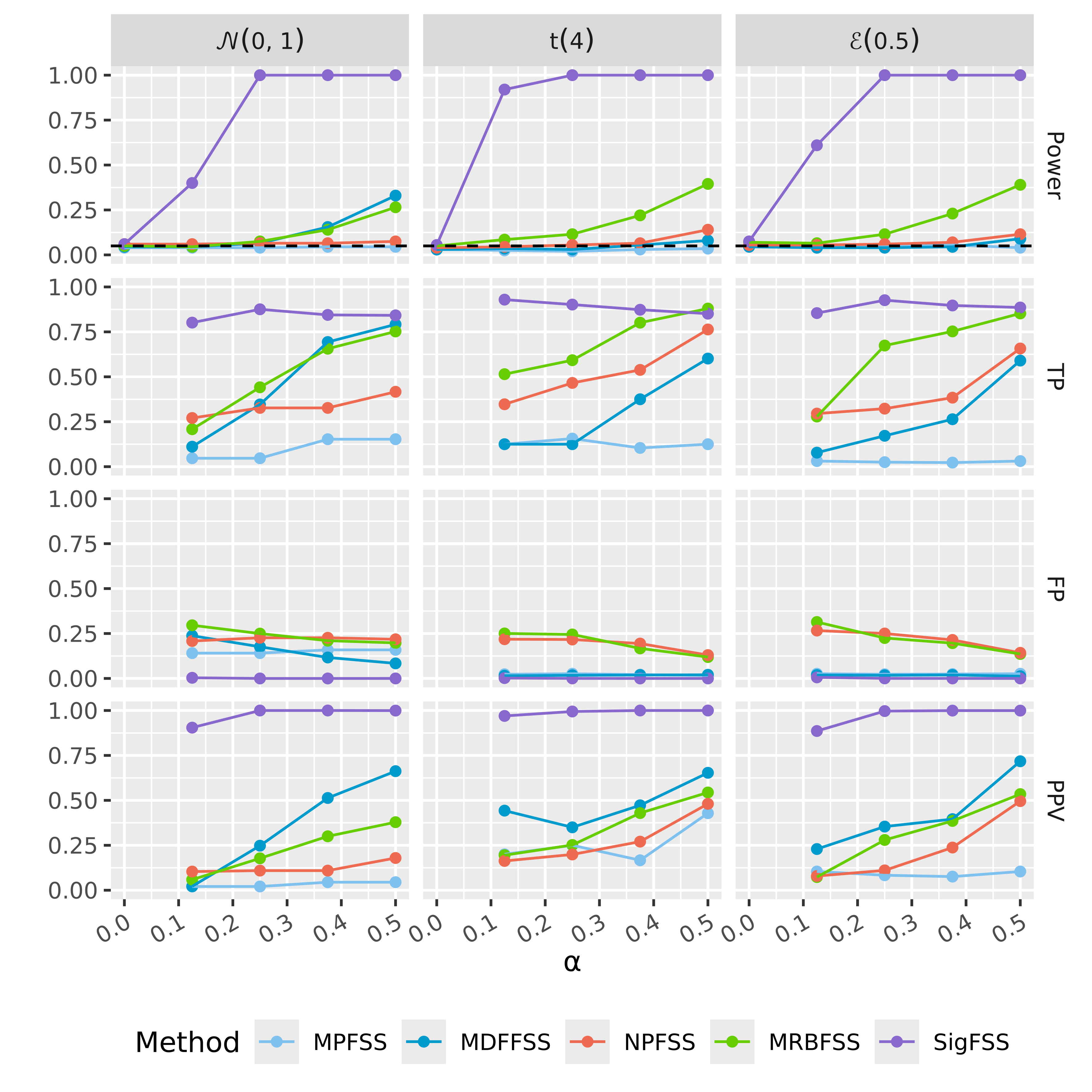}
\end{minipage}
\caption{The simulation study (multivariate case): a comparison of the SigFSS, NPFSS, MDFFSS, MRBFSS, and MPFSS methods for detection of the spatial cluster as the MLC, when $\rho = 0.2$ and considering a fixed threshold of 99\% for the cumulative inertia in the SigFSS. $\alpha$ is the parameter that controls the cluster intensity.}
\label{fig:rho0.2_99}
\end{figure}

\begin{figure}[H]
\begin{minipage}{0.49\linewidth}
\centering $\Delta_1$
\includegraphics[width=\linewidth]{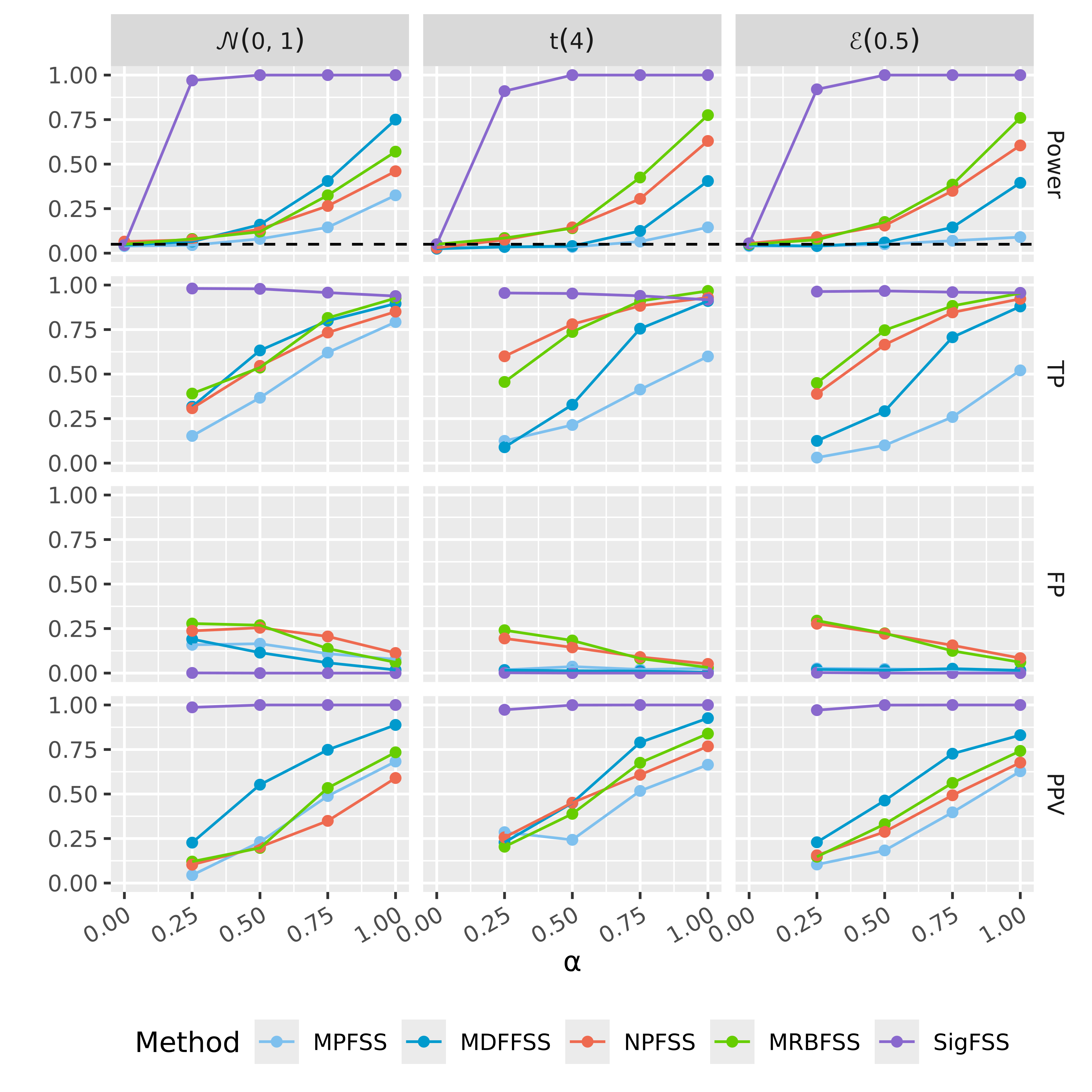}
\end{minipage}
\begin{minipage}{0.49\linewidth}
\centering $\Delta_2$
\includegraphics[width=\linewidth]{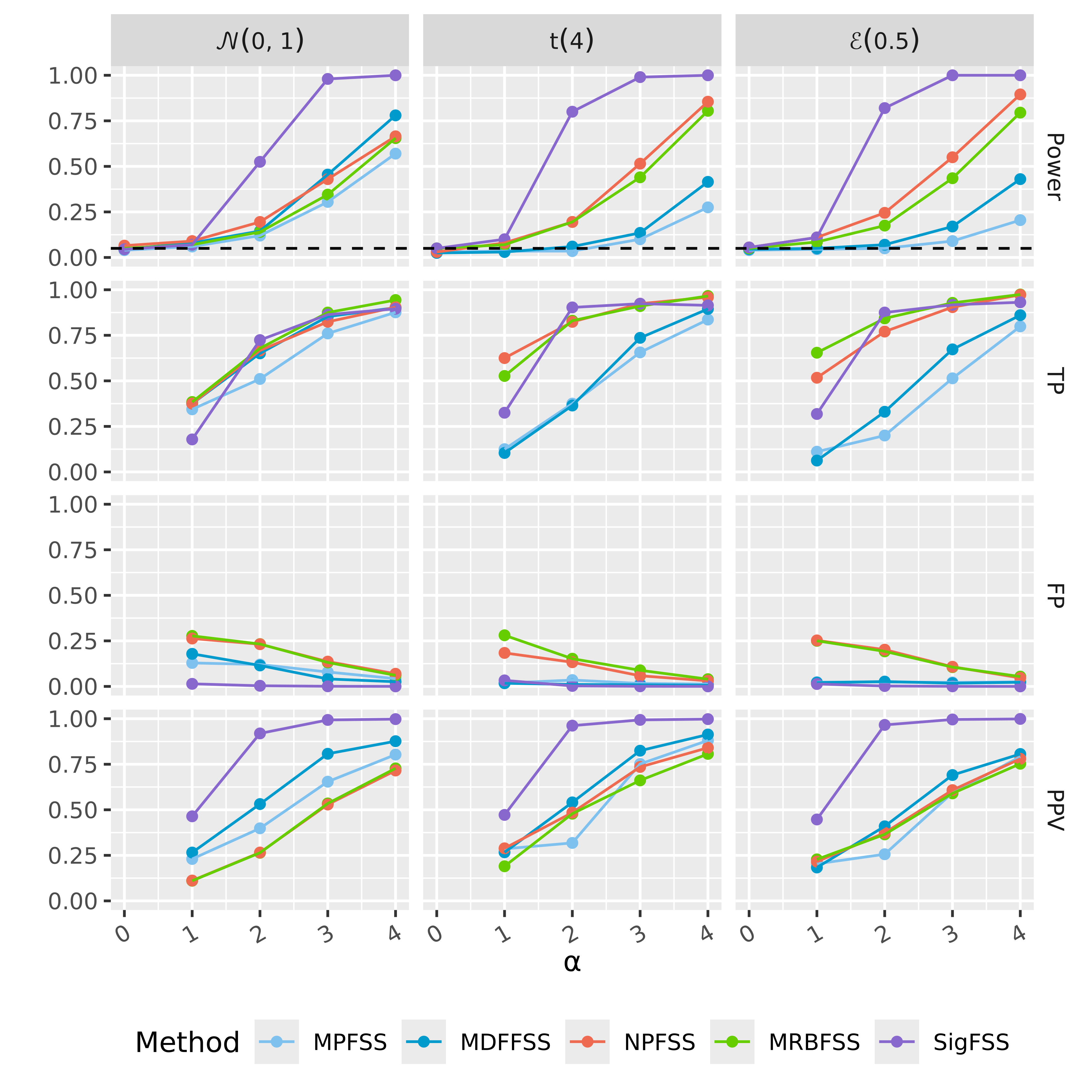}
\end{minipage}
\begin{minipage}{0.49\linewidth}
\centering $\Delta_3$
\includegraphics[width=\linewidth]{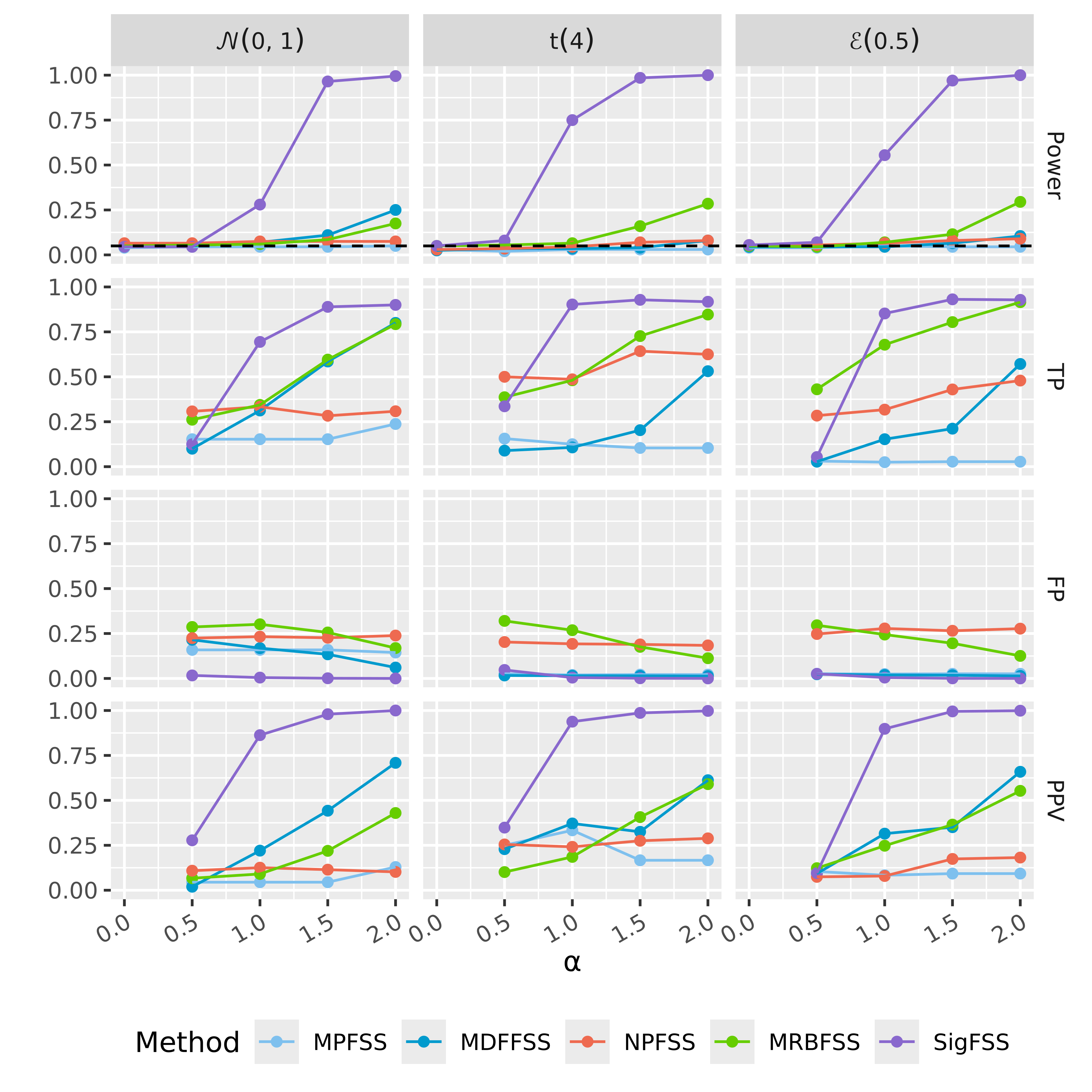}
\end{minipage}
\begin{minipage}{0.49\linewidth}
\centering $\Delta_4$
\includegraphics[width=\linewidth]{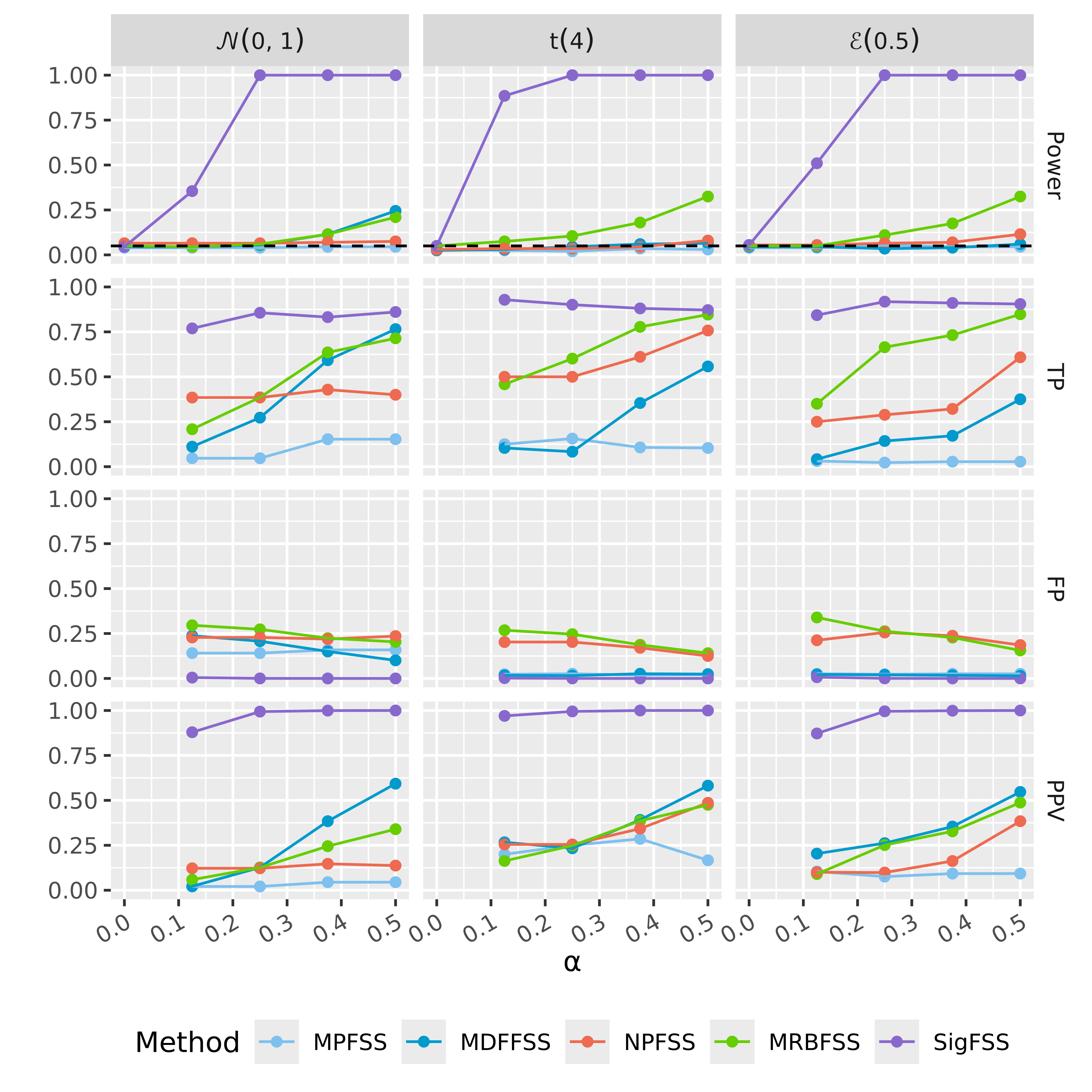}
\end{minipage}
\caption{The simulation study (multivariate case): a comparison of the SigFSS, NPFSS, MDFFSS, MRBFSS, and MPFSS methods for detection of the spatial cluster as the MLC, when $\rho = 0.5$ and considering a fixed threshold of 99\% for the cumulative inertia in the SigFSS. $\alpha$ is the parameter that controls the cluster intensity.}
\label{fig:rho0.5_99}
\end{figure}

\begin{figure}[H]
\begin{minipage}{0.49\linewidth}
\centering $\Delta_1$
\includegraphics[width=\linewidth]{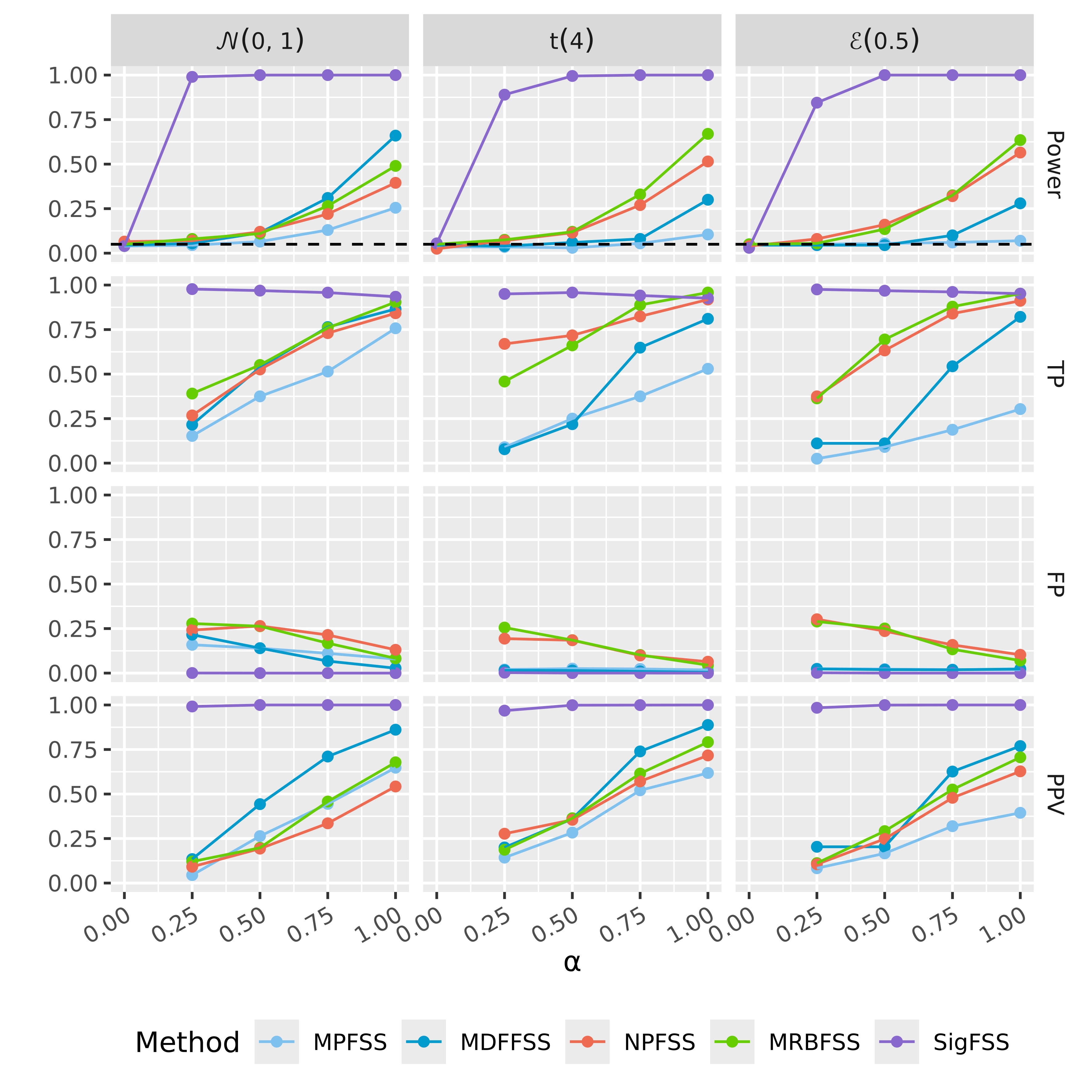}
\end{minipage}
\begin{minipage}{0.49\linewidth}
\centering $\Delta_2$
\includegraphics[width=\linewidth]{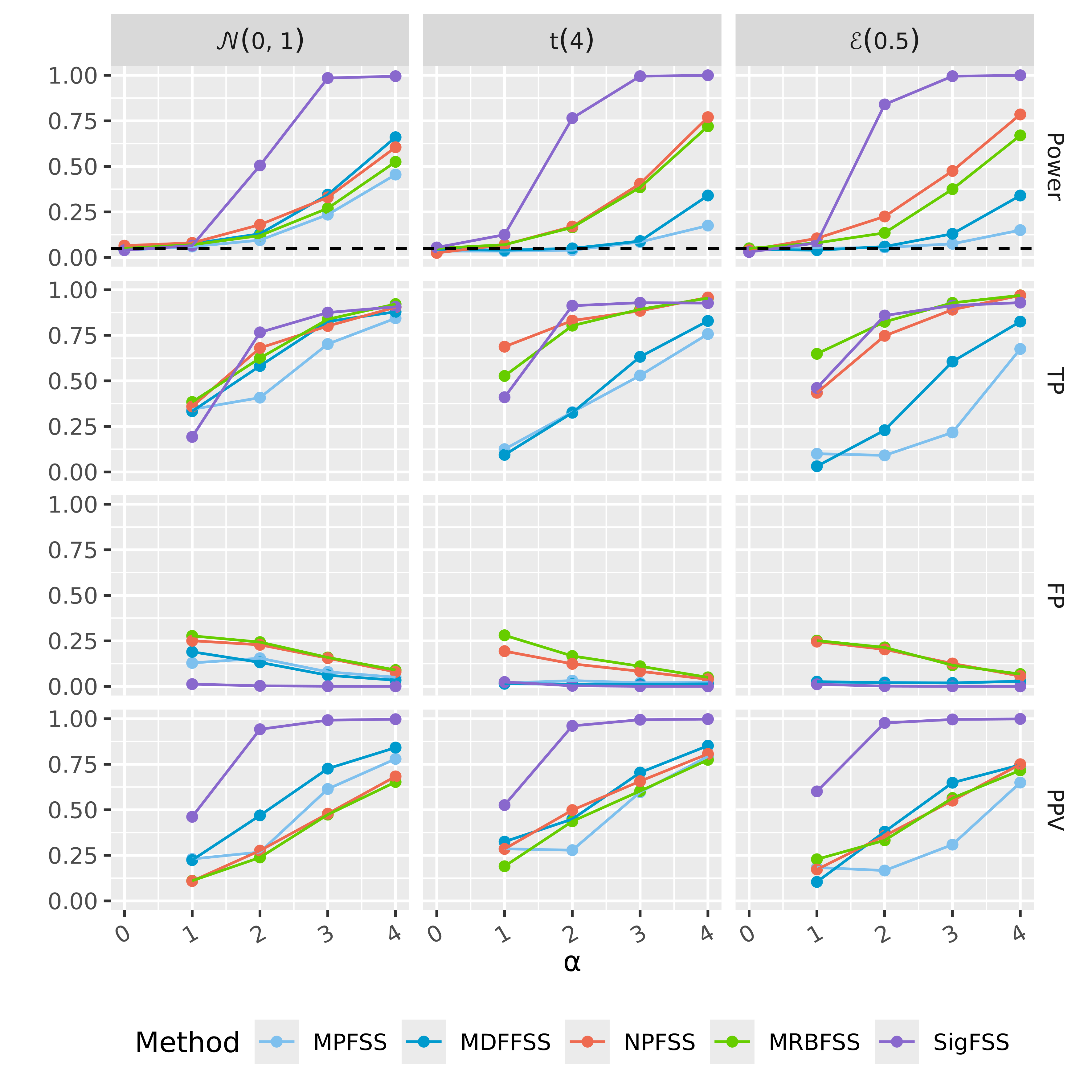}
\end{minipage}
\begin{minipage}{0.49\linewidth}
\centering $\Delta_3$
\includegraphics[width=\linewidth]{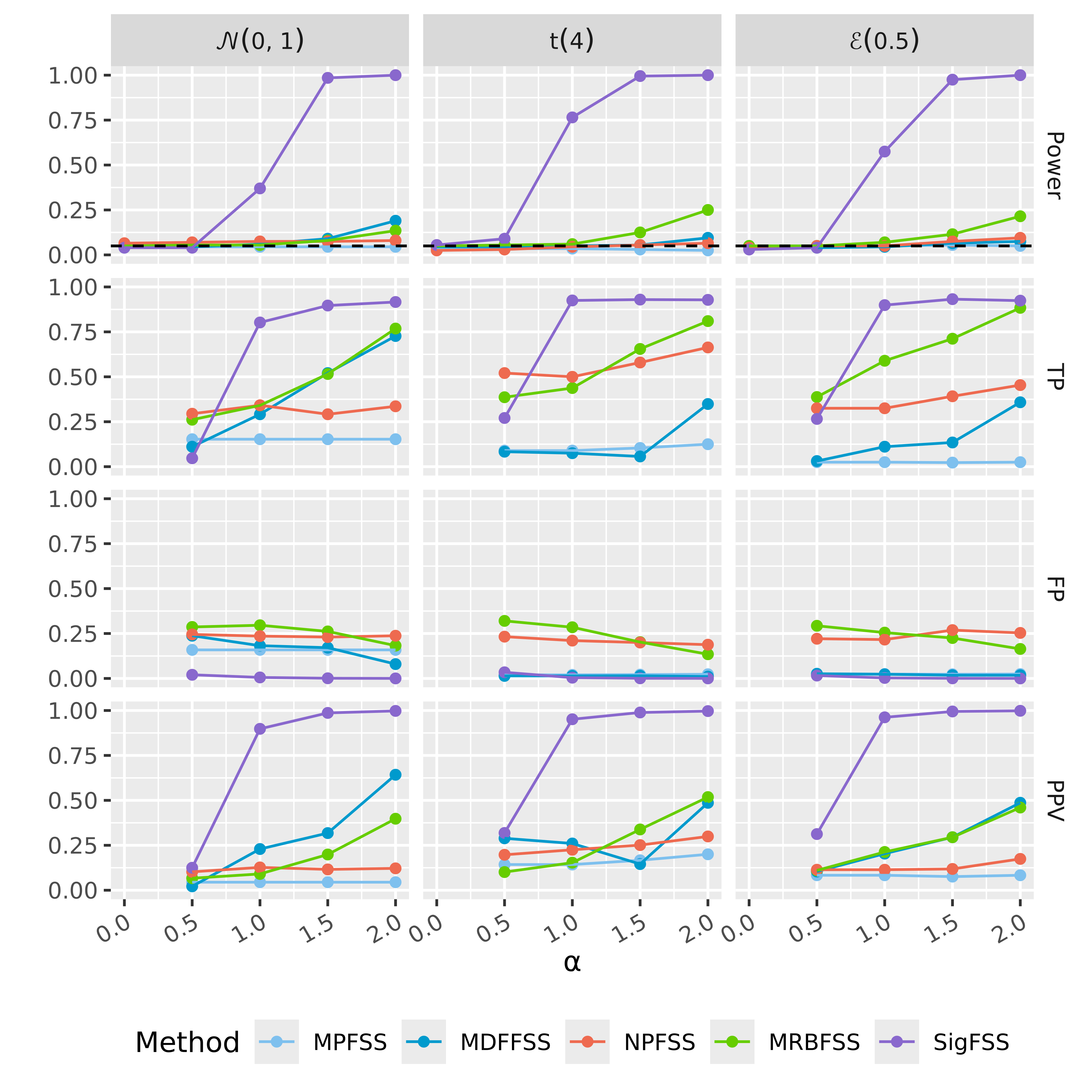}
\end{minipage}
\begin{minipage}{0.49\linewidth}
\centering $\Delta_4$
\includegraphics[width=\linewidth]{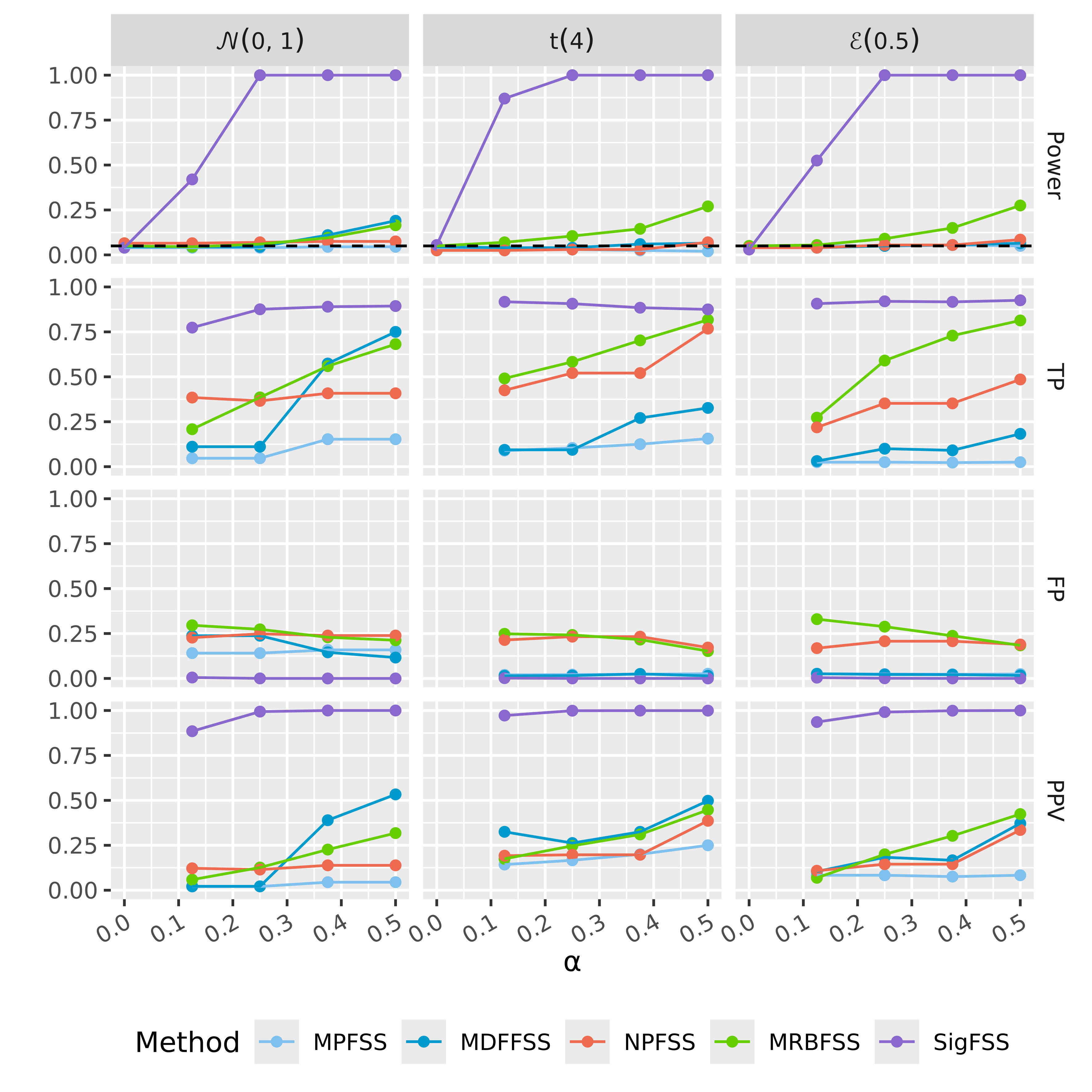}
\end{minipage}
\caption{The simulation study (multivariate case): a comparison of the SigFSS, NPFSS, MDFFSS, MRBFSS, and MPFSS methods for detection of the spatial cluster as the MLC, when $\rho = 0.8$ and considering a fixed threshold of 99\% for the cumulative inertia in the SigFSS. $\alpha$ is the parameter that controls the cluster intensity.}
\label{fig:rho0.8_99}
\end{figure}

\section{Supplementary materials for the application}

\begin{figure}[H]
\centering
\begin{minipage}{0.32\linewidth}
\includegraphics[width=\linewidth]{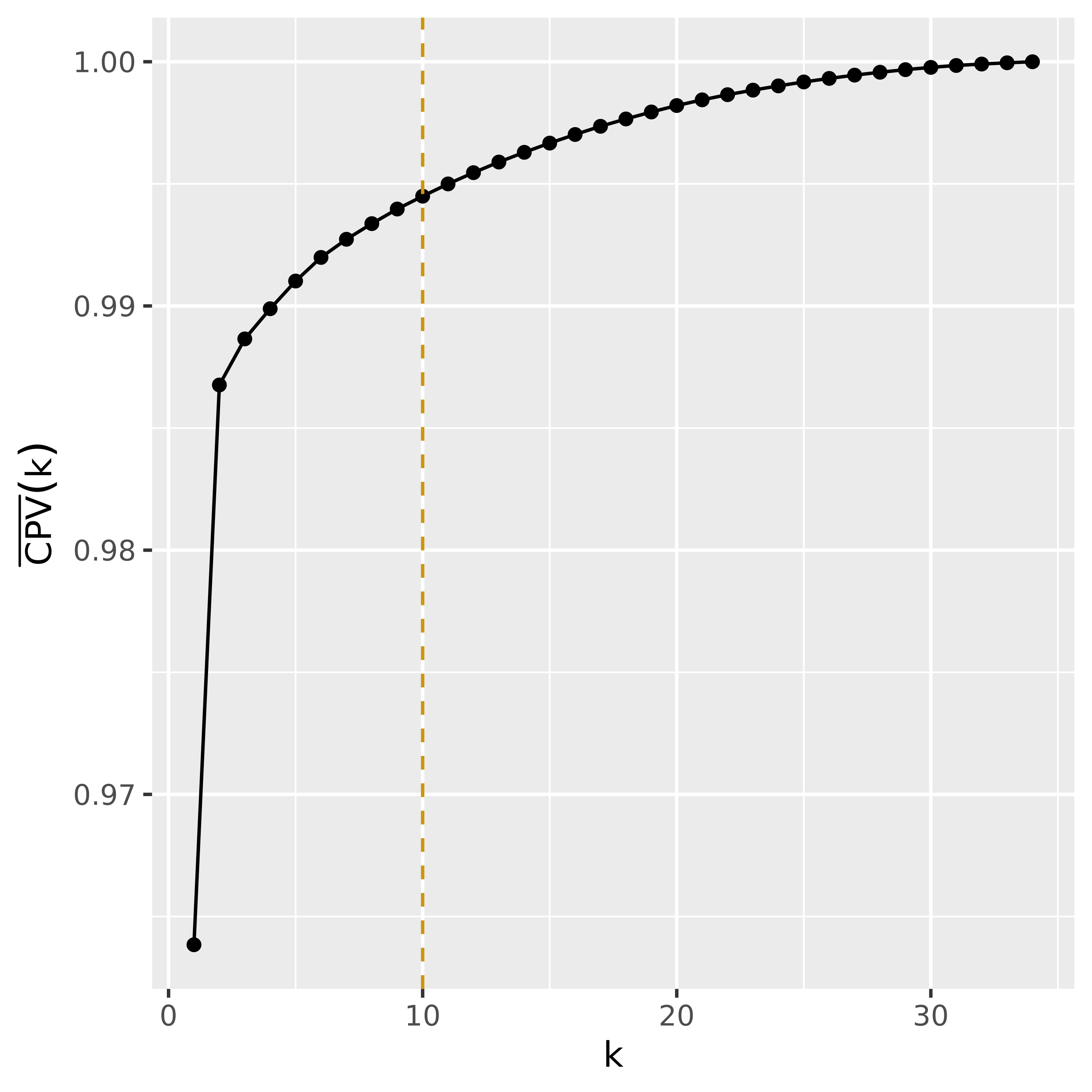}
\end{minipage} \hfill
\begin{minipage}{0.32\linewidth}
\includegraphics[width=\linewidth]{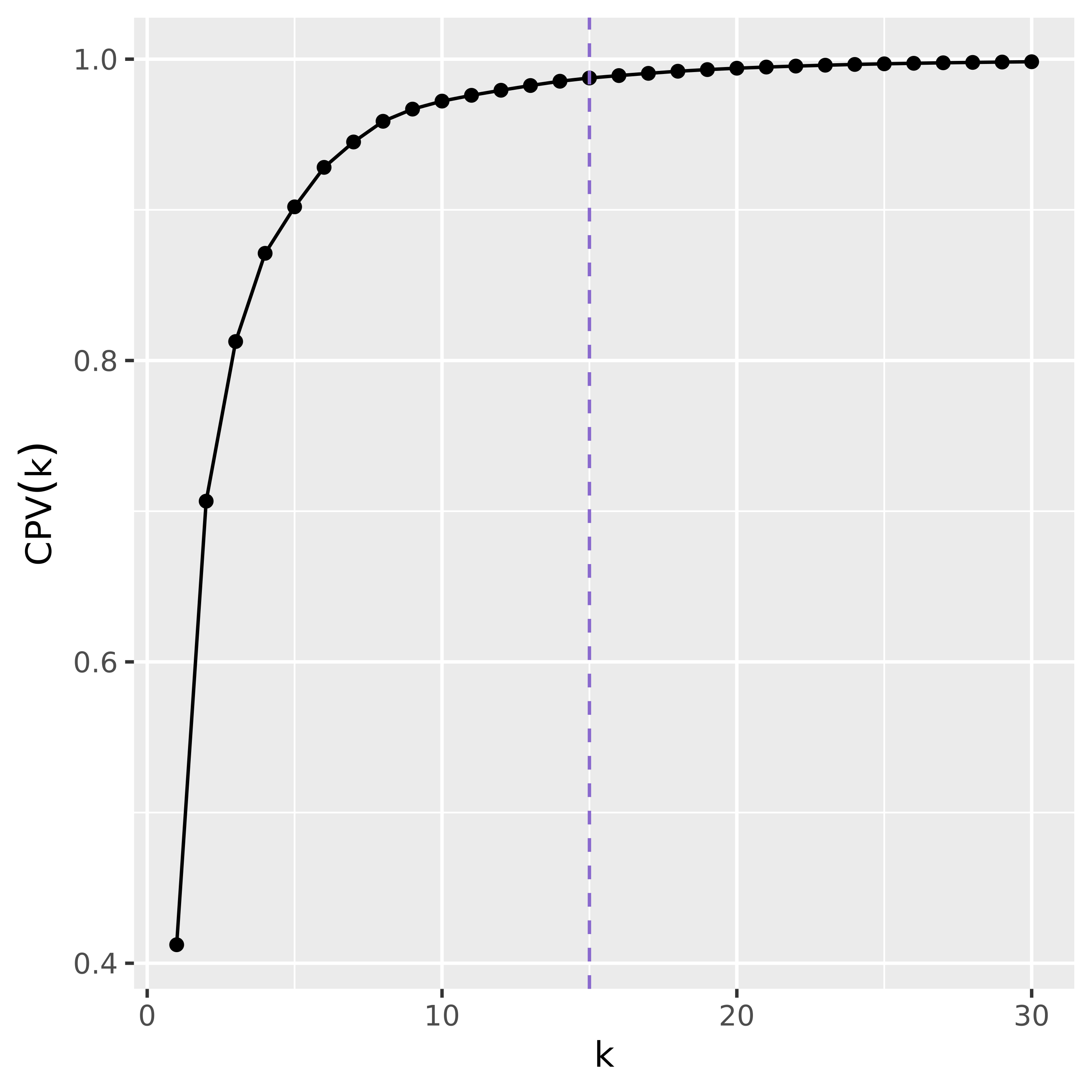}
\end{minipage} \hfill
\begin{minipage}{0.32\linewidth}
\includegraphics[width=\linewidth]{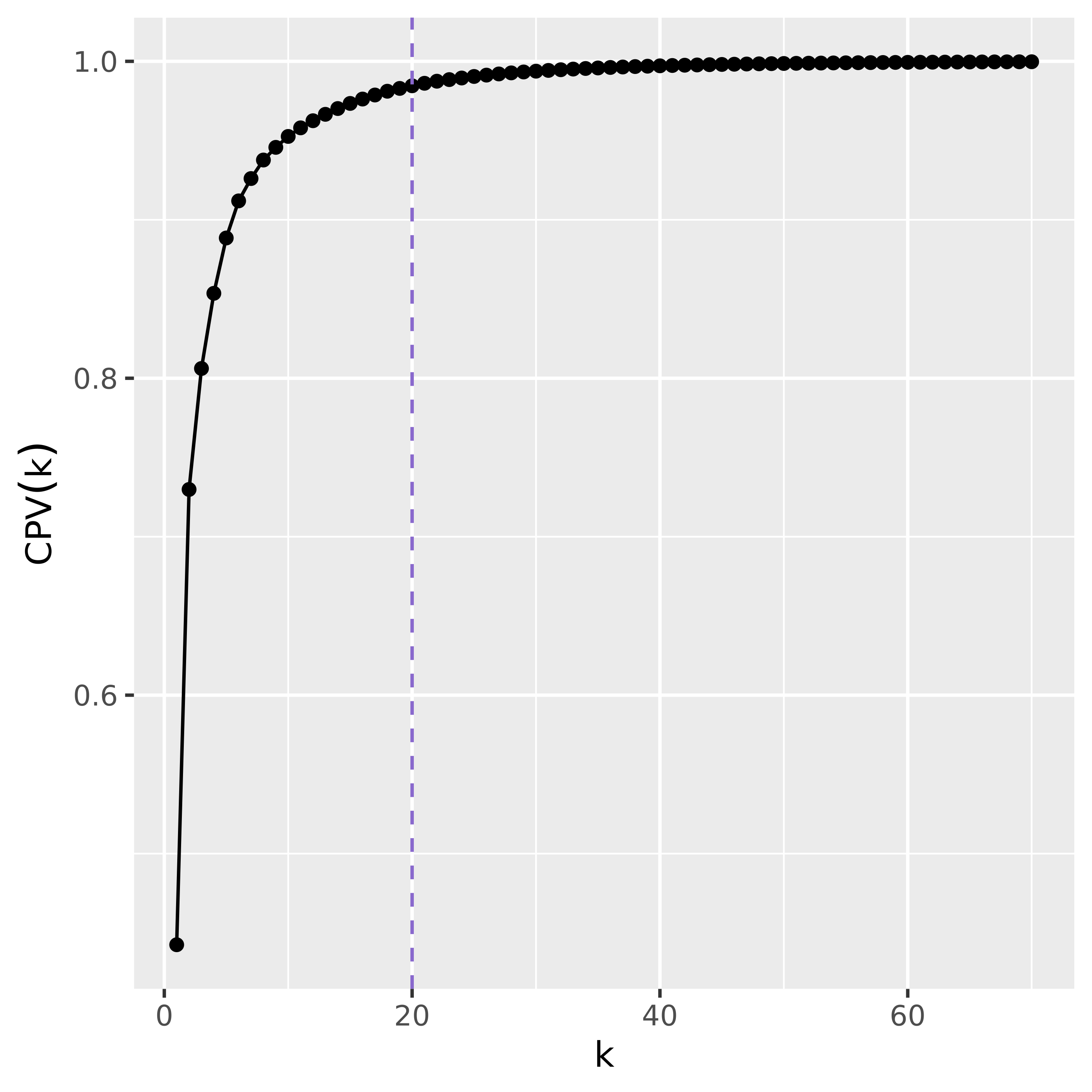}
\end{minipage}
\caption{The cumulative inertia for the choice of $K$ in the HFSS (left panel), in the SigFSS for the univariate case (middle panel), and in the SigFSS for the multivariate case (right panel). The selected value is highlighted by a vertical line.}
\label{fig:choiceK}
\end{figure}

\begin{table}[H]
\caption{Statistically significant spatial clusters with a higher or lower mortality rate.}
\label{tab:clusteruni_supp}
\centering
\begin{tabular}{cccc}
\hline
Method & Cluster & Number of \textit{départements} & p-value \\ \hline
HFSS (95\% for the cumulative inertia) & Most likely cluster & 37 & 0.001 \\
                                    & Secondary cluster & 9 & 0.001 \\ \hline
HFSS (99\% for the cumulative inertia) & Most likely cluster & 35 & 0.001 \\
                                    & Secondary cluster & 8 & 0.001 \\ \hline
SigFSS (95\% for the cumulative inertia) & Most likely cluster & 31 & 0.002 \\
                                    & Secondary cluster & 4 & 0.002 \\ \hline
SigFSS (99\% for the cumulative inertia) & Most likely cluster & 4 & 0.003 \\
                                    & Secondary cluster & 23 & 0.010 \\ \hline
\end{tabular}
\end{table}

 \begin{figure}[H]
 \centering
 \includegraphics[width=\linewidth]{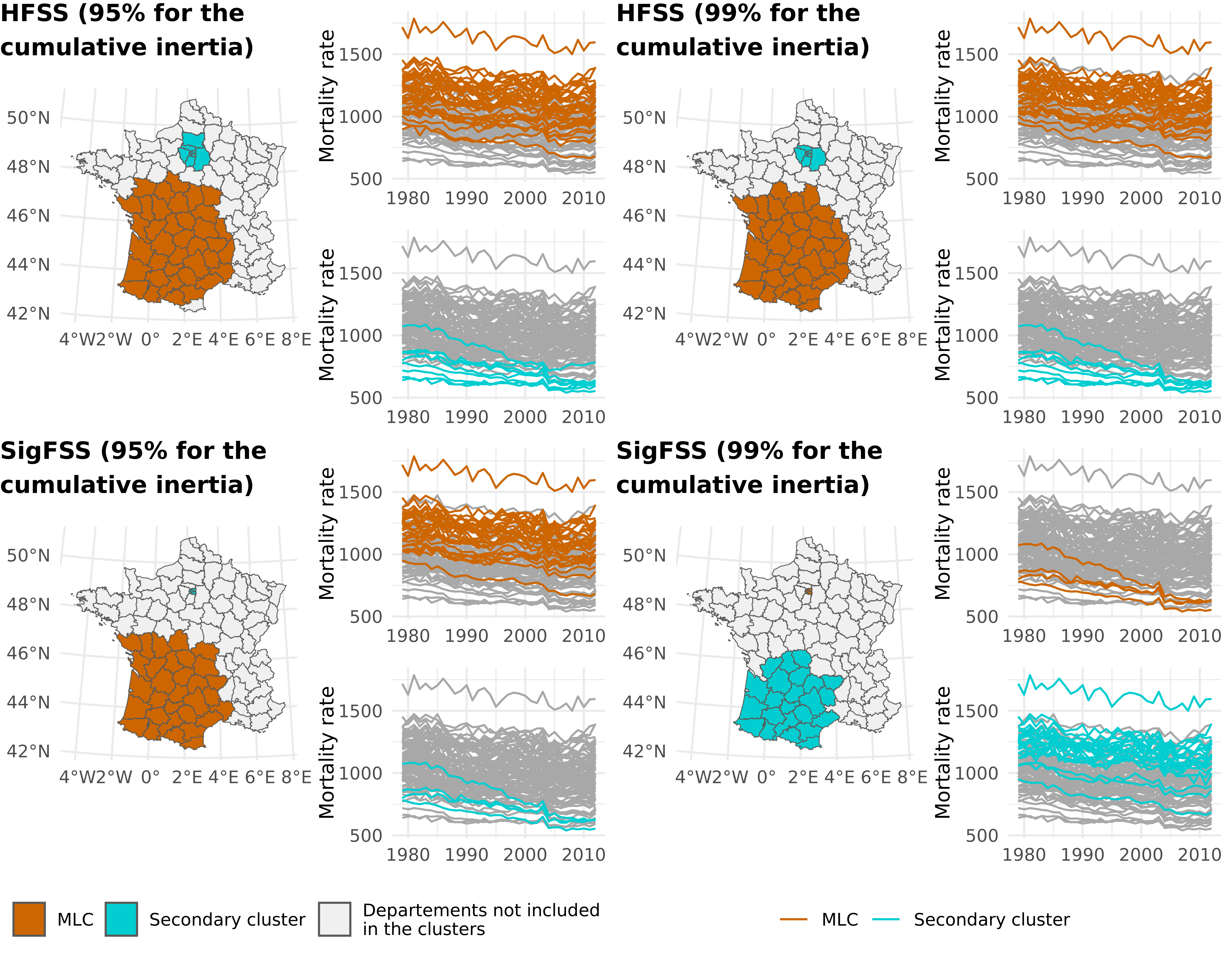}
 \caption{Statistically significant spatial clusters with a higher or lower mortality rate (per 100,000 people).}
 \label{fig:clusterunisupp}
 \end{figure}

\begin{table}[H]
\caption{Statistically significant spatial clusters of abnormal mortality rate due to circulatory system diseases and respiratory diseases.}
\label{tab:clustermulti_supp}
\centering
\begin{tabular}{cccc}
\hline
Method & Cluster & Number of \textit{départements} & p-value \\ \hline
SigFSS (95\% for the cumulative inertia) & Most likely cluster & 19 & 0.002 \\
                                    & Secondary cluster & 13 & 0.004 \\ \hline
SigFSS (99\% for the cumulative inertia) & Most likely cluster & 8 & 0.002 \\
                                    & Secondary cluster & 25 & 0.004 \\ \hline
\end{tabular}
\end{table}

 \begin{figure}[H]
 \centering
 \includegraphics[width=\linewidth]{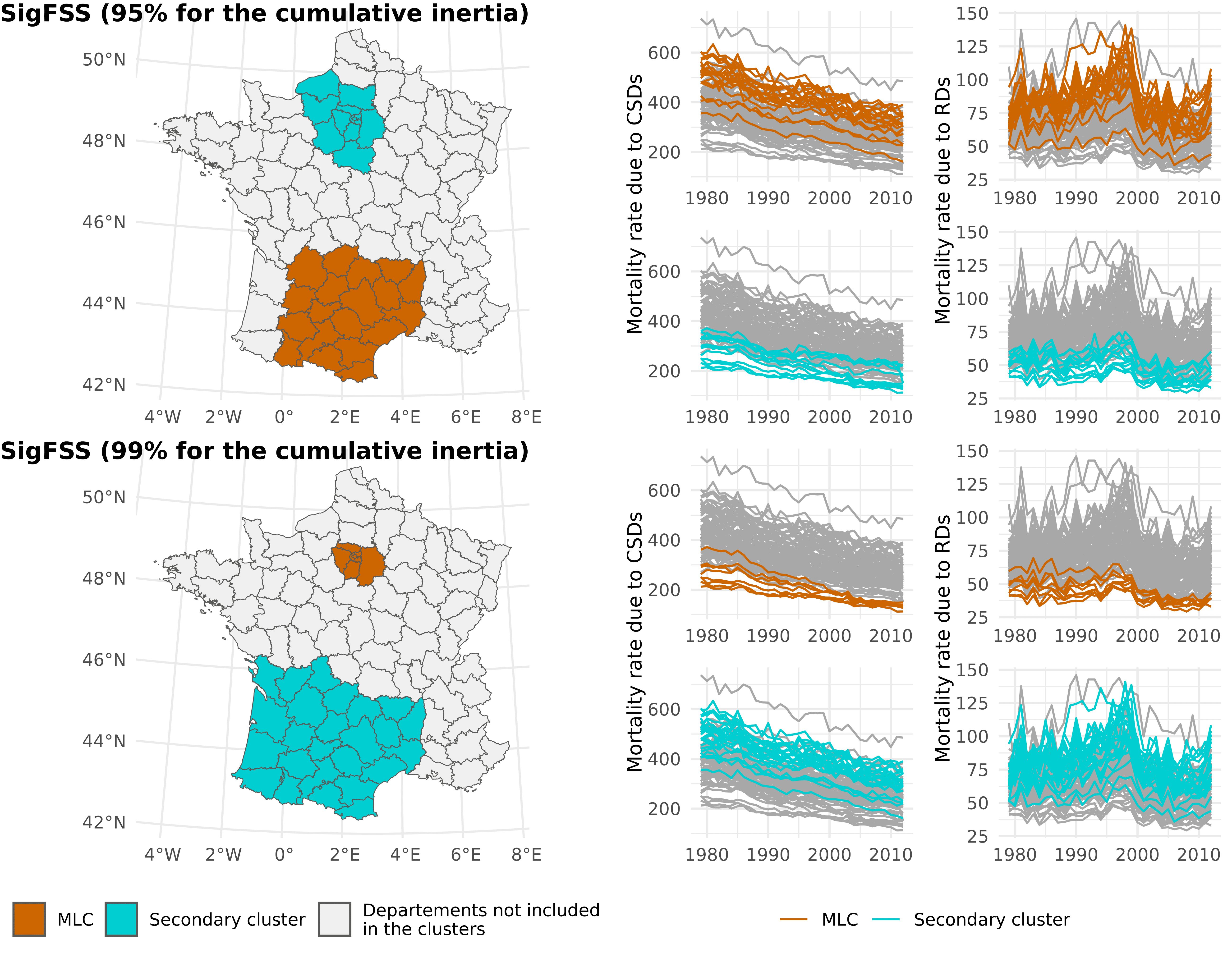}
 \caption{Statistically significant spatial clusters of abnormal mortality rate (per 100,000 people) due to circulatory system diseases (CSDs) and respiratory diseases (RDs).}
 \label{fig:clustermultisupp}
 \end{figure}

\end{document}